\documentclass[11pt,letterpaper]{article}
\usepackage{jheppub}

\usepackage{graphicx}
\usepackage{bbm}
\usepackage{amsmath}
\usepackage{amssymb}
\usepackage{mathtools}
\usepackage{amsthm}
\usepackage{mathrsfs}
\usepackage{marvosym}
\usepackage{dsfont}
\usepackage[labelformat=simple]{subcaption}
\usepackage{xcolor}
\usepackage{braket}
\usepackage{cleveref}
\usepackage{comment}
\usepackage{float}
\usepackage{fancybox}
\usepackage{slashed}
\usepackage[skins,theorems]{tcolorbox}
\tcbset{highlight math style={enhanced,
		colframe=black,colback=white,arc=0pt,boxrule=1pt}}
\usepackage{overpic}
\usepackage{booktabs}

\allowdisplaybreaks

\definecolor{dark-gray}{gray}{0.20}
\definecolor{gray}{gray}{0.30}
\definecolor{light-gray}{gray}{0.80}
\definecolor{dark-red}{rgb}{0.7,0,0}
\definecolor{dark-green}{rgb}{0.1,0.4,0}
\definecolor{dark-blue}{rgb}{0.3,0.3,0.7}
\definecolor{light-blue}{rgb}{0.8,0.8,1}
\definecolor{swamp}{RGB}{240, 199, 197}

\newcommand{\be}{\begin{equation}}
	\newcommand{\ee}{\end{equation}}

\def\be{\begin{equation}}
	\def\ee{\end{equation}}
\def\bea{\begin{eqnarray}}
	\def\eea{\end{eqnarray}}

\newcommand{\beq}{\begin{equation}}  \newcommand{\eeq}{\end{equation}}
\newcommand{\bal}{\begin{aligned}}   \newcommand{\eal}{\end{aligned}}
\def\beqa{\begin{eqnarray}}
	\def\eeqa{\end{eqnarray}}

\newcommand{\C}{\mathbb{C}}
\newcommand{\R}{\mathbb{R}}

\DeclareMathOperator{\sech}{sech}

\DeclareMathOperator{\arcosh}{arcosh}
\DeclareMathOperator{\arsinh}{arsinh}
\DeclareMathOperator{\artanh}{artanh}

\definecolor{thalassa}{rgb}{0.2823529411764706,0.6313725490196078,0.6274509803921569}

\DeclareMathOperator{\U}{U}
\DeclareMathOperator{\SL}{SL}
\DeclareMathOperator{\SO}{SO}

\numberwithin{equation}{section}

\def\simleq{\; \raise0.3ex\hbox{$<$\kern-0.75em
		\raise-1.1ex\hbox{$\sim$}}\; }
\def\simgeq{\; \raise0.3ex\hbox{$>$\kern-0.75em
		\raise-1.1ex\hbox{$\sim$}}\; }

\numberwithin{equation}{section}

\hypersetup{
	colorlinks=true,
	linkcolor=dark-blue,
	citecolor=dark-red,
	urlcolor=dark-green,
	linktoc=page
}

\theoremstyle{remark}

\newtheoremstyle{named}{}{}{\itshape}{}{\bfseries}{.}{.5em}{#3}
\theoremstyle{named}

\title{\centering  Complex Worldline Instantons in (Anti-)de Sitter Space}

\author{Alberto Castellano$^{1,2,3}$,} 
\author{Jinwei Chu$^{1}$,}
\author{Eliseu Kloster$^{2,4}$,}
\author{Jamie A.P. Law-Smith$^{1,2,4}$}
\affiliation{$^1$Enrico Fermi Institute \& Leinweber Institute for Theoretical Physics,\\
University of Chicago, Chicago, IL 60637, USA}
\affiliation{$^2$Kavli Institute for Cosmological Physics,\\
University of Chicago, Chicago, IL 60637, USA}
\affiliation{$^3$Jefferson Physical Laboratory, Harvard University,\\
Cambridge, MA 02138, USA}
\affiliation{$^4$Department of Astronomy \& Astrophysics,\\
University of Chicago, Chicago, IL 60637, USA}

\emailAdd{acastellanomora@fas.harvard.edu, jinweichu@uchicago.edu, ekloster@uchicago.edu, jamielawsmith@uchicago.edu}

\abstract{We study complex saddle points of the worldline path integral for massive charged particles in two-dimensional maximally symmetric backgrounds threaded by homogeneous electric fields. Within the worldline formalism, we show that the semiclassical saddles generically require complexifying both the target spacetime and the proper-time parameter. The resulting solutions comprise two slices of a single complex family, distinguished by the direction of proper-time evolution. The one evolving in imaginary proper time carries an action that typically controls the exponential suppression of non-perturbative effects, while that evolving in real proper time may encode additional complex phases responsible for quantum interference between saddles. Treating this family as a whole, we give a prescription for constructing and classifying all topologically inequivalent instanton sectors, and we find that the worldline action depends only on their topological data, namely winding numbers around the compact directions and certain isolated punctures. Moreover, in Anti-de Sitter spacetime, the qualitative structure of the solutions is governed by the dimensionless ratio $R |eE|/m$ ($R$ being the curvature radius), which distinguishes sub-extremal, extremal, and super-extremal regimes. By contrast, no analogous distinction occurs in de Sitter---where the instantons are insensitive to this ratio---or in Minkowski spacetime, which lacks a curvature scale altogether. We also recast the action as a sum over residues of the relevant (non-conserved) conjugate momentum, making transparent which analytic data control each contribution. Finally, we discuss the subtleties that arise when solutions built in various local patches are lifted to global (Anti-)de Sitter space. These impose further necessary conditions on which saddles can in principle contribute to the one-loop path integral.
}

\begin{document}

	\makeatletter
	\let\old@fpheader\@fpheader
	\renewcommand{\@fpheader}{\vspace*{-0.1cm} \hfill EFI-26-5}
	\makeatother
	\maketitle
	\setcounter{page}{1}
	\pagenumbering{roman}

\newcommand{\remove}[1]{\textcolor{red}{\sout{#1}}}

\newpage
\pagenumbering{arabic} 

\section{Introduction}
\label{s:intro}

Semiclassical methods are one of the few systematic windows into the non-perturbative structure of quantum theories.
In the standard approach, one analytically continues the theory to Euclidean signature, finds real solutions of the Euclidean equations of motion with finite action (the \emph{instantons}), and expands the path integral around them. Tunneling, vacuum decay, and particle production are controlled by these instantons, weighted by $e^{-S_E}$, where $S_E$ is the Euclidean action \cite{Coleman:1977py,Callan:1977pt,Coleman:1985rnk}. The archetype of this paradigm is the Schwinger effect \cite{Sauter:1931zz,Heisenberg:1936nmg,Weisskopf:1936hya,Schwinger:1951nm}: a background electric field $E$ destabilizes the vacuum and induces the production of particle--antiparticle pairs of mass $m$ and charges $\pm e$, at a rate proportional to $e^{-\pi m^2/|eE|}$. This effect is invisible to all orders in perturbation theory.
The worldline formulation of quantum field theory \cite{Feynman:1950ir,Strassler:1992zr,Schubert:2001he} makes its instanton origin transparent: the one-loop effective action of a charged field in a background gauge field becomes a sum over closed paths of a relativistic point particle, and its non-perturbative part is dominated by the closed classical trajectories (the \emph{worldline instantons}) \cite{Affleck:1981bma,Dunne:2005sx,Dunne:2006st}. For a constant field, these are circles in the Euclidean plane, and their action reproduces the exponent of Schwinger's formula.

However, real Euclidean saddles do not contain all of the semiclassical data of a path integral \cite{Witten:2010zr,Witten:2010cx,Cherman:2014ofa,Tanizaki:2014xba,Behtash:2015zha,Behtash:2015loa,Alexandru:2020wrj}.
Already in flat space, pair production in more general time-dependent fields is governed by worldline instantons that are intrinsically complex: they are not Wick rotations of real trajectories, and their actions have both an exponential suppression and the phases responsible for quantum interference between saddles \cite{Dumlu:2011cc,Ilderton:2015lsa}.
Beyond a constant field in flat space, two basic questions remain open. What is the complete set of complex saddles of the worldline path integral? And what determines their action?

\smallskip

In this paper we address
both questions for a massive charged particle in two-dimensional maximally symmetric spacetimes threaded by a homogeneous electric field: Minkowski space ($\R^{1,1}$), Anti-de Sitter space (AdS$_2$), and de Sitter space (dS$_2$). Besides being the simplest non-trivial settings that remain exactly solvable, these backgrounds are of physical interest. AdS$_2\times \mathbf{S}^2$ with an electromagnetic flux is the near-horizon geometry of an extremal charged black hole, and dS$_2 \times \mathbf{S}^2$ is the Nariai limit of a charged black hole in de Sitter space---in both cases the gauge field felt by a probe particle comes from the black hole charge \cite{Gibbons:1975kk}.
dS$_2$ is also the simplest arena in which to study pair production in an expanding universe \cite{Garriga:1994bm,Frob:2014zka,Kobayashi:2014zza}. Exact pair-production rates and one-loop effective actions have been computed in AdS$_2$ and dS$_2$ \cite{Garriga:1994bm,Comtet:1984mm,Villalba:1995za,Pioline:2005pf,Kim:2008xv,Kim:2011hu,Cai:2014qba}. In AdS$_2$ the vacuum is unstable only if the particle is super-extremal, i.e., if its charge-to-mass ratio measured in units of the AdS radius is greater than one, $R|eE|/m>1$.\footnote{The terminology of sub-extremal, extremal, and super-extremal is borrowed from the black-hole literature. For an extremal 4D Reissner--Nordstr\"om black hole, the near-horizon geometry is AdS$_2\times \mathbf{S}^2$ with $R=r_+$ and $E=Q/r_+^2$, so that $R|eE|/m=|eQ|/(m r_+)$ equals the particle's charge-to-mass ratio in units of the black hole's extremal value.}
This is the worldline counterpart of the statement that only super-extremal particles can discharge an extremal black hole \cite{Gibbons:1975kk}, which underlies the Weak Gravity Conjecture \cite{Arkanihamed:2006dz,Harlow:2022gzl}.

The sub-extremal regime produces a puzzle. States propagating in the supersymmetric near-horizon geometry of a BPS black hole are never super-extremal \cite{Castellano:2025yur,Castellano:2025rvn}.
These particles admit no real Euclidean worldline instanton, since the would-be trajectories run off to the conformal boundary of AdS$_2$, so the standard semiclassical analysis predicts no non-perturbative correction at all.
However, the exact one-loop effective action of these particles does contain exponentially small corrections, whose exponents are set by the electric and magnetic charges of the probe \cite{Castellano:2026ojb}; the possibility that they originate from complex saddles of the worldline path integral was raised in \cite{Castellano:2025yur}.
Resolving this puzzle requires knowing all of the saddles.

\medskip

The complex extension of the worldline instanton method \cite{Affleck:1981bma,Dunne:2005sx,Dunne:2006st,Dunne:2006ur} is by now well developed in flat space.
For suitable time-dependent fields the instantons become complex and interfere with one another \cite{Dumlu:2011cc}; for fields with lightlike inhomogeneities the pair-production exponent is itself a contour integral along the complex instanton, evaluated by residues at the poles it encircles \cite{Ilderton:2014mla,Ilderton:2015lsa}; at finite temperature, or in compact directions, the instantons acquire winding numbers \cite{Medina:2015qzc,Brown:2015kgj,Gould:2017fve}; and instantons with negative Euclidean action were found early on in strong backgrounds \cite{Lavrelashvili:1989he}.
Closely related ideas, in which the time contour is deformed into the complex plane, underlie several recent approaches to tunneling in real time \cite{Turok:2013dfa,Cherman:2014sba,Bramberger:2016yog,Andreassen:2016cff,Ai:2019fri}.
In curved spacetime, by contrast, semiclassical treatments of pair production have relied on real Euclidean worldline instantons \cite{Castellano:2025yur,Pioline:2005pf} or on exact mode functions and WKB phase integrals \cite{Kim:2008xv,Kim:2015azt}; complex worldline instantons for particle production by space- and time-dependent gravitational fields have been constructed only recently \cite{Semren:2025dix}, and complex geodesics are known to control correlation functions in de Sitter and Nariai backgrounds \cite{Aalsma:2026kqj,Aalsma:2022eru}.
What is missing is a systematic construction of the complete set of complex saddles in a given background, and an understanding of what determines their action.

\smallskip

We carry out this construction for a relativistic particle of mass $m$ and charge $e$, described by the worldline action (\ref{eq:worldlineAction}) in the proper-time gauge, whose closed solutions we refer to as instantons. Three general lessons emerge.
First, the relevant saddles are complex in two ways: the closed trajectories live in the complexified spacetime, and the worldline parameter $\tau$ must itself be allowed to take complex values. Solutions evolving in imaginary proper time (which we refer to as the \emph{instanton phase}) and solutions evolving in real proper time (the \emph{interference phase}) turn out to be just two slices of a single complex family. Neither is complete on its own---in Minkowski space the real-time solutions are open hyperbolic motions, whereas in AdS$_2$ the Euclidean trajectories of a sub-extremal particle fail to close. Only analytic continuation to the full complex proper time plane unifies them.

\begin{figure}
    \centering
    \subcaptionbox{\label{sfig:AdSCylinder}}{
		\includegraphics[scale=0.6]{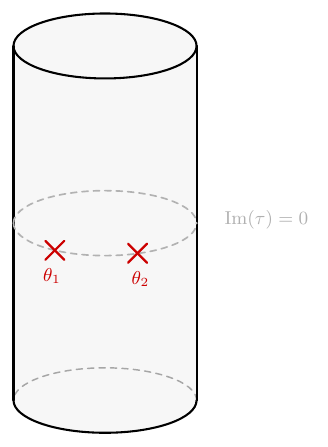}}
        \hspace{0.6cm}
        \subcaptionbox{\label{sfig:dSCylinder}}{\includegraphics[scale=0.6]{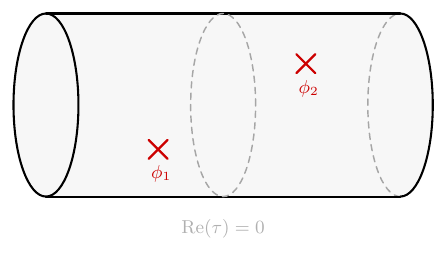}}
		\caption{\small Sketch of the proper-time cylinder for the $\textbf{(a)}$ sub-extremal AdS$_2$ and $\textbf{(b)}$ dS$_2$ instantons, in the Poincar\'e and flat-slicing patches respectively. The dashed circle marks the reference contour $\text{Im}(\tau)=0$ in $\textbf{(a)}$ and $\text{Re}(\tau)=0$ in $\textbf{(b)}$; the red crosses denote the punctures.}
    \label{fig:CylinderSketch}
\end{figure}

Second, once proper time is regarded as a complex coordinate, the construction and classification of all instantons becomes a topology problem. The solutions are periodic in $\tau$, with a period along the real or imaginary direction depending on the background, so that the proper-time plane is a cylinder. The maps from proper time to spacetime have isolated singular points, which we call \emph{punctures} (see Figure~\ref{fig:CylinderSketch}). Any closed contour on the punctured cylinder is mapped to a closed spacetime trajectory, i.e., to an instanton, and contours that can be deformed to one another yield the same action. The distinct instanton sectors are labeled by winding numbers around the compact direction and the punctures.
The worldline action depends on these integers alone, not on the shape of the contour or the integration constants of the solution.

%%%%%%%%%%%%%%%%%%%%%%%%%%%%%%%%%%%%%%%%%%%%%
%%%%%%%%%%%%%%%%%%%%%%%%%%%%%%%%%%%%%%%%%%%%%

Third, the same result can be obtained directly in spacetime. On shell, the action reduces to a contour integral of the non-conserved conjugate momentum, an algebraic function of the corresponding coordinate; in curved space its residues at the conformal boundary and at infinity reproduce the winding and the puncture terms, respectively, whereas in flat space the point at infinity supplies Schwinger's exponent. The residue computation makes transparent which analytic data control each contribution, but it requires a case-by-case bookkeeping of branches and orientations that the proper-time picture avoids. In this sense, the proper-time plane encodes the semiclassical data more economically than the spacetime trajectories themselves. The independence of the action from the integration constants has a symmetry origin as well: constant complex shifts of $\tau$ act on the complexified spacetime as certain SL(2,$\C$) transformations, i.e., as complexified isometries whose fixed points explain, as a byproduct, where the trajectories tend to localize.

\smallskip

In Minkowski space (Section~\ref{s:Minkowski}), which we use to introduce ideas and to review the known Euclidean and complex instantons \cite{Affleck:1981bma,Dunne:2005sx,Dumlu:2011cc,Ilderton:2015lsa}, the proper-time cylinder has no punctures and the classification reduces to a single winding number, so that the sectors are the familiar multi-instantons, each winding contributing Schwinger's $\pi m^2/|eE|$ to the imaginary part of the action. In AdS$_2$ (Section~\ref{s:AdS2}), the qualitative behavior of the solutions is governed by the ratio $R|eE|/m$, and the three regimes are related by analytic continuation. 
For sub-extremal particles (Section~\ref{ss:SubExtremal}), the compact direction is the real proper time axis: the closed complex trajectories are interference-phase solutions with a real action, i.e., pure phases with no Euclidean counterpart, while the Euclidean trajectories run off to the conformal boundary (in agreement with the absence of finite-action Euclidean instantons noted above).
This regime is the only one in which the action has a real part---consistent with the  real non-perturbative corrections found in the exact calculation of \cite{Castellano:2026ojb}.
For super-extremal particles (Section~\ref{ss:Superextremal}), the roles of real and imaginary proper time are exchanged: the Euclidean solutions become periodic with finite action, and the family containing the real Euclidean instanton reproduces the Schwinger exponent in AdS$_2$ of \cite{Pioline:2005pf}. 
At extremality (Section~\ref{ss:Extremal}), the period diverges, the cylinder collapses into the plane and only the punctures survive.
In dS$_2$ (Section~\ref{s:dS2}), the behavior is insensitive to the ratio $R|eE|/m$ and mirrors the super-extremal AdS$_2$ case through the analytic map relating the two backgrounds (cf.\ \eqref{eq:dS2AdS2Mapping}); the family containing the familiar instanton-phase solutions reproduces the pair-production exponent of \cite{Garriga:1994bm,Kim:2008xv}. In all cases, the exponent of the exactly known rate is recovered as the action of one particular sector, and higher windings reproduce the familiar multi-instanton contributions. The sectors that wind around the punctures, as well as the real action sectors of the sub-extremal regime, have to the best of our knowledge not appeared before.

The above analysis is carried out in Poincar\'e coordinates for AdS$_2$ and flat slicing for dS$_2$, which cover only a part of the spacetimes---one may ask whether a trajectory that closes in one patch also closes in the full geometry (Section~\ref{s:Global(A)dS}).
A subtlety then emerges: once the coordinates are complexified, the map from the patch to global coordinates becomes multi-valued: encircling a single branch point shifts the global time by $\pi$ and reflects the spatial coordinate as $\chi\to -\chi \pm i\pi$, while leaving the embedding coordinates of the hyperboloid unchanged. Hence, in AdS$_2$ the trajectory closes on the hyperboloid but not on its universal cover (where the global time is unwrapped). This already imposes some non-trivial conditions on what type of closed contours can be considered in the proper-time plane, which have no counterpart in the local (i.e., Poincar\'e) analysis. By contrast, in global de Sitter---which is actually defined by the hyperboloid---all possible windings are in principle allowed, since the corresponding trajectories obtained via the spacetime maps always close, regardless of the number and type of punctures enclosed. 

\medskip

We deliberately leave two natural questions for future work. First, the \emph{selection problem}: given an observable, which saddles actually contribute, and with what weights? Answering this requires embedding our classification into Picard--Lefschetz theory \cite{Witten:2010zr,Cherman:2014ofa,Tanizaki:2014xba,Behtash:2015zha,Behtash:2015loa}, analysis of the associated Stokes phenomena, and computation of the one-loop fluctuation determinants \cite{Dunne:2006st}; this is the subject of \cite{PicardLefschetz}.
Second: what is the physical meaning of the punctures? Their positions and weights determine so much of the classification, yet their interpretation in the original spacetime picture remains to be clarified; this is the subject of \cite{Letter}.
Further directions, including the resurgent organization of the instanton sectors \cite{Dunne:2015eaa,Aniceto:2018bis} and the extension to the four-dimensional geometries AdS$_2\times \mathbf{S}^2$ and dS$_2\times \mathbf{S}^2$ \cite{Chen:2023swn,Castellano:2025ljk,Castellano:2026pwj}, are briefly discussed in Section~\ref{s:Conclusions}.

\medskip

The paper is organized as follows. 
Section \ref{s:Minkowski} reviews the worldline formalism and introduces two tools used throughout the paper: closed contours on the complex proper-time cylinder (Section~\ref{ss:MixedSolutionsMink}) and the Hamiltonian constraint method together with the residue evaluation of the instanton action (Section~\ref{ss:HamiltonianConstraintMink}).
Sections \ref{s:AdS2}-\ref{s:Global(A)dS} treat AdS$_2$, dS$_2$, and their lift to global coordinates. Section \ref{s:Conclusions} concludes and provides an outlook.       Appendix \ref{ap:SL2C&ComplexShifts} relates the invariance of the action under complex proper-time shifts to SL(2,$\C$) transformations.

\smallskip

Table~\ref{tab:summary} summarizes the main results of the paper.
\begin{table}[htbp]
\centering
%\footnotesize
\begin{tabular}{l l l l l l}
\toprule
Background              & period              & punctures             & instanton action                  & exact results                    \\
\midrule
$\R^{1,1}$               & $2\pi i/|eE|$       & none                  & $-iw\pi m^2/|eE|$        & \cite{Schwinger:1951nm}          \\
AdS$_2$, sub-extremal   & $2\pi R^2/\beta$    & two, on real axis     & $-2\pi(w\beta + inq_e)$ & \cite{Castellano:2026ojb}        \\
AdS$_2$, extremal       &         $\infty$            & two, on real axis     & $-2\pi inq_e$           & \cite{Castellano:2026ojb}        \\
AdS$_2$, super-extremal & $2\pi i R^2/\kappa$ & two, off both axes    & $2\pi i(w\kappa-nq_e)$ & \cite{Pioline:2005pf,Kim:2008xv} \\
dS$_2$                  & $2\pi i R^2/\Omega$ & two, off both axes    & $-2\pi i(w\Omega-nq_e)$  & \cite{Garriga:1994bm,Kim:2008xv} \\
\bottomrule
\end{tabular}
\caption{
Summary of results (in local coordinates).
The period column lists the instanton periodicity along the compact direction of the proper-time cylinder.
Closed contours on this cylinder are labeled by winding numbers around the compact direction, $w$, and the punctures, $n$.
The winding number $w$ is defined to be positive in the direction of increasing proper time, i.e., increasing $\text{Re}(\tau)$ for sub-extremal AdS$_2$ and increasing $\text{Im}(\tau)$ otherwise, while $n$ is defined to be positive for counterclockwise orientation.
Each saddle contributes to the worldline path integral with weight proportional to $e^{iS_{\rm inst}}$.
The exact results column lists the known one-loop exponents (reproduced here as one particular sector).
In global coordinates each puncture splits into two independent branch points with half weight, for four branch points total.
Notation: $\tilde{m}=mR$, $q_e=eER^2$, $\beta=\sqrt{\tilde{m}^2-q_e^2}$, $\kappa=\sqrt{q_e^2-\tilde{m}^2}$, $\Omega=\sqrt{\tilde{m}^2+q_e^2}$.
}
\label{tab:summary}
\end{table}

\section{Instanton Solutions in $\R^{1,1}$}\label{s:Minkowski}

Throughout this work, we consider the classical trajectories of a relativistic point particle of mass $m$ and electric charge $e$, propagating along a one-dimensional worldline parametrized by $\tau \in \gamma$. Its corresponding action is
\begin{equation}\label{eq:worldlineAction}
 S_{\rm wl} = - \int_\gamma d\tau\, \frac{m^2 h(\tau)}{2}+\int_\gamma d\tau\, \frac{g_{\mu \nu}(x) \dot{x}^\mu \dot{x}^\nu}{2 h(\tau)} +e \int_\gamma d\tau\, \dot{x}^\mu A_\mu (x)\, ,
\end{equation}
where $A=A_\mu dx^\mu$ is the background one-form $\U(1)$ gauge field, $g_{\mu\nu}(x)$ is the target-spacetime metric, $h(\tau)=\sqrt{-g_{\tau\tau}}$ denotes the worldline einbein, and $\dot{x}^\mu = dx^\mu/d\tau$. For convenience, we henceforth work in the proper-time gauge $h(\tau)=1$. The equation of motion (EOM) associated with the einbein then reduces to the mass-shell (or Hamiltonian) constraint
\begin{equation}\label{eq:HamiltinianConstraint}
g_{\mu \nu} \dot{x}^\mu \dot{x}^\nu+m^2=0 \, .
\end{equation}
On the other hand, varying \eqref{eq:worldlineAction} with respect to $x^\mu$ yields the EOM
\begin{equation}\label{eq:EOMNambuGotoReal}
\ddot{x}^\mu
+\Gamma^\mu_{\nu\rho}\dot{x}^\nu\dot{x}^\rho
= e g^{\mu\nu}F_{\nu\rho}\dot{x}^\rho\, ,
\end{equation}
where $F_{\mu\nu}=\partial_\mu A_\nu-\partial_\nu A_\mu$ is the field strength, and
\begin{equation}
\Gamma^\mu_{\nu\rho}
=\frac{1}{2}g^{\mu\sigma}
\left(
\partial_\nu g_{\sigma\rho}
+\partial_\rho g_{\nu\sigma}
-\partial_\sigma g_{\nu\rho}
\right)\, ,
\end{equation}
are the Christoffel symbols associated with the target-spacetime metric.

\medskip

In this paper, we study classical motions satisfying (\ref{eq:EOMNambuGotoReal}) that describe closed trajectories in spacetime. As explained in Section~\ref{s:intro}, such solutions, which we refer to as \textit{instantons}, usually capture certain non-perturbative effects in the relevant physical couplings. Furthermore, to obtain these, it is often necessary to analytically continue the worldline proper-time parameter~$\tau$---originally real---to imaginary values. In those cases, we may define $\tau=-i\sigma$, with $\sigma\in\mathbb{R}$, and the EOM (\ref{eq:EOMNambuGotoReal}) then become
\begin{equation}\label{eq:EOMNambuGotoImaginary}
\ddot{x}^\mu
+\Gamma^\mu_{\nu\rho}\dot{x}^\nu\dot{x}^\rho
=-ie g^{\mu\nu}F_{\nu\rho}\dot{x}^\rho\, ,
\end{equation}
where the dots now denote derivatives with respect to $\sigma$. In addition, we see that (\ref{eq:EOMNambuGotoImaginary}) typically renders the spacetime trajectories complex, motivating the consideration of more general instanton solutions evolving in real (interference phase), imaginary (instanton phase), or complex proper time; see, e.g., \cite{Lavrelashvili:1989he,Dumlu:2011cc}. Such complex worldline instantons, and a complete characterization thereof, will constitute the central focus of this work.

\medskip

To elucidate the physical mechanism of interest, we begin in this section with the simpler setting of 2D flat spacetimes, with coordinates $(t,z)$ and metric $\eta_{\mu\nu}=\operatorname{diag}(-1,1)$. The only nonvanishing component of the field strength is therefore the electric field, $E\equiv F_{zt}=-F_{tz}$, which in general may depend on $x^\mu$. The equations of motion (\ref{eq:EOMNambuGotoImaginary}) then reduce to\footnote{In this section we first solve the Wick-rotated equations of motion and then continue the solutions to complex proper time. For (Anti-)de Sitter we instead start from the real form \eqref{eq:EOMNambuGotoReal} and continue from there.}
\begin{equation}
\label{eomtz}
\ddot{t}= -ieE\dot{z}\, ,\qquad \ddot{z}= -ieE\dot{t}\, .
\end{equation}
In the following subsections, we provide a brief review of known results concerning Euclidean instanton solutions in Minkowski backgrounds with constant electromagnetic fields, as well as their complex generalizations. In doing so, we develop a unifying picture that will prove useful when extending the analysis to curved spacetimes in later sections. Our new results are presented mainly in Sections~\ref{ss:MixedSolutionsMink} and \ref{ss:HamiltonianConstraintMink}.

\subsection{Solving the worldline equations of motion}\label{ss:solutionsMink}

To solve (\ref{eomtz}), it is convenient to introduce the light-cone coordinates, defined by $x^\pm=t\pm z$. In terms of these, the EOM simplify to
\begin{equation}
\label{eq:EOMNMinkLightfront}
\ddot{x}^\pm=\mp ieE\dot{x}^\pm\, .
\end{equation}
In what follows, we restrict our attention to constant electric fields. A convenient gauge choice is $A_t=Ez$ and $A_z=0$. This case is exactly solvable and closely parallels the (Anti-)de Sitter setups discussed later in Sections~\ref{s:AdS2} and \ref{s:dS2}, respectively.

\medskip

The solutions to (\ref{eq:EOMNMinkLightfront}) read
\begin{equation}\label{eq:MinkSolsLightfront}
x^\pm (\sigma)=x_0^\pm \pm r_\pm e^{\mp ieE\sigma}\, ,
\end{equation}
where $x_0^\pm\in \mathbb R$ and $r_\pm\in \mathbb C$ are integration constants.\footnote{We point out that since the sum over zero modes of the one-loop partition function yields an overall $\mathbb{R}^{1,1}$ volume factor (see e.g., \cite{Castellano:2026ojb}), physically we should demand that $x_0^\pm$ are both real constants.} Notice that, despite $x^\pm$ being originally real coordinates, in general (\ref{eq:MinkSolsLightfront}) describes complex profiles and hence requires complexifying the configuration space. In particular, each light-cone coordinate traces a circle within its corresponding complex plane, centered at $x_0^\pm$ and with radius equal to $|r_\pm|$. Moreover, the two constants $r_\pm$ are not independent, since the mass-shell constraint (\ref{eq:HamiltinianConstraint}), when expressed in imaginary proper time, takes the form
\begin{equation}
\eta_{\mu \nu} \frac{dx^\mu}{d\sigma} \frac{dx^\nu}{d\sigma} = m^2\, ,
\end{equation}
and thus requires
\begin{equation}
\label{rpmcon}
r_+r_-=\left(\frac{m}{eE}\right)^2\, .
\end{equation}
The solution (\ref{eq:MinkSolsLightfront}) is periodic in $\sigma$, with period $2\pi/|eE|$. Therefore, for it to describe a closed trajectory in the complexified Minkowski space, i.e., an instanton, the worldline parameter $\sigma$ must cover an integer number of periods. Consequently, up to an arbitrary constant shift, we may take $\sigma\in[0,2\pi w_\sigma/|eE|]$, $w_\sigma\in\mathbb{Z}$.

\smallskip

Substituting the above instanton paths into the worldline action (\ref{eq:worldlineAction}), we obtain the classic Schwinger exponent \cite{Schwinger:1951nm}
\begin{equation}
\label{SinstMin}
S_{\rm inst}=i\frac{w_\sigma \pi m^2}{|eE|}\, ,
\end{equation}
which depends on the winding number $w_\sigma$ but is independent of the integration constants appearing in (\ref{eq:MinkSolsLightfront}). In particular, we note that rescaling $r_\pm\to e^{\pm\varphi}r_\pm$ leaves $S_{\rm inst}$ invariant. This is easily understood from the fact that it corresponds to Lorentz boosts of the form \cite{Ilderton:2015lsa}
\begin{equation}
x^\pm-x_0^\pm\to e^{\pm\varphi}\left(x^\pm-x_0^\pm\right)\, .
\end{equation}
Interestingly, the same transformation can also be formally realized as an imaginary shift of the worldline parameter, namely\footnote{The generator $\varphi$ may in fact be complex, whose imaginary part induces an ordinary real translation of $\sigma$.}
\begin{equation}\label{eq:complexreparametrization}
\sigma\to\sigma+\frac{i\varphi}{eE}\, .
\end{equation}

Notice that, when the electric field is turned off, the only remaining periodic motions are $x^\pm(\sigma)=x_0^\pm$. Such solutions exist at every point on $\mathbb{R}^{1,1}$ and correspond to the trivial saddle configurations about which the perturbative one-loop partition function is constructed.

\smallskip

Equivalently, in terms of $(t,z)$, the instanton solutions take the form
\begin{equation}
\label{soltz}
t=t_0+\frac{r_+e^{-ieE\sigma}-r_-e^{ieE\sigma}}{2}\, ,
\qquad
z=z_0+\frac{r_+e^{-ieE\sigma}+r_-e^{ieE\sigma}}{2}\, ,
\end{equation}
where $x_0^\pm=t_0\pm z_0$. We stress that for generic values of $r_\pm$, both $t(\sigma)$ and $z(\sigma)$ are complex. A distinguished solution is obtained for $r_+=r_-=m/(eE)$. In this case, \eqref{soltz} reduces to
\begin{equation}
\label{soltzr+=r-}
t=t_0-\frac{i m}{eE} \sin \left( eE \sigma\right)\, ,
\qquad
z=z_0+\frac{m}{eE} \cos \left( eE \sigma\right)\, ,
\end{equation}
such that $t(\sigma)-t_0$ and $z(\sigma)-z_0$ become purely imaginary and real, respectively. In fact, upon Wick rotating the target-space time according to $t-t_0=-it_E$, the trajectory becomes entirely real and coincides with the more familiar Euclidean worldline instantons \cite{Affleck:1981bma,Dunne:2005sx,Dunne:2006st}.

\smallskip

\begin{figure}[t!]
    \centering
\subcaptionbox{\label{sfig:TcoordinateImTau}}{
		\includegraphics[scale=1]{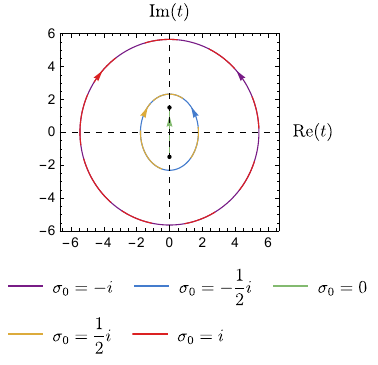}}
        \hspace{0.6cm}
        \subcaptionbox{\label{sfig:ZcoordinateImTau}}{\includegraphics[scale=1]{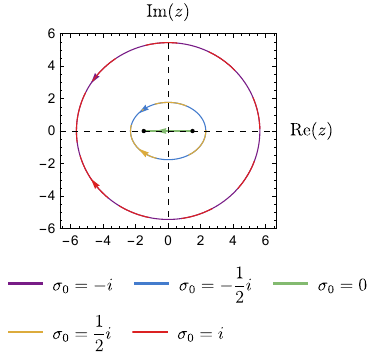}}
		\caption{\small The solutions to the imaginary proper time EOM \eqref{eomtz} are all closed, with periodicity $2\pi/ |eE|$. We plot \eqref{tztau} parametrized by \(\sigma = i \tau \in \mathbb{R}\) and set \(\sigma_0 = i \tau_0\in \mathbb{C}\). Their behavior with respect to $\text{Im}(\sigma_0)$ is symmetric, such that changing its sign preserves the trajectory while reversing its orientation. To make the plots we chose $m=3$, $eE=2$, $z_0=0$, and $t_0=0$.} 
		\label{fig:TrajectoriesImaginaryTauMinkowski}
\end{figure}

More generally, for $r_\pm$ satisfying the constraint (\ref{rpmcon}), as shown in Figure~\ref{fig:TrajectoriesImaginaryTauMinkowski}, the projections of the classical trajectory onto the complex $z$- and $t$-planes enclose the corresponding projections of the Euclidean instanton with the same $t_0$ and $z_0$. Interestingly, as the hierarchy between $|r_+|$ and $|r_-|$ increases, the projections become increasingly circular, with a radius that grows with the larger of $|r_+|$ and $|r_-|$.

%%%%%%%%%%%%%%%%%%%%%%%%%%%%%%%%%%%%%%%%%%%%%
%%%%%%%%%%%%%%%%%%%%%%%%%%%%%%%%%%%%%%%%%%%%%

\subsection{Generalization}\label{ss:MixedSolutionsMink}

We can extend the analysis beyond the instanton phase by allowing $\sigma$ to take complex values. Let us first consider purely imaginary $\sigma$, which, under the relation $\tau=-i\sigma$, amounts to returning to real proper time $\tau$. We henceforth refer to this regime as the \textit{interference phase}. In terms of $\tau$, the solution (\ref{soltz}) reads as follows\footnote{Note that these trajectories correspond to the ordinary hyperbolic motion of a charged particle in a constant electric field $(z-z_0)^2-(t-t_0)^2=m^2/(eE)^2$.}
\begin{equation}
\label{tztau}
t=t_0+\frac{m}{eE} \sinh\left(eE(\tau+\tau_0)\right)\, ,
\qquad
z=z_0+\frac{m}{eE}\cosh\left(eE(\tau+\tau_0)\right)\, ,
\end{equation}
where we have conveniently parametrized $r_\pm$ as
\begin{equation}
r_\pm=\frac{m}{eE}e^{\pm eE\tau_0}\, .
\end{equation}
We note that, since $r_\pm\in\mathbb{C}$, the integration constant $\tau_0$ is generically complex.

\smallskip

\begin{figure}[t!]
    \centering
\subcaptionbox{\label{sfig:TcoordinateRealTau}}{
		\includegraphics[scale=1]{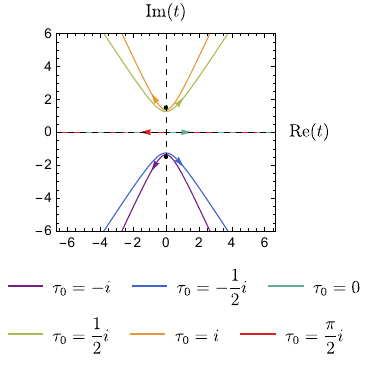}}
        \hspace{0.6cm}
        \subcaptionbox{\label{sfig:ZcoordinateRealTau}}{\includegraphics[scale=1]{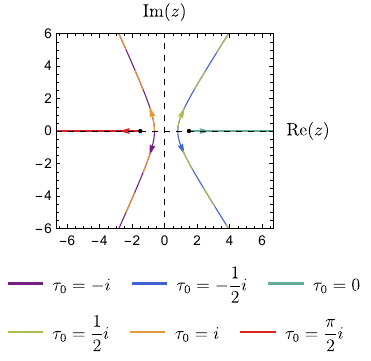}}
		\caption{\small All solutions to the real proper time EOM for $z(\tau)$ and $t(\tau)$ are open. Their actual shape depends on the precise value of the imaginary part of the integration constant $\tau_0$. We plot different trajectories in $\textbf{(a)}$ $t$- and $\textbf{(b)}$ $z$-directions for the particular case of $m=3$, $eE=2$, $z_0=0$, and $t_0=0$.} 
		\label{fig:TrajectoriesRealTauMinkowski}
\end{figure}

The spacetime trajectories described by (\ref{tztau}) are open and thus should not be considered instantons. Nevertheless, their AdS analogues, analyzed in Section~\ref{s:AdS2}, close and become genuine instantons, which is our motivation for studying the Minkowski counterparts. In addition, these trajectories exhibit an interesting structure that depends on the value of~$\tau_0$. First, it can be readily seen that $\tau_0\sim\tau_0+\frac{2\pi i}{eE}$, reflecting the periodicity inherited from the instanton-phase solutions. Moreover, $t(\tau)$ and $z(\tau)$ are real-valued whenever
\begin{equation}
\tau_0\in\mathbb{R}+\frac{i\pi}{eE}\mathbb{Z}\, .
\end{equation}
By contrast, they both become purely imaginary (up to the constant shifts $t_0$ and $z_0$) when
\begin{equation}
\tau_0\in\mathbb{R}+\frac{i\pi}{eE}\left(\mathbb{Z}+\frac{1}{2}\right)\, .
\end{equation}
These properties are exhibited in Figure~\ref{fig:TrajectoriesRealTauMinkowski}, where the interference-phase trajectories are plotted in the complex $t$- and $z$-planes for several representative values of $\tau_0$.

%%%%%%%%%%%%%%%%%%%%%%%%%%%%%%%%%%%%%%%%%%%%%
%%%%%%%%%%%%%%%%%%%%%%%%%%%%%%%%%%%%%%%%%%%%%

\medskip

\begin{figure}[H]
    \centering
\subcaptionbox{\label{sfig:NoLoopTauMink}}{
		\includegraphics[scale=1]{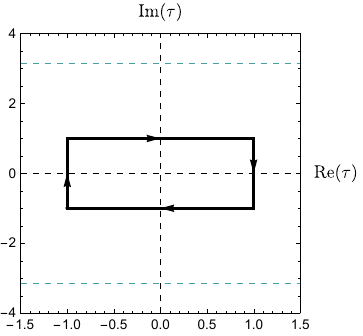}}
        \hspace{0.6cm}
        \subcaptionbox{\label{sfig:LoopTauMink}}{\includegraphics[scale=1]{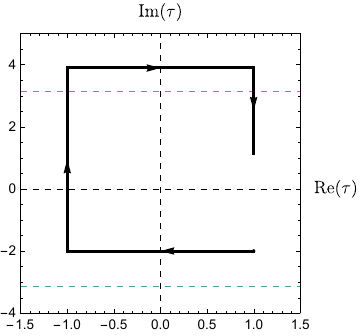}}
		\caption{\small We obtain different instanton trajectories by constructing closed loops in the complex proper-time cylinder. These are grouped into topologically distinct sectors depending on the winding number $w \in \mathbb{Z}$ of the $\tau$-contour. The latter, in turn, controls the resulting worldline action through~\eqref{Sinscomplex}. We chose $m=3$, $eE=2$, $z_0=t_0 =0$, as well as $\textbf{(a)}$ $w=0$ and $\textbf{(b)}$ $w=1$.} 
		\label{fig:ClosedPathsTauMink}
\end{figure}

\begin{figure}[H]
    \centering
\subcaptionbox{\label{sfig:NoLoopTMink}}{
		\includegraphics[scale=1]{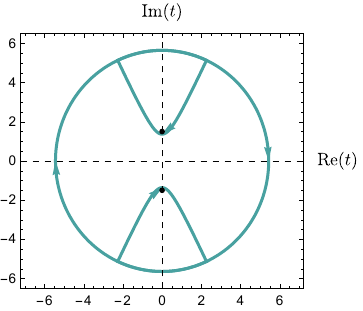}}
        \hspace{0.6cm}
        \subcaptionbox{\label{sfig:NoLoopZMink}}{\includegraphics[scale=1]{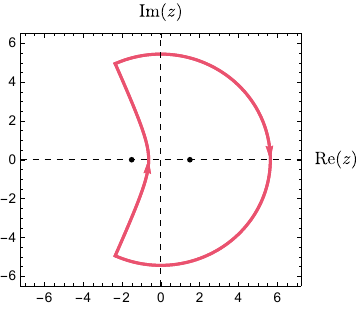}}
		\caption{\small Time $\textbf{(a)}$ and space $\textbf{(b)}$ trajectories associated with the 0-winding path from Figure~\ref{sfig:NoLoopTauMink}.} 
		\label{fig:ImageNoLoopSpacetimeMink}
\end{figure}

\begin{figure}[H]
    \centering
\subcaptionbox{\label{sfig:LoopTMink}}{
		\includegraphics[scale=1]{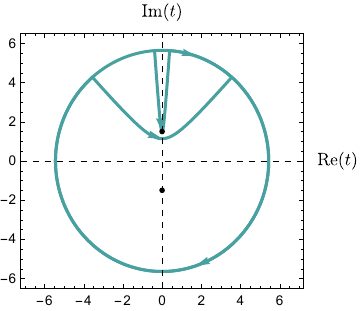}}
        \hspace{0.6cm}
        \subcaptionbox{\label{sfig:LoopZMink}}{\includegraphics[scale=1]{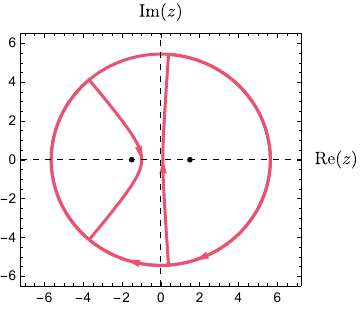}}
		\caption{\small Time $\textbf{(a)}$ and space $\textbf{(b)}$ trajectories corresponding to the $w=1$ path from Figure~\ref{sfig:LoopTauMink}.} 
		\label{fig:ImageLoopSpacetimeMink}
\end{figure}

Next, we consider the general case by analytically continuing (\ref{tztau}) to the complex $\tau$-plane. Without loss of generality, we set $t_0=z_0=\tau_0=0$ and take
\begin{equation}
\label{tztau0}
t= \frac{m}{eE} \sinh\left(eE\,\tau\right)\, ,
\qquad
z= \frac{m}{eE}\cosh\left(eE\, \tau\right)\, .
\end{equation}
As already mentioned above, this solution is periodic under the identification
\begin{equation}
\tau\sim\tau+\frac{2\pi i}{eE}\, .
\end{equation}
Therefore, if a contour $\gamma$ in the complex $\tau$-plane begins and ends at two points separated by $2\pi i w/|eE|$, with $w\in\mathbb{Z}$, the solution obtained via (\ref{tztau0}) corresponds to a complex instanton; see, for instance, Figures~\ref{fig:ClosedPathsTauMink}-\ref{fig:ImageLoopSpacetimeMink}.

%%%%%%%%%%%%%%%%%%%%%%%%%%%%%%%%%%%%%%%%%%%%%
%%%%%%%%%%%%%%%%%%%%%%%%%%%%%%%%%%%%%%%%%%%%%
%%%%%%%%%%%%%%%%%%%%%%%%%%%%%%%%%%%%%%%%%%%%%
%%%%%%%%%%%%%%%%%%%%%%%%%%%%%%%%%%%%%%%%%%%%%
%%%%%%%%%%%%%%%%%%%%%%%%%%%%%%%%%%%%%%%%%%%%%
%%%%%%%%%%%%%%%%%%%%%%%%%%%%%%%%%%%%%%%%%%%%%
Evaluating the action on the complex instanton solution, we obtain
\begin{equation}
\label{Sinscomplex}
\begin{split}
S_{\rm inst}
&=\int_\gamma d\tau\,\left(-m^2+eE\dot{t}z\right) =m^2 \int_\gamma d\tau\, \sinh^2\left(eE\,\tau\right) =-i\frac{w\pi m^2}{|eE|}\, .
\end{split}
\end{equation}
Thus, the instanton action depends only on the winding number $w$ and not on the detailed shape of the contour in the complex $\tau$-plane. This result is in perfect agreement with (\ref{SinstMin}),\footnote{\label{fnote:SignDiffWinding}The sign difference between \eqref{Sinscomplex} and \eqref{SinstMin} arises because \eqref{Sinscomplex} counts winding along the direction of increasing $\text{Im}(\tau)$, whereas \eqref{SinstMin} counts winding along the direction of increasing $\text{Re}(\sigma)$. Since $\sigma=i\tau$, one has that $\text{Re}(\sigma)=-\text{Im}(\tau)$, so the two conventions differ by an orientation reversal and are related via $w=-w_\sigma$.} where $\tau$ was restricted to the imaginary axis up to an arbitrary real shift. In addition, we see that this restriction can be relaxed without changing the resulting worldline action.

\subsection{A complementary approach}\label{ss:HamiltonianConstraintMink}

We now present an alternative, novel derivation of the complex instanton solutions and their action, complementary to the analysis above. First, with the gauge choice $(A_t,A_z)=(Ez,0)$, we see that the Lagrangian (\ref{eq:worldlineAction}) is left invariant under time translations. Consequently, the momentum conjugate to $t$,
\begin{equation}
\label{pt}
p_t=-\dot t+eEz\, ,
\end{equation}
is conserved. Therefore, rather than solving the two second-order differential equations (\ref{eomtz}), we are able to consider the following first-order system:
\begin{equation}
\label{1eomtz}
\begin{split}
-\dot t^2+\dot z^2+m^2&=0\, ,\\
\dot t - eEz+p_t&=0\, .
\end{split}
\end{equation}
%
%%%%%%%%%%%%%%%%%%%%%%%%%%%%%%%%%%%%%%%%%%%%%
\begin{figure}[t!]
    \centering
\subcaptionbox{\label{sfig:RealTau}}{
		\includegraphics[scale=1]{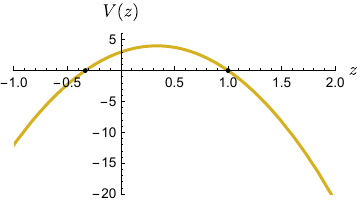}}
        \hspace{0.6cm}
        \subcaptionbox{\label{sfig:ImaginaryTau}}{\includegraphics[scale=1]{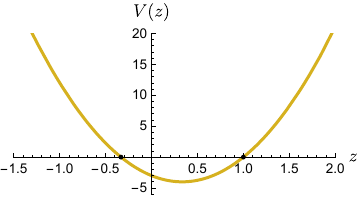}}
		\caption{\small Effective potential $V(z)$ controlling the spatial dynamics in $\mathbb{R}^{1,1}$ for the cases where the proper time is $\textbf{(a)}$ real, and $\textbf{(b)}$ imaginary. The turning points $z_\pm= (p_t\pm m)/eE$ are marked as black. To make the plots, we set \(m=2\), \(eE=3\), and \(p_t = 1\).} 
		\label{fig:EffectivePotMinkowski}
\end{figure}
%%%%%%%%%%%%%%%%%%%%%%%%%%%%%%%%%%%%%%%%%%%%%
Combining the two equations in (\ref{1eomtz}), we obtain a first-order differential equation for $z(\tau)$:
\begin{equation}
\label{zdot}
\dot z=\sqrt{-V(z)}\equiv \sqrt{(p_t- eEz)^2-m^2}\, .
\end{equation}
One can see that $\dot z$ vanishes whenever the effective potential $V(z)$ does, namely at
\begin{equation}
z_\pm=\frac{p_t\pm m}{eE}\, .
\end{equation}
We refer to $z_\pm$ as the \textit{turning points}. See also Figure~\ref{fig:EffectivePotMinkowski} for representative plots of the effective potential for real and imaginary proper-time evolution, the latter being nothing but the negative of the former.

\smallskip

Integrating (\ref{zdot}) thus gives
\begin{equation}
\label{zpt}
z=\frac{m}{eE}\left(\cosh \left( e E (\tau+\tau_0)\right)+\frac{p_t}{m}\right)\, ,
\end{equation}
with $\tau_0$ some complex constant. Substituting this into the second equation in (\ref{1eomtz}), we find
\begin{equation}
\label{tpt}
t=t_0+\frac{m}{eE}\sinh \left( e E (\tau+\tau_0)\right)\, ,
\end{equation}
where $t_0$ denotes the integration constant. Note that the solutions (\ref{zpt}) and (\ref{tpt}) agree with (\ref{tztau}), derived in the previous subsection from (\ref{eomtz}), with the identification $z_0=p_t/(eE)$. Therefore, motions with different $p_t$ are simply related to each other by a translation in $z$.

\medskip

Concerning the instanton action, if we use the conjugate momentum $p_t$, together with
\begin{equation}
\label{pz}
p_z=\dot z\, ,
\end{equation}
we can evaluate (\ref{eq:worldlineAction}) as follows. First, we express the Lagrangian as
\begin{equation}
\label{Lpq}
L=p_t\dot t+p_z\dot z\, .
\end{equation}
Indeed, the Legendre transformation gives
\begin{equation}
H=p_t\dot t+p_z\dot z-L\, ,
\end{equation}
and the constraint (\ref{eq:HamiltinianConstraint}), obtained upon varying the einbein before imposing the gauge choice $h(\tau)=1$, implies $H=0$. It then follows that the on-shell Lagrangian takes the form (\ref{Lpq}). One may also verify (\ref{Lpq}) directly by substituting (\ref{pt}) and (\ref{pz}) into its right-hand side, thereby recovering the original expression presented at the beginning of Section~\ref{s:Minkowski}.

\smallskip

Using (\ref{Lpq}), the instanton action becomes
\begin{equation}\label{eq:instantonactionMink}
 S_{\rm{inst}}= \oint p_tdt+\oint p_zdz =\oint p_z dz\ .
\end{equation}
The first term vanishes because $p_t$ is constant and can be pulled out of the integral, leaving the trivial contribution $\oint dt=0$. We now evaluate the remaining integral using the Cauchy residue theorem. To this end, we first write the integrand as (cf.\ (\ref{zdot}))
\begin{equation}
\label{pzVz}
    p_z=\sqrt{-V(z)}=\sqrt{ (eE)^2(z-z_+)(z-z_-)}\ .
\end{equation}
Thus, the two turning points $z_+$ and $z_-$ are precisely the branch points of the square root.

\medskip

A useful feature of the solution (\ref{zpt}) is that the trajectory of $z (\tau)$ along the instanton phase never crosses the line connecting $z_+$ and $z_-$.\footnote{Unless $\tau+\tau_0 \in i\mathbb{R}$, in which case $z(\tau)$ runs along the segment $[z_-, z_+]$ itself, and the two sign branches of \eqref{zdot} must be joined consistently at the turning points $z_\pm$ to evaluate the action integral.} This enables us to choose the branch cut to lie along this line segment. The path can then be deformed in the complex $z$-plane to a similar one encircling infinity $\tilde w$ times. Depending on the contour $\gamma$ in the complex $\tau$-plane, we have either $\tilde w=w$ or $\tilde w=-w$, where $w$ is the winding number in proper time defined in the previous subsection. Furthermore, it follows from the form of (\ref{zpt}) that, when $\tilde w=\pm w$, the corresponding branch of (\ref{pzVz}) should be chosen such that $\sqrt{-V(z)}\sim\pm |eE| z$ as $z\to\infty$. The associated residue is then
\begin{equation}
\text{Res}_{z=\infty}\left(\sqrt{-V(z)}\right)
=\pm\frac{m^2}{2|eE|}\, .
\end{equation}
Hence, the two sign choices compensate each other, and we obtain
\begin{equation}\label{eq:ActionMinkResidues}
S_{\rm inst}
=\oint\sqrt{-V(z)}\,dz
=-2\pi i\tilde w\,\text{Res}_{z=\infty}\left(\sqrt{-V(z)}\right)
=-i\frac{w\pi m^2}{|eE|}\, ,
\end{equation}
in agreement with our previous result (\ref{Sinscomplex}).

\smallskip

One subtlety worth emphasizing is that, as seen above, $\tilde w$ does not necessarily coincide with $w$, and it is the latter that controls the sign of the instanton action. In particular, if $\tau$ runs along the purely imaginary direction, then sending $\tau\to \tau- 2 \text{Re}(\tau_0)$ leaves the trajectory $z(\tau)$ unchanged up to an orientation reversal (i.e., $\tilde{w}\to-\tilde{w}$), while preserving the orientation of the proper-time contour (hence $w\to w$); consequently, $S_{\rm inst}$ remains invariant. By contrast, flipping $\tau \to-\tau$ reverses the orientation in both the proper-time plane and in spacetime ($w\to-w$ and $\tilde{w}\to-\tilde{w}$), thus yielding a sign change in the action~\eqref{eq:ActionMinkResidues}.

\section{Instanton Solutions in AdS$_2$}\label{s:AdS2}

In this section, we consider a relativistic point particle propagating in two-dimensional Anti-de Sitter spacetime with radius $R$, threaded by a `constant' (more precisely, homogeneous) electric field $E$. The line element and gauge curvature in Poincaré coordinates take the form
\begin{subequations}\label{eq:BackgroundAdS2}
\begin{equation}\label{eq:PoincareMetricAdS}
 ds^2= \frac{R^2}{\rho^2} (-dt^2+d\rho^2)=g_{\mu \nu} dx^\mu dx^\nu\, ,\\
\end{equation}
\begin{equation}\label{eq:GaugeFieldPoincareAdS}
 F=E\, \omega_{\rm{AdS}_2} = E \left( \frac{R}{\rho}\right)^2 dt \wedge d\rho\, , 
\end{equation}
\end{subequations}
where $t\in (-\infty, \infty)$ is the Poincaré time, $\rho \in (0, \infty)$ denotes the radial coordinate, and e.g., $A= E R^2 \rho^{-1} dt$. The action (\ref{eq:worldlineAction}) then reads (here again we set $h(\tau)=1$)
\begin{equation}\label{SwlAdS}
 S_{\rm wl} = \int_\gamma d\tau\,\left(- \frac{m^2}{2}-\frac{R^2\dot t^2}{2 \rho^2}+\frac{R^2\dot \rho^2}{2 \rho^2} +\frac{eER^2\dot t}{\rho}  \right)\, ,
\end{equation}
with the Hamiltonian constraint
\begin{equation}
\label{H0AdS}
   -\frac{R^2\dot t^2}{\rho^2}+ \frac{R^2\dot\rho^2}{\rho^2}+m^2=0\ .
\end{equation}
Due to the non-triviality of the spacetime metric, the worldline EOM \eqref{eq:EOMNambuGotoReal} become
\begin{equation}\label{eq:EOMAdS2}
 \Ddot{\rho}-\frac{\dot{\rho}^2+\dot{t}^2}{\rho}=-eE \dot{t}\, ,\qquad \Ddot{t}-2\frac{\dot{t} \dot{\rho}}{\rho}=-eE \dot{\rho}\, .
\end{equation}

In order to make the symmetries of the problem fully explicit, let us introduce light-cone coordinates as before, namely $x^\pm= t\pm \rho$. In terms of these, \eqref{eq:EOMAdS2} reduces to
\begin{equation}
 \Ddot{x}^\pm\mp \frac{2}{x^+-x^-} \left( \dot{x}^\pm\right)^2=\mp eE \dot{x}^\pm\, .
\end{equation}
This set of coordinates transform nicely under $\SL(2,\mathbb{R})$, namely
\begin{equation}\label{eq:SL2Maps}
\mathcal{A}=\left(\begin{array}{cc}
        a\ & b \\
        c\ &  d
    \end{array}\right) \in \SL(2, \mathbb{R})\, ,\qquad x^\pm \to \mathcal{A} \cdot x^\pm \equiv \frac{a x^\pm +b}{c x^\pm +d}\, ,
\end{equation}
which indeed leaves invariant the 2D metric as well as the gauge connection $A=A_t dt$---in this case up to a gauge transformation $A \to A+ d \Lambda$, with $\Lambda=ER^2 \log \left[ (cx^-+d)(cx^++d)^{-1}\right]$.

\medskip

In the following sections, we derive and discuss in detail the solutions to \eqref{eq:EOMAdS2} for both real and imaginary proper time. As we will see, in contrast to the Minkowski case, the qualitative behavior of the resulting integral curves depends crucially on whether $R |eE|/m$ is greater than (super-extremal), equal to (extremal), or smaller than (sub-extremal) one.
Here, however, we proceed by directly solving the first-order differential equation obtained from the constraint \eqref{H0AdS}, as this approach turns out to be considerably simpler than solving the corresponding second-order equations. We will also draw connections to---and emphasize the differences from---the analogous flat-spacetime problem studied thoroughly in Section~\ref{s:Minkowski}.

\subsection{The sub-extremal case}\label{ss:SubExtremal}

\subsubsection{Solving the worldline equations of motion}\label{sss:solutionsSubAdS}

To obtain the semiclassical saddles in AdS$_2$, we follow the same procedure as in Section~\ref{ss:HamiltonianConstraintMink}. First, we note that the momentum conjugate to the time coordinate, derived from (\ref{SwlAdS}),
\begin{equation}
\label{ptAdS}
p_t=- \frac{R^2\dot{t}}{\rho^2}+ \frac{eE R^2}{\rho}\, ,
\end{equation}
is conserved. This enables us to rewrite the Hamiltonian constraint (\ref{H0AdS}) as the following first-order differential equation\footnote{The sign ambiguity in \eqref{eq:DiffEqAdS2} amounts to a proper-time reversal of the form $\tau \to -\tau$; see below.}
\begin{equation}\label{eq:DiffEqAdS2}
\frac{R^2 d\rho}{\rho \sqrt{(p_t \rho-eER^2)^2-(mR)^2}}=d\tau\, ,
\end{equation}
whose general solution is
\begin{equation}\label{eq:RhoSolutionRealTau}
p_t \rho (\tau)= -\frac{\beta^2}{q_e-\tilde{m} \sin \left[ R^{-2}\beta (\tau+\tau_0)\right]}\, .
\end{equation}
Here, we defined the real, dimensionless quantities\footnote{Throughout this subsection, we take the positive real root, $\beta>0$, with sub-extremality requiring $\tilde{m}^2 >q_e^2$.} $\tilde{m}=mR$, $q_e= eE R^2$, and $\beta = \sqrt{\tilde{m}^2-q_e^2}$. Furthermore, plugging the solution $\rho(\tau)$ in (\ref{ptAdS}), one finds
\begin{equation}\label{eq:TSolutionRealTau}
p_t t(\tau)= p_t t_0 -\frac{\tilde{m}\beta\cos \left[ R^{-2}\beta (\tau+\tau_0)\right]}{q_e-\tilde{m} \sin \left[ R^{-2}\beta (\tau+\tau_0)\right]}\, ,
\end{equation}
with $t_0 \in \mathbb{R}$ the integration constant. 

\smallskip
%%%%%%%%%%%%%%%%%%%%%%%%%%%%%%%%%%%%%%%%%%%%%
\begin{figure}[t!]
    \centering
\subcaptionbox{\label{sfig:TcoordinateReTauAdS}}{
		\includegraphics[scale=1]{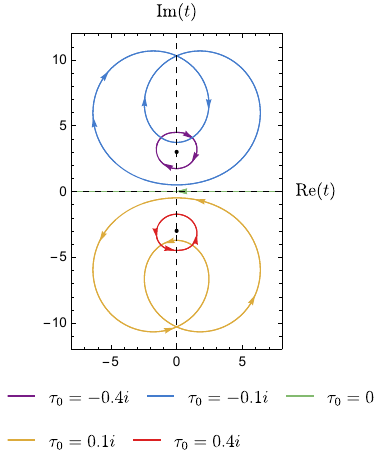}}
        \hspace{0.6cm}
        \subcaptionbox{\label{sfig:RhocoordinateReTauAdS}}{\includegraphics[scale=1]{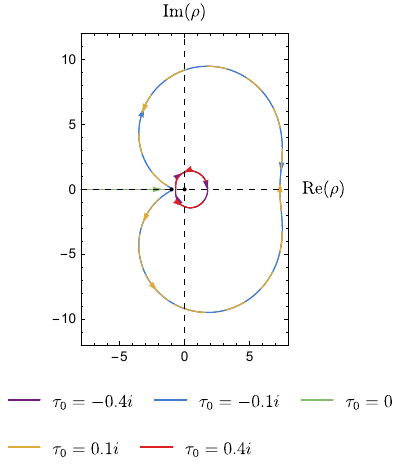}}
		\caption{\small The solutions to the real proper-time EOM \eqref{eq:EOMAdS2} are complex and closed, with periodicity equal to $2\pi R^2/\sqrt{\tilde{m}^2-q_e^2}$, except for those with $\tau_0\in \mathbb{R}$, which are real-valued. Their behavior is symmetric under $\text{Im}(\tau_0) \to -\text{Im}(\tau_0)$: in the complex $\rho$-plane, this transformation preserves the trajectory while reversing its orientation, whereas in the complex $t$-plane, it reflects the trajectory across the real axis. We plot several cases with $\tilde{m}=5$, $q_e=4$, $p_t=R=1$, $t_0=0$, and different $\tau_0$.} 
		\label{fig:TrajectoriesRealTauAdS}
\end{figure}
%%%%%%%%%%%%%%%%%%%%%%%%%%%%%%%%%%%%%%%%%%%%%
One can easily verify that the functions \eqref{eq:RhoSolutionRealTau} and \eqref{eq:TSolutionRealTau} satisfy the second-order EOM \eqref{eq:EOMAdS2} derived from the worldline action. We stress that the solutions above evolve naturally in real proper time, corresponding to the interference phase. In Figure \ref{fig:TrajectoriesRealTauAdS}, we plot the spacetime trajectories for several representative values of the complex integration constant $\tau_0$. One can see that when $\tau_0 \in \mathbb{R}$, the solutions are open and purely real. Remarkably, once $\tau_0$ acquires a nonzero imaginary part, the trajectories become complex and closed, with period $2\pi R^2/\beta$. They therefore describe genuine instantons provided that the worldline parameter spans an integer number of periods. Accordingly, we may take $\tau \in [0, 2\pi R^2 w/\beta]$, with $w \in \mathbb{Z}$, up to arbitrary constant shifts. Evaluating the associated action (\ref{SwlAdS}) yields
\begin{equation}
\label{SinstSubAdS}
S_{\rm inst}=-2\pi w \beta\, ,
\end{equation}
which, as in the Minkowski case (cf.\ (\ref{SinstMin})), does not depend on the particular value of the integration constants $\tau_0$ and $t_0$. This is related to the fact that both types of shifts induce certain conformal symmetries in Anti-de Sitter space; see Appendix \ref{ap:SL2C&ComplexShifts} for details. Notice that, unlike their Minkowski counterparts, these instantons have a real action \eqref{SinstSubAdS}, a direct consequence of $\beta$ being real in the sub-extremal regime \cite{PicardLefschetz}. 

\smallskip

\begin{figure}[t!]
	\begin{center}
		\includegraphics[scale=1]{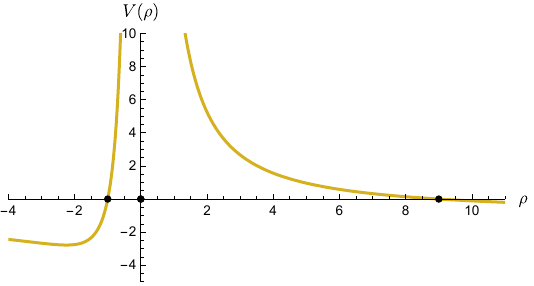}
		\caption{\small Effective potential $V(\rho)$ controlling the spatial dynamics in AdS$_2$ for real proper time. The turning points appear as black dots at $\rho=\{0, (q_e\pm \tilde{m})/p_t \}$. We set $q_e=4$, $\tilde{m}=5$, $p_t=1$.} 
		\label{fig:EffPotentialRealTauAdS}
	\end{center}
\end{figure}

\noindent In addition, it can be seen from Figure~\ref{fig:TrajectoriesRealTauAdS} that in the complex $\rho$-plane the traversed paths are invariant, up to a reversal of orientation, under the involution $\tau_0 \to \bar{\tau}_0$. For ${\rm Im }\tau_0\neq0$, these closed trajectories wind around the conformal boundary at $\rho=0$, while leaving the two roots $\rho_\pm=(q_e\pm \tilde m)/p_t$ of the effective potential (cf.\ Figure~\ref{fig:EffPotentialRealTauAdS})
\begin{equation}\label{eq:EffectivePotentialAdS}
V(\rho)= \frac{\tilde{m}^2}{\rho^2}-\left( p_t-\frac{q_e}{\rho} \right)^2\, ,
\end{equation}
outside. As $\tau_0 \to \pm i \infty$, they become asymptotically circular and shrink towards $\rho=0$. Similarly, in the complex $t$-plane one finds periodic orbits that self-intersect for small values of $|\text{Im}(\tau_0)|$ and tend to localize around 
\begin{equation}
\label{tpm}
t_\pm=t_0 \pm i\, \frac{\beta}{p_t}\, ,
\end{equation}
as $\tau_0 \to \mp i \infty$.
%%%%%%%%%%%%%%%%%%%%%%%%%%%%%%%%%%%%%%%%%%%%%
%%%%%%%%%%%%%%%%%%%%%%%%%%%%%%%%%%%%%%%%%%%%%

\subsubsection{Generalization}\label{ss:MixedSolutionsSubAdS}

\begin{figure}[t!]
    \centering
\subcaptionbox{\label{sfig:TcoordinateImTauAdS}}{
		\includegraphics[scale=0.83]{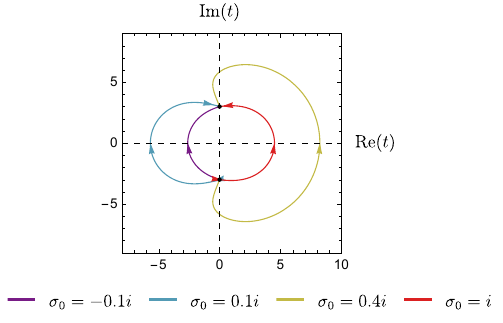}}
        \hspace{0.6cm}
        \subcaptionbox{\label{sfig:RhocoordinateImTauAdS}}{\includegraphics[scale=0.825]{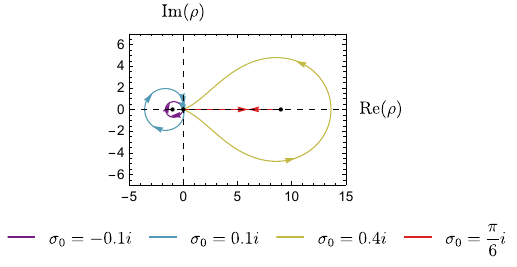}}
		\caption{\small The solutions for $\rho(\sigma)$ and $t(\sigma)$ in imaginary proper time are closed and open, respectively, with $\sigma \in (-\infty, \infty)$. These can be divided into two families depending on the branch point $\rho_\pm$ that is enclosed, which in turn controls the growth exhibited by the worldline instanton action \eqref{SwlAdS}. We plot various representative cases with $\tilde{m}=5$, $q_e=4$, $p_t= R=1$, $t_0=0$, and different values~of~$\sigma_0$.} 
		\label{fig:TrajectoriesImaginaryTauAdS}
\end{figure}

To extend the analysis beyond the interference phase, we can consider generic evolution in complex $\tau$. The simplest such instance corresponds to classical motion in imaginary proper time, which may be obtained from \eqref{eq:RhoSolutionRealTau}-\eqref{eq:TSolutionRealTau} via the Wick rotation $\tau \to -i\sigma$, with $\sigma \in \mathbb{R}$:
\begin{subequations}\label{eq:InstantonSolsAdS}
\begin{equation}\label{eq:RhoSolAdS}
p_t \rho (\sigma)= -\frac{\beta^2}{q_e+i\tilde{m} \sinh \left[ R^{-2}\beta (\sigma+\sigma_0)\right]}\, ,
\end{equation}
\begin{equation}\label{eq:TSolAdS}
p_t t(\sigma)= p_t t_0 -\frac{\tilde{m}\beta\cosh \left[ R^{-2}\beta (\sigma+\sigma_0)\right]}{q_e+i\tilde{m} \sinh \left[ R^{-2}\beta (\sigma+\sigma_0)\right]}\, .
\end{equation}
\end{subequations}
We have plotted several representative solutions, differing only in the value assigned to $\sigma_0$, in Figure~\ref{sfig:TcoordinateImTauAdS} for the $t$-coordinate and Figure~\ref{sfig:RhocoordinateImTauAdS} for the corresponding $\rho$-coordinate. These solutions exhibit several interesting features. First, we notice that, in the complex $\rho$-plane, the trajectories are closed, but they start and end at the conformal boundary of AdS$_2$, located at $\rho=0$. As a consequence, completing the loop requires an infinite range of (imaginary) proper time, namely $\sigma \in (-\infty, \infty)$, thus giving rise to an a priori divergent contribution to the worldline action. On the other hand, in the complex $t$-plane, one finds that the paths are open and connect the points $t_\pm$ defined in \eqref{tpm}. These solutions therefore do not, by themselves, describe instantons.

\smallskip
%%%%%%%%%%%%%%%%%%%%%%%%%%%%%%%%%%%%%%%%%%%%%
%%%%%%%%%%%%%%%%%%%%%%%%%%%%%%%%%%%%%%%%%%%%%
Furthermore, the behavior of $t(\sigma)$ and $\rho(\sigma)$ with respect to the complex integration constant $\sigma_0$ is quite interesting and parallels the situation already encountered in $\mathbb{R}^{1,1}$ for real $\tau$. In particular, we note that
\begin{equation}
\sigma_0 \sim \sigma_0 + \frac{2\pi i R^2}{\beta}\, ,
\end{equation}
a periodicity inherited from the interference solutions in Section~\ref{sss:solutionsSubAdS}. In addition, one readily sees that the paths \eqref{eq:InstantonSolsAdS} are generically complex, except for special values of $\sigma_0$, namely
\begin{equation}
\label{sigma0pi2}
\sigma_0\in\mathbb{R}+\frac{\pi iR^2}{\beta}\left(\mathbb{Z}+\frac12\right)\, .
\end{equation}
In those cases, $\rho(\sigma)$ becomes purely real, whilst $t(\sigma)$ is purely imaginary (up to the constant shift $t_0$). These solutions moreover resemble the familiar Euclidean instantons, which become entirely real upon Wick rotating the target-space time according to $t(\sigma)-t_0=-it_E$. Along the $\rho$-direction, such trajectories begin at the boundary $\rho=0$, reach one of the two turning points $\rho_\pm$, and then return to the boundary; see~\cite{Castellano:2025yur} and references therein.

For generic $\sigma_0$, the solutions fall into two distinct families, distinguished by which root $\rho_\pm$ of $V(\rho)$ is enclosed in the $\rho$-plane. One may move continuously within a given family by shifting $\sigma_0$ by a purely imaginary constant, which leaves the instanton action invariant.\footnote{The action of each individual saddle is formally divergent, but the difference between two members of the same family corresponds to a closed contour enclosing zero net winding, and hence vanishes exactly; cf.\ \eqref{eq:InstantonActionAdSMixed}.} Furthermore, the transition between the two families occurs when the $\rho$-trajectory passes through infinity, approaching it along a direction parallel to the imaginary axis. This happens for
\begin{equation}\label{eq:FamilyJumpSubExtremal}
\sigma_0 \in \mathbb R+\frac{iR^2}{\beta}\left(2\pi\mathbb Z +\frac{\pi}{2}\pm\arccos\left( \frac{q_e}{\tilde{m}}\right)\right)\, .
\end{equation}
Interestingly, the orientation with which the loop in the complex $\rho$-plane is traversed reverses from clockwise to counterclockwise, or vice versa, whenever the imaginary part of $\sigma_0$ crosses any of the critical values in (\ref{sigma0pi2}) and (\ref{eq:FamilyJumpSubExtremal}). Similarly, in the complex $t$-plane, the point at which the trajectory crosses the real axis moves from one side of $t_0$ to the other.

\medskip

Next, we consider the general case by analytically continuing \eqref{eq:RhoSolutionRealTau} and \eqref{eq:TSolutionRealTau} to the full complex $\tau$-plane. Without loss of generality, we henceforth set $t_0=\tau_0=0$, and thus take
\begin{equation}\label{eq:RealTauInstantonSolsSubExtremal}
p_t \rho (\tau)= -\frac{\beta^2}{q_e-\tilde{m} \sin \left( R^{-2}\beta\, \tau\right)}\, ,\qquad  p_t t(\tau)= -\frac{\tilde{m}\beta\cos \left( R^{-2}\beta\, \tau\right)}{q_e-\tilde{m} \sin \left( R^{-2}\beta\, \tau\right)}\, .
\end{equation}
As in the Minkowski case discussed in Section~\ref{s:Minkowski}, we provide a prescription for constructing and classifying all topologically inequivalent instanton sectors, which can in turn be characterized by their worldline action. To do so, we consider contours in the complexified proper-time plane whose images under \eqref{eq:RealTauInstantonSolsSubExtremal} describe the closed spacetime trajectories. Due to the periodicity
\begin{equation}\label{eq:PeriodSubExtremal}
\tau \sim \tau+ \frac{2\pi R^2}{\beta}\, ,
\end{equation}
paths in the complex $\tau$-plane need only be closed up to the above identification. The complex proper-time plane modulo this identification may therefore be regarded as an infinite cylinder, whose compact direction corresponds to the real axis. Thus, given the single-valuedness of the instanton maps, it is sufficient to consider loops that are closed on this cylinder in order to construct cyclic spacetime trajectories. 

\smallskip

\begin{figure}[H]
    \centering
\subcaptionbox{\label{sfig:NoLoopTauAdSsub}}{
		\includegraphics[scale=1]{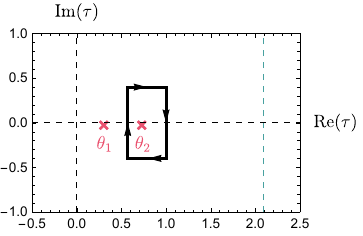}}
        \hspace{0.6cm}
        \subcaptionbox{\label{sfig:LoopTauAdSsub}}{\includegraphics[scale=1]{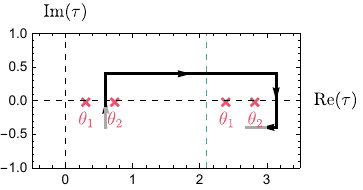}}
		\caption{\small We obtain different instanton trajectories by constructing closed loops in the complex proper-time cylinder. These can be grouped into distinct sectors depending on integral numbers $w, n=n_{\theta_1}-n_{\theta_2}$ associated with the $\tau$-contour. The latter control the worldline action through \eqref{eq:InstantonActionAdSMixed}. To make the plots, we chose $\tilde{m}=5$, $q_e=4$, $p_t = R=1$, with $\textbf{(a)}$ $w=0, n=1$, and $\textbf{(b)}$ $w=n=1$.} 
		\label{fig:ClosedPathsTauAdSsub}
\end{figure}

\begin{figure}[H]
    \centering
\subcaptionbox{\label{sfig:NoLoopTAdSsub}}{
		\includegraphics[scale=1]{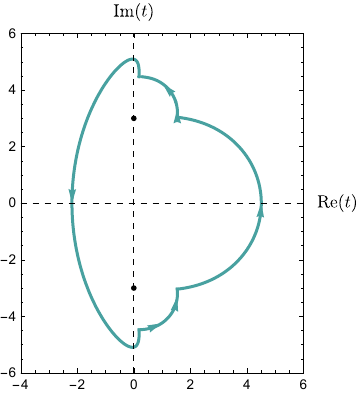}}
        \hspace{0.6cm}
        \subcaptionbox{\label{sfig:NoLoopRhoAdSsub}}{\includegraphics[scale=1]{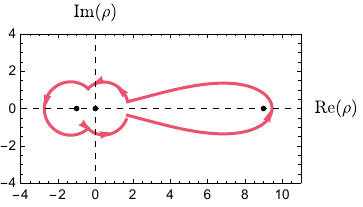}}
		\caption{\small Time $\textbf{(a)}$ and space $\textbf{(b)}$ trajectories corresponding to the path shown in Figure~\ref{sfig:NoLoopTauAdSsub}.} 
		\label{fig:ImageNoLoopSpacetimeAdSsub}
\end{figure}

\begin{figure}[H]
    \centering
\subcaptionbox{\label{sfig:LoopTAdSsub}}{
		\includegraphics[scale=1]{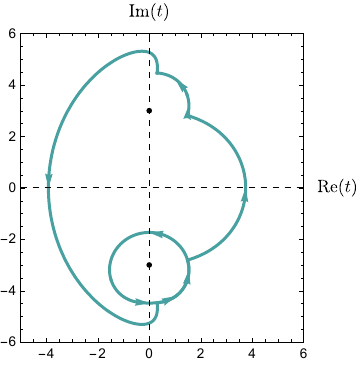}}
        \hspace{0.6cm}
        \subcaptionbox{\label{sfig:LoopRhoAdSsub}}{\includegraphics[scale=1]{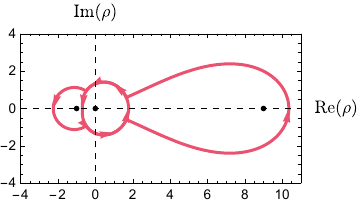}}
		\caption{\small Time $\textbf{(a)}$ and space $\textbf{(b)}$ motions associated with the $w=n=1$ path in Figure~\ref{sfig:LoopTauAdSsub}.} 
		\label{fig:ImageLoopSpacetimeAdSsub}
\end{figure}

These loops are further distinguished by the number of times they wind around the periodic direction, as well as by their winding numbers around two \emph{punctures} located (up to the periodicity \eqref{eq:PeriodSubExtremal}) at
\begin{equation}\label{eq:PunctureSsubExtremal}
\tau=\theta_1 = \frac{R^2}{\beta}\, \arcsin \left( \frac{q_e}{\tilde{m}}\right)\, ,\qquad \tau=\theta_2 = \frac{R^2}{\beta} \left(\pi-\arcsin \left( \frac{q_e}{\tilde{m}}\right)\right)\, .
\end{equation}
The punctures correspond to the points where the spacetime trajectories become singular (cf.\ \eqref{eq:RealTauInstantonSolsSubExtremal}), and they carry opposite weights---as one may easily check. Thus, encircling both gives a vanishing net winding number. This defines equivalence classes of paths according to whether they can be continuously deformed into one another; see Figures \ref{fig:ClosedPathsTauAdSsub}--\ref{fig:ImageLoopSpacetimeAdSsub}. In fact, one readily finds that all closed paths in complexified proper time that wind $w \in \mathbb{Z}$ times around the periodic (i.e., instanton-like) direction and $n_{\theta_i}\in \mathbb{Z}$ times around each puncture \eqref{eq:PunctureSsubExtremal} (with $i=1,2$) yield a worldline action of the form
\begin{equation}\label{eq:InstantonActionAdSMixed}
 S_{\rm{inst}}= \int_\gamma L\, d\tau =-2\pi \left[ w\beta+ i (n_{\theta_1}-n_{\theta_2}) q_e\right]\, .
\end{equation}
The winding numbers $n_{\theta_i}$ are defined to be positive (negative) for counterclockwise (clockwise) orientation. We stress that the result is independent of the particular representative path chosen and of whether the latter transitions between different branches of, e.g., the instanton-phase solutions (see the discussion around \eqref{eq:FamilyJumpSubExtremal}).

%%%%%%%%%%%%%%%%%%%%%%%%%%%%%%%%%%%%%%%%%%%%%
%%%%%%%%%%%%%%%%%%%%%%%%%%%%%%%%%%%%%%%%%%%%%
%%%%%%%%%%%%%%%%%%%%%%%%%%%%%%%%%%%%%%%%%%%%%
%%%%%%%%%%%%%%%%%%%%%%%%%%%%%%%%%%%%%%%%%%%%%
%%%%%%%%%%%%%%%%%%%%%%%%%%%%%%%%%%%%%%%%%%%%%
%%%%%%%%%%%%%%%%%%%%%%%%%%%%%%%%%%%%%%%%%%%%%

\subsubsection{Instanton action from Cauchy's residue theorem}\label{sss:CauchyAdSSubExtremal}

We now show that, exactly as in the flat-spacetime analysis of Section~\ref{ss:HamiltonianConstraintMink}, the worldline action associated with closed contours in the complexified proper-time plane can be recast in terms of the residue theorem. Beyond providing an independent check of \eqref{eq:InstantonActionAdSMixed}, this reformulation makes it transparent which analytic data of the problem control the two main contributions to the instanton action in Anti-de Sitter space.

\medskip

As before, the on-shell Lagrangian takes the first-order form $L=p_\mu\dot x^\mu$ by virtue of the Hamiltonian constraint $H=0$. From (\ref{SwlAdS}), the momentum conjugate to $\rho$ is
\begin{equation}
\label{pRhoAdS}
p_\rho= \frac{R^2\dot{\rho}}{\rho^2}\, ,
\end{equation}
while $p_t$ was defined in \eqref{ptAdS}. Since $p_t$ is conserved and the instanton paths are closed in spacetime, the time component does not contribute:
\begin{equation}
S_{\rm inst} = \oint p_\mu \dot x^\mu\, d\tau
= \oint p_t\, dt + \oint p_\rho\, d\rho
= \oint p_\rho\, d\rho\,,
\label{eq:Sinst-ads-contour}
\end{equation}
where $\oint p_t\, dt = p_t\oint dt = 0$. The action is thus entirely determined by the spatial term $\oint p_\rho\,d\rho$, whose evaluation we now reduce to a residue calculation.

\smallskip

To do so, we first use (\ref{H0AdS}), (\ref{ptAdS}), and express (\ref{pRhoAdS}) as
\begin{equation}
p_\rho^2=-V(\rho)
=\frac{p_t^2}{\rho^2}(\rho-\rho_+)(\rho-\rho_-)\,,
\qquad
\rho_\pm=\frac{q_e\pm\tilde m}{p_t}\,.
\label{eq:prho-factored}
\end{equation}
Thus, on either sheet of the associated two-sheeted Riemann surface, we may write
\begin{equation}
p_\rho(\rho)
=\frac{p_t}{\rho}
\sqrt{(\rho-\rho_+)(\rho-\rho_-)}\,,
\label{eq:prho-branch}
\end{equation}
where the branch of the remaining square root is to be specified according to the trajectory under consideration. The turning points $\rho_\pm$ are then the only branch points, as in Minkowski. The new feature of the AdS problem is the additional factor $1/\rho$, which gives rise to a simple pole at the conformal boundary $\rho=0$. Notice, moreover, that on a branch normalized by
\begin{equation}
\sqrt{(\rho-\rho_+)(\rho-\rho_-)}
\sim \rho
\qquad \text{as}\quad \rho\to\infty\,,
\end{equation}
the radial conjugate momentum behaves as
\begin{equation}
p_\rho
\xrightarrow{\ \rho\to\infty\ }
p_t-\frac{q_e}{\rho}+\mathcal{O}(\rho^{-2})\,.
\end{equation}
Near the conformal boundary, on the other hand,
\begin{equation}
p_\rho
\xrightarrow{\ \rho\to0\ }
\pm\,\frac{i\beta}{\rho}+\mathcal{O}(1)\,,
\end{equation}
where the sign is fixed by the analytic continuation of the chosen branch along the trajectory under consideration. The residue calculation can therefore probe two distinguished points of the analytically continued radial momentum, namely the boundary pole at $\rho=0$ and the point at infinity. As we will see, these account for the $\beta$ and $q_e$ terms of \eqref{eq:InstantonActionAdSMixed}, respectively, which are themselves associated with the two distinct phases of the classical solutions \eqref{eq:RhoSolutionRealTau}. We treat each of them in turn.

\paragraph{Interference phase.}
For real $\tau$ with $\tau_0\in i\mathbb{R}\setminus\{0\}$, the trajectory of $\rho(\tau)$ closes without crossing the half-lines extending outward from $\rho_+$ and $\rho_-$ along the real axis. We may therefore choose the branch cuts so as to leave the interval between the two turning points, and in particular the conformal boundary, in the cut-free region. The closed trajectory can then be continuously deformed, on the relevant sheet, to a contour encircling the pole at $\rho=0$. Let $\tilde w$ denote the corresponding winding number in the $\rho$-plane. Depending on the proper-time contour $\gamma$, its orientation is related to the proper-time winding number $w$ introduced in the previous subsection by $\tilde w=\pm w$.

\smallskip

The solution \eqref{eq:RhoSolutionRealTau} fixes which sheet of \eqref{eq:prho-branch} is followed by the trajectory. In particular, for $\tilde w=\pm w$, the corresponding analytic continuation requires
\begin{equation}
p_\rho
\sim
\pm\,\frac{i\beta}{\rho}
\qquad (\rho\to0)\,,
\end{equation}
with the sign correlated with the orientation of the radial trajectory. Hence
\begin{equation}
\operatorname{Res}_{\rho=0}\left( \sqrt{-V(\rho)}\right)
=\pm\,i\beta\,.
\end{equation}
The simultaneous reversal of the radial winding and of the sheet therefore leaves the result invariant, and Cauchy's residue theorem unambiguously gives
\begin{equation}
S_{\rm inst}
=\oint p_\rho\,d\rho
=2\pi i\,\tilde w\,
\operatorname{Res}_{\rho=0}\left(\sqrt{-V(\rho)}\right)
=-2\pi w\beta\,,
\label{eq:Sinst-ads-res0}
\end{equation}
in agreement with the interference-phase result and, in particular, with the $w\beta$ term of \eqref{eq:InstantonActionAdSMixed}.

\paragraph{Instanton phase.}
In imaginary proper time, the analysis most closely parallels the flat-space computation of Section~\ref{ss:HamiltonianConstraintMink}. As
discussed in Section~\ref{ss:MixedSolutionsSubAdS}, the individual saddles do not close even over $\sigma\in(-\infty,\infty)$. An elementary closed spacetime trajectory is instead obtained by stitching together a pair of motions of opposite orientation, while more general closed trajectories may be constructed by concatenating such elementary stitched motions. On each constituent, the radial trajectory does not cross the interval between the two turning points $\rho_-$ and $\rho_+$.\footnote{The exceptional cases are those with $\sigma_0 \in \mathbb{R}+i \pi R^2 \beta^{-1} (\mathbb{Z}+1/2)$, for which the trajectory lies precisely along the branch cut.} We may thus choose the branch cut to lie along this segment, as in the Minkowski analysis. On the corresponding Riemann surface, the resulting closed contour can then be deformed to a path winding around the point at infinity an arbitrary integer number of times. Let $\tilde n\in \mathbb{Z}$ denote this oriented winding number. Depending on the proper-time contour $\gamma$, one has $\tilde n=\pm n$, where
\begin{equation}
n\equiv n_{\theta_1}-n_{\theta_2}\, ,
\end{equation}
is the net winding number around the punctures \eqref{eq:PunctureSsubExtremal}, as defined in the previous subsection. In particular, a single stitched pair generates an elementary winding, whereas repeated concatenation gives rise to sectors with arbitrary $|n|>1$.

\smallskip

As before, it follows from the solution that, when $\tilde n=\pm n$, the branch of \eqref{eq:prho-factored} should be chosen such that $\sqrt{-V(\rho)}\sim \pm\,(p_t-q_e/\rho)$ as $\rho\to\infty$. The associated residue is then
\begin{equation}
\mathrm{Res}_{\rho =\infty}\!\left(\sqrt{-V(\rho)}\right) = \pm\, q_e\,,
\end{equation}
so that once again the two sign choices compensate, and we find
\begin{equation}
S_{\rm inst} = \oint p_\rho\, d\rho
= -2\pi i\, \tilde n\,\mathrm{Res}_{\rho=\infty}\!\left(\sqrt{-V(\rho)}\right)
= -2\pi in q_e\,,
\label{eq:Sinst-ads-resinf}
\end{equation}
in agreement with the direct evaluation and with the $q_e$ term of \eqref{eq:InstantonActionAdSMixed}.

\medskip

Taken together, the two phases display the full residue structure of $p_\rho$. Indeed, the boundary pole at $\rho=0$ supplies the real $w\beta$ contribution \eqref{eq:Sinst-ads-res0}, whereas the point at infinity gives rise to the purely imaginary $q_e$ contribution \eqref{eq:Sinst-ads-resinf}. Their sum reproduces
\begin{equation}
S_{\rm inst} = \oint p_\rho\, d\rho
= -2\pi\left( w\beta + in q_e\right)\,,
\end{equation}
thereby providing a derivation entirely in terms of Cauchy's residue theorem, in complete parallel with the Minkowski analysis of Section~\ref{ss:HamiltonianConstraintMink}.

\subsection{The super-extremal case}\label{ss:Superextremal}

\subsubsection{Real proper time solutions}\label{sss:InterferenceSolsSuperExtremal}

In the previous subsection, we derived the explicit form of the exact solutions to the classical EOM \eqref{eq:EOMAdS2} and analyzed their properties under the assumption $mR>|q_e|$. Here we consider instead the \emph{super-extremal} regime, wherein the charged relativistic particle satisfies the opposite inequality, $mR<|q_e|$. Crucially, since the relevant qualitative features of the solutions are controlled by the dimensionless combination $\beta=\sqrt{\tilde m^2-q_e^2}$, the super-extremal motions may be obtained directly from the sub-extremal ones through the analytic continuation
\begin{equation}
\beta \to i\kappa\,,\qquad \kappa\equiv\sqrt{q_e^2-\tilde m^2}\,,
\label{eq:beta-to-kappa}
\end{equation}
applied to the instanton maps \eqref{eq:RhoSolutionRealTau} and \eqref{eq:TSolutionRealTau}. This yields
\begin{subequations}\label{eq:RealTauInstantonSolsSuperExtremal}
\begin{align}
p_t\,\rho(\tau)
&= \frac{\kappa^2}{\,q_e-i\tilde m\,\sinh\!\left[R^{-2}\kappa(\tau+\tau_0)\right]}\,,
\\[4pt]
p_t\,t(\tau)
&= p_t t_0 -\frac{i\tilde m\kappa\,\cosh\!\left[R^{-2}\kappa(\tau+\tau_0)\right]}
{\,q_e-i\tilde m\,\sinh\!\left[R^{-2}\kappa(\tau+\tau_0)\right]}\,,
\end{align}
\end{subequations}
with $\tau_0$ an arbitrary complex constant.

\medskip

%%%%%%%%%%%%%%%%%%%%%%%%%%%%%%%%%%%%%%%%%%%%%
\begin{figure}[t!]
    \centering
\subcaptionbox{\label{sfig:TcoordinateReTauAdSSuperExt}}{
		\includegraphics[scale=1]{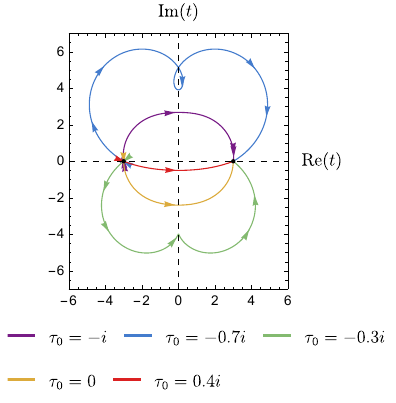}}
        \hspace{0.6cm}
        \subcaptionbox{\label{sfig:RhocoordinateReTauAdSSuperExt}}{\includegraphics[scale=1]{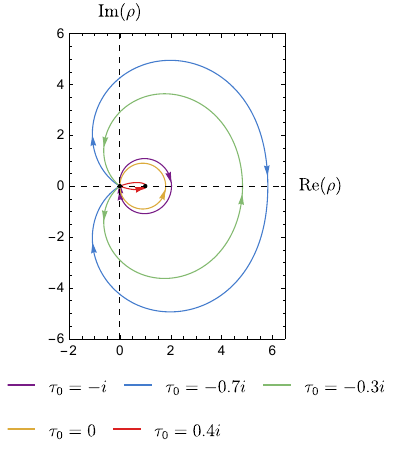}}
		\caption{\small The solutions for $\rho(\tau)$ and $t(\tau)$ in real proper time are closed and open, respectively, with $\tau \in (-\infty, \infty)$. The full trajectories always enclose the root $\rho_{-\varepsilon}$, leaving the other one outside. We plot various representative cases with $\tilde{m}=4$, $q_e=5$, $p_t =R=1$, $t_0=0$, and different values~of~$\tau_0$.} 
		\label{fig:TrajectoriesRealTauAdSSuperExt}
\end{figure}
%%%%%%%%%%%%%%%%%%%%%%%%%%%%%%%%%%%%%%%%%%%%%
When evolving in real proper time, these paths exhibit a qualitative behavior that closely resembles that of the sub-extremal solutions in the \emph{instanton} phase, rather than the interference phase (cf.\ \eqref{eq:InstantonSolsAdS}). This is the first manifestation of a general feature of the super-extremal regime: the roles of the real and imaginary proper-time evolutions are effectively exchanged, as anticipated by the continuation \eqref{eq:beta-to-kappa}. In Figure~\ref{fig:TrajectoriesRealTauAdSSuperExt}, we plot the spacetime motion for several representative values of $\tau_0$. One can see that the trajectories close in the complex $\rho$-plane only over the infinite range $\tau\in(-\infty,\infty)$, along which they start and end at the conformal boundary $\rho=0$.\footnote{These orbits always enclose the root of the effective potential $V(\rho)$, defined in \eqref{eq:EffectivePotentialAdS}, that lies nearest the origin, namely $\rho_{-\varepsilon}=(q_e-\varepsilon\tilde m)/p_t$, where $\varepsilon=\text{sgn}(q_e)$, while leaving the other root outside; cf.\ Section~\ref{ss:MixedSolutionsSubAdS}.} Correspondingly, in the complex $t$-plane the paths are open and connect the points
\begin{equation}
t_\pm = t_0 \pm \frac{\kappa}{p_t}\,,
\end{equation}
which---in contrast to the sub-extremal analogues $t_0\pm i\beta/p_t$---now lie on the \emph{real} time axis, and are reached as $\tau\to\pm\infty$. Indeed, from \eqref{eq:RealTauInstantonSolsSuperExtremal} one finds $p_t t\to p_t t_0\pm\kappa$ in these limits, where the hyperbolic sine and cosine grow without bound.

\medskip

Concerning the dependence on the integration constant $\tau_0$, one first notices that \eqref{eq:RealTauInstantonSolsSuperExtremal} are left invariant under
\begin{equation}
\tau_0 \to \tau_0+\frac{2\pi i R^2}{\kappa}\,,
\label{eq:tau0-periodicity-super}
\end{equation}
a feature inherited from the periodicity of the instanton-phase solutions (see Section~\ref{ss:MixedSolutionsSuperAdS}). The spacetime trajectories are generically complex. However, for the special values\footnote{Recall that the real part of $\tau_0$ can always be absorbed in $\tau$ via ordinary proper-time shifts.}
\begin{equation}
	\label{eq:tau0-special-super}
\tau_0\in\mathbb{R}+\frac{\pi i R^2}{\kappa}\left(\mathbb Z+\frac12\right)\,,
\end{equation}
they degenerate into purely real motions. Modulo the periodicity \eqref{eq:tau0-periodicity-super}, these fall into two inequivalent classes. For $\tau_0=(2+\varepsilon)\,i\pi R^2/(2\kappa)$, where $\varepsilon=\text{sgn}(q_e)$, the denominator in \eqref{eq:RealTauInstantonSolsSuperExtremal} vanishes at a finite real $\tau$, so that the trajectory is open and divergent, passing through $\rho=\pm\infty$ and reaching $\rho=\rho_{\varepsilon}$ on the other side. By contrast, for $\tau_0=(2-\varepsilon)\,i\pi R^2/(2\kappa)$, the denominator has no real zero, and the path is thus bounded, running along the segment joining $\rho=0$ with $\rho_{-\varepsilon}$. We emphasize that in both cases the worldline action diverges.

\subsubsection{Generalization}\label{ss:MixedSolutionsSuperAdS}

\begin{figure}[t!]
    \centering
\subcaptionbox{\label{sfig:TcoordinateImTauAdSSuperExt}}{
		\includegraphics[scale=1]{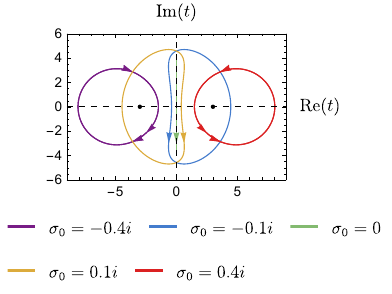}}
        \hspace{0.6cm}
        \subcaptionbox{\label{sfig:RhocoordinateImTauAdSSuperExt}}{\includegraphics[scale=1]{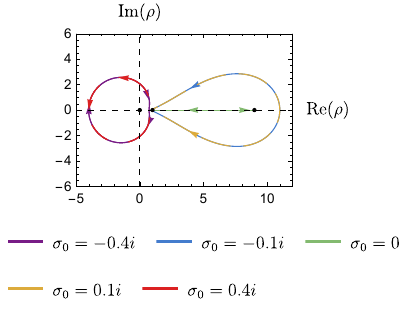}}
		\caption{\small The solutions for $\rho(\sigma)$ and $t(\sigma)$ in imaginary proper time are generically complex and closed, with period $2\pi R^2/\sqrt{q_e^2-\tilde{m}^2}$. The exceptional cases with $\sigma_0\in \mathbb{R}$ degenerate to open Euclidean trajectories, where $\rho(\sigma)$ becomes real and $t(\sigma)-t_0$ purely imaginary. The trajectories are symmetric under $\text{Im}(\sigma_0)\to-\text{Im}(\sigma_0)$: in the complex $\rho$-plane, this transformation preserves the path while reversing its orientation, whereas in the complex $t$-plane it reflects the motion across the line $t_0+i \mathbb{R}$. We plot various representative cases with $\tilde{m}=4$, $q_e=5$, $p_t R=1$, $t_0=0$, and different values of $\sigma_0$.} 
		\label{fig:TrajectoriesImaginaryTauAdSSuperExt}
\end{figure}

We can again extend the analysis above by considering instanton solutions that evolve along generic directions in the complex $\tau$-plane. The simplest of these are obtained by Wick rotating proper time in \eqref{eq:RealTauInstantonSolsSuperExtremal} according to $\tau \to -i\sigma$, with $\sigma \in \mathbb{R}$. This yields
\begin{subequations}\label{eq:InstantonSolsAdSSuperExt}
\begin{equation}
p_t \rho (\sigma)= \frac{\kappa^2}{q_e-\tilde{m} \sin \left[ R^{-2}\kappa (\sigma+\sigma_0)\right]}\, ,
\end{equation}
\begin{equation}
p_t t(\sigma)= p_t t_0 -\frac{i\tilde{m}\kappa\cos \left[ R^{-2}\kappa (\sigma+\sigma_0)\right]}{q_e-\tilde{m} \sin \left[ R^{-2}\kappa (\sigma+\sigma_0)\right]}\, .
\end{equation}
\end{subequations}
Features of these solutions are illustrated in Figure~\ref{fig:TrajectoriesImaginaryTauAdSSuperExt}, where we plot the resulting target-space trajectories for several representative values of the complex integration constant $\sigma_0$. These motions mirror the sub-extremal \emph{interference} phase of Section~\ref{sss:solutionsSubAdS}, thus completing the exchange of roles between real and imaginary proper-time evolution anticipated by the continuation \eqref{eq:beta-to-kappa}. In particular, whereas the real-$\tau$ trajectories run off to
the conformal boundary and require the infinite range $\tau\in(-\infty,\infty)$ to close, the maps \eqref{eq:InstantonSolsAdSSuperExt} are periodic in $\sigma$, with period $2\pi R^2/\kappa$, and hence lead to genuinely cyclic orbits. Accordingly, for $\rho(\sigma)$ and $t(\sigma)$ to describe an instanton, the worldline parameter must complete an integer number of periods. We may therefore take
\begin{equation}
\sigma\in \left[0, \frac{2\pi w_\sigma R^2}{\kappa} \right]\,,\qquad w_\sigma\in \mathbb{Z}\, , 
\end{equation}
up to arbitrary constant shifts, with the sign of $w_\sigma$ specifying the orientation. The resulting worldline action is now \emph{finite} and, as in the previous cases, independent of the integration constant $t_0$, whereas its dependence on $\sigma_0$ is more subtle (cf.\ \eqref{eq:Sinst-super-twofamilies} below).

%%%%%%%%%%%%%%%%%%%%%%%%%%%%%%%%%%%%%%%%%%%%%
%%%%%%%%%%%%%%%%%%%%%%%%%%%%%%%%%%%%%%%%%%%%%
\smallskip

\begin{figure}[t!]
	\begin{center}
		\includegraphics[scale=1]{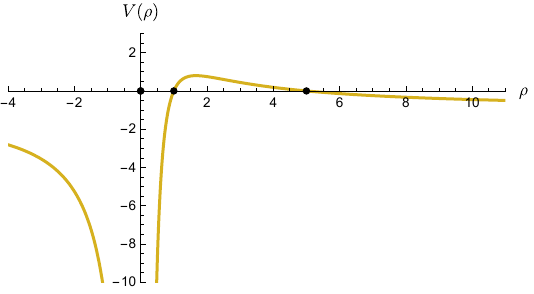}
		\caption{\small Effective potential $V(\rho)$ controlling the super-extremal dynamics in AdS$_2$ with real proper time. The black dots denote the turning points and we set $q_e=3$, $\tilde{m}=2$, $p_t=1$.} 
		\label{fig:EffPotentialRealTauAdSSuperExt}
	\end{center}
\end{figure}

\noindent The spacetime trajectories are generically complex. For the special values $\sigma_0\in\mathbb R$, however, $\rho(\sigma)$ becomes purely real while $t(\sigma)-t_0$ is purely imaginary, and the solutions reduce to the familiar Euclidean instantons---which become entirely real upon Wick rotating the target-space time coordinate. In the $\rho$-direction, these trajectories degenerate to the real segment joining the two turning points $\rho_\pm=(q_e\pm\tilde m)/p_t$ of the effective potential \eqref{eq:EffectivePotentialAdS} (see Figure~\ref{fig:EffPotentialRealTauAdSSuperExt}).

%%%%%%%%%%%%%%%%%%%%%%%%%%%%%%%%%%%%%%%%%%%%%
%%%%%%%%%%%%%%%%%%%%%%%%%%%%%%%%%%%%%%%%%%%%%

\medskip

\noindent Concerning the dependence on $\sigma_0$, the motions behave symmetrically under $\sigma_0\to \bar{\sigma}_0$: for the $\rho$-projection, this transformation preserves the trajectory while reversing its orientation, whereas for the complex $t$-projection it reflects the path across the line $t_0+ i \mathbb{R}$. In addition, as $\sigma_0\to\pm i\infty$, the loops become asymptotically circular. More precisely, in the complex $\rho$-plane, they shrink towards the boundary $\rho=0$, whereas in the complex $t$-plane they collapse onto $t_\pm = t_0 \pm \frac{\kappa}{p_t}$, respectively.

\medskip 

Finally, the closed orbits in the complex $\rho$-plane fall into two inequivalent families, which are distinguished by whether the loop encircles the boundary pole $\rho=0$ or, instead, the two branch
points $\rho_\pm$ of $V(\rho)$. Both enclosures are of course available in the sub-extremal case as well. There, however, they are realized in \emph{different} proper-time phases---the loop around
$\rho=0$ arises in the interference phase and yields the residue \eqref{eq:Sinst-ads-res0}, whereas the contour around the turning points occurs only in the instanton phase and contributes as in
\eqref{eq:Sinst-ads-resinf}. The distinctive feature of the super-extremal regime is that, because $\rho =0 \notin [\rho_-, \rho_+]$,\footnote{Here we have assumed $\rho_-<\rho_+$. Note that this ordering depends on the signs of $q_e$ and $p_t$ and may be reversed. Either way, one always has that $\text{sgn} (\rho_- \rho_+)=1$, so the conclusion remains unchanged.} the two enclosures now occur within a single phase: the same family of solutions \eqref{eq:InstantonSolsAdSSuperExt} wraps either the pole or the roots, depending on the imaginary part of the integration constant $\sigma_0$. Thus, for small $|\mathrm{Im}(\sigma_0)|$ the orbit stays close to the Euclidean segment joining the roots and encircles the branch points, whereas for large $|\mathrm{Im}(\sigma_0)|$ it wraps the boundary pole---consistently with the collapse onto $\rho=0$ found above as $\sigma_0\to\pm i\infty$. The two regimes are separated by the critical value $|\mathrm{Im}(\sigma_0)|=\psi$, with
\begin{equation}
\psi \equiv \frac{R^2}{\kappa}\,\arcosh\!\left(\frac{|q_e|}{\tilde m}\right)\,.
\label{eq:psi-transition-super}
\end{equation}
Correspondingly, the two families carry different worldline actions,
\begin{equation}
S_{\rm inst} =
\begin{cases}
\;2\pi i w_\sigma\bigl(|q_e| - \kappa\bigr)\,, & |\mathrm{Im}(\sigma_0)| \in [0,\psi)\,,\\[6pt]
-2\pi i w_\sigma\kappa\,, & |\mathrm{Im}(\sigma_0)| \in (\psi,\infty)\,.
\end{cases}
\label{eq:Sinst-super-twofamilies}
\end{equation}
Notice that the second line is precisely the analytic continuation via \eqref{eq:beta-to-kappa} of the sub-extremal interference result \eqref{SinstSubAdS}, now rendered purely imaginary by the super-extremal kinematics. The first one, on the other hand, differs from it by the contribution of the winding around the point at infinity in the $\rho$-plane. Notably, upon identifying the pair-production exponent with $i S_{\rm inst}$, the $ w_\sigma=1$ term of the first line reproduces the known AdS$_2$ Schwinger rate \cite{Pioline:2005pf}. In this sense, the enclosed singularities directly control the value of the action, in complete analogy with the residue structure uncovered in Section~\ref{sss:CauchyAdSSubExtremal}.

\smallskip

The transition between the two families moreover occurs exactly at the critical value \eqref{eq:psi-transition-super}, i.e., when $\sigma_0=\pm i\psi$. There, the orbit degenerates, and both $\rho(\sigma)$ and $t(\sigma)$ develop divergences at
\begin{equation}
\sigma= \frac{\pi R^2}{\kappa}\left[\frac{\text{sgn}(q_e) }{2}+2 k\right]\,,\qquad \text{with}\quad k\in\mathbb{Z}\,.
\end{equation}
In the $\rho$-plane the closed loop pinches at the boundary, and the winding around the point at infinity jumps by $ w_\sigma\in \mathbb{Z}$ units when switching between the two sets of \eqref{eq:Sinst-super-twofamilies}.

\medskip

\begin{figure}[H]
    \centering
\subcaptionbox{\label{sfig:NoLoopTauAdSsuper}}{
		\includegraphics[scale=1]{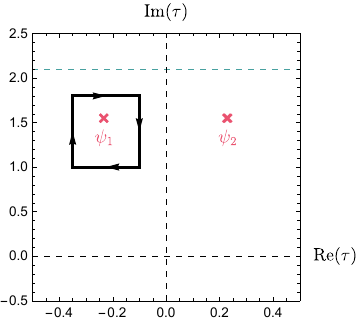}}
        \hspace{0.6cm}
        \subcaptionbox{\label{sfig:LoopTauAdSsuper}}{\includegraphics[scale=1]{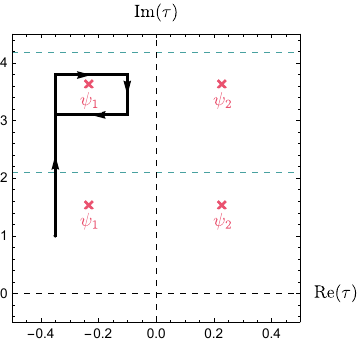}}
		\caption{\small We obtain different instanton trajectories by constructing closed loops in the complex proper-time cylinder. These are grouped into distinct sectors depending on two integers, $w$ and $n=n_{\psi_1}-n_{\psi_2}$, characterizing the $\tau$-contour. The latter control the worldline action through \eqref{eq:InstantonActionAdSMixedSuper}. To make the plots, we chose $\tilde{m}=5$, $q_e=4$, $p_t =R=1$, with $\textbf{(a)}$ $w=0, n=-1$, and $\textbf{(b)}$ $w=-n=1$.} 
		\label{fig:ClosedPathsTauAdSsuper}
\end{figure}

\begin{figure}[H]
    \centering
\subcaptionbox{\label{sfig:NoLoopTAdSsuper}}{
		\includegraphics[scale=1]{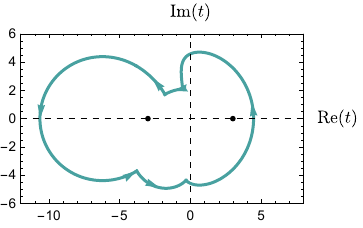}}
        \hspace{0.6cm}
        \subcaptionbox{\label{sfig:NoLoopRhoAdSsuper}}{\includegraphics[scale=1]{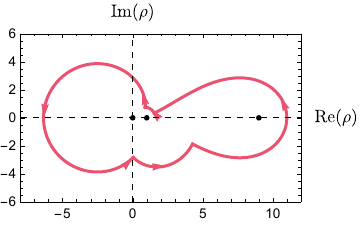}}
		\caption{\small Time $\textbf{(a)}$ and space $\textbf{(b)}$ trajectories corresponding to the path shown in Figure~\ref{sfig:NoLoopTauAdSsuper}.} 
		\label{fig:ImageNoLoopSpacetimeAdSsuper}
\end{figure}

\begin{figure}[H]
    \centering
\subcaptionbox{\label{sfig:LoopTAdSsuper}}{
		\includegraphics[scale=1]{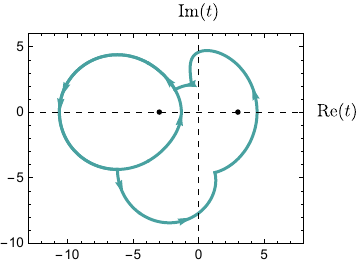}}
        \hspace{0.6cm}
        \subcaptionbox{\label{sfig:LoopRhoAdSsuper}}{\includegraphics[scale=1]{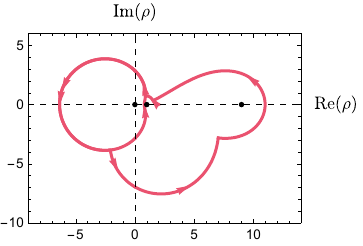}}
		\caption{\small Time $\textbf{(a)}$ and space $\textbf{(b)}$ motions associated with the $w=-n=1$ path in Figure~\ref{sfig:LoopTauAdSsuper}.} 
		\label{fig:ImageLoopSpacetimeAdSsuper}
\end{figure}

Next, we turn to the general case, obtained by analytically continuing \eqref{eq:RealTauInstantonSolsSuperExtremal} to the full complex proper-time plane. For convenience, we set $t_0=\tau_0=0$, so that
\begin{equation}\label{eq:InstantonSolsSuperExtremalNoTau0}
p_t \rho (\tau)=\frac{\kappa^2}{\,q_e-i\tilde m\,\sinh\!\left(R^{-2}\kappa\, \tau\right)}\, ,\qquad  p_t t(\tau)= -\frac{i\tilde m\kappa\,\cosh\!\left(R^{-2}\kappa\, \tau\right)}
{\,q_e-i\tilde m\,\sinh\!\left(R^{-2}\kappa\, \tau\right)}\, .
\end{equation}
As in the Minkowski and sub-extremal analyses of Sections~\ref{ss:MixedSolutionsMink} and \ref{ss:MixedSolutionsSubAdS}, we now provide a prescription for constructing and classifying all topologically inequivalent super-extremal instanton sectors, each characterized by
its worldline action. To this end, we consider closed paths in the complex $\tau$-plane, whose images under \eqref{eq:InstantonSolsSuperExtremalNoTau0} describe the corresponding spacetime trajectories. Moreover, we note that, owing to the periodicity
\begin{equation}
\tau \sim \tau + \frac{2\pi i R^2}{\kappa}\,,
\label{eq:periodicity-super-general}
\end{equation}
of the maps \eqref{eq:InstantonSolsSuperExtremalNoTau0}, the complex proper-time plane is again more appropriately regarded as an infinite cylinder---though now, in contrast to the sub-extremal case, it is the \emph{imaginary} axis that becomes compact. Thus, given the single-valuedness of the instanton maps, it suffices to consider contours that close on this cylinder in order to generate cyclic spacetime trajectories. Such loops are distinguished by the number of times they wind around the periodic direction and by their winding numbers around two \emph{punctures} located at
\begin{equation}\label{eq:PuncturesSuperExtremal}
\tau=\psi_{1} = -\varepsilon\psi +\frac{(2+\varepsilon)\pi i R^2}{2\kappa}\,,\qquad
\tau=\psi_{2} = \varepsilon \psi +\frac{(2+\varepsilon)\pi i R^2}{2\kappa}\,,
\end{equation}
together with their periodic images. Here, we set $\varepsilon=\text{sgn} (q_e)$, while $\psi$ was defined in \eqref{eq:psi-transition-super}. These are precisely the points at which the spacetime trajectories become singular. As in the sub-extremal case, the two punctures carry opposite weights, so that encircling both yields a vanishing net weighted contribution to the action. This defines equivalence classes of paths according to whether they can be continuously deformed into one another; see Figures~\ref{fig:ClosedPathsTauAdSsuper}--\ref{fig:ImageLoopSpacetimeAdSsuper}.

%%%%%%%%%%%%%%%%%%%%%%%%%%%%%%%%%%%%%%%%%%%%%
%%%%%%%%%%%%%%%%%%%%%%%%%%%%%%%%%%%%%%%%%%%%%
%%%%%%%%%%%%%%%%%%%%%%%%%%%%%%%%%%%%%%%%%%%%%
%%%%%%%%%%%%%%%%%%%%%%%%%%%%%%%%%%%%%%%%%%%%%
%%%%%%%%%%%%%%%%%%%%%%%%%%%%%%%%%%%%%%%%%%%%%
%%%%%%%%%%%%%%%%%%%%%%%%%%%%%%%%%%%%%%%%%%%%%

\smallskip

\noindent The essential novel feature of the super-extremal analysis is the location of these punctures. In Section~\ref{ss:MixedSolutionsSubAdS}, the analogous $\theta_{1,2}$ were seen to lie on the real $\tau$-axis, aligned with the periodic direction, and a single winding $w$ sufficed to label the instanton-phase sectors (cf.\ Figure~\ref{fig:ClosedPathsTauAdSsub}). Here, by contrast, the punctures \eqref{eq:PuncturesSuperExtremal} are displaced off the real axis to $\mathrm{Im}(\tau)=(2+\varepsilon)\pi R^2/2\kappa$ and sit at points with opposite real parts $\pm\psi$, thereby splitting the cylinder into two branches. It would then seem that one must keep track of two independent instanton windings $w_i\in\mathbb Z$, counting the wraps around the periodic direction within the $i^{\rm th}$ branch,\footnote{These branches are nothing but the two families of solutions previously found in the instanton phase, which were characterized by their worldline action according to \eqref{eq:Sinst-super-twofamilies}.} in addition to the winding numbers $n_{\psi_i}\in\mathbb Z$ around each puncture. With these data, every closed contour yields a worldline action of the form
\begin{equation}\label{eq:InstantonActionAdSMixedSuper}
S_{\rm inst}= \oint p_\rho\, d\rho
= 2\pi i \left[\,(w_1+w_2)\,\kappa - \bigl(n_{\psi_1}-n_{\psi_2}+\text{sgn}(q_e)w_2\bigr)\,q_e\,\right]\,,
\end{equation}
where, as before, the $n_{\psi_i}$ are positive (negative) for counterclockwise (clockwise) orientation in the complex $\tau$-plane, and---as in the sub-extremal case---the result is independent of the chosen representative, hence purely \emph{topological}.

\medskip

The two windings $w_i$ are not, however, independent labels. Because the punctures are displaced from the periodic axis, a loop that winds once around the imaginary direction on the second branch cannot be deformed to one that winds around the first branch without sweeping across a puncture: the two are homologous only up to an additional unit of winding around $\psi_1$ or $\psi_2$. In other words, a unit of $w_2$ can always be traded for a unit of $w_1$ together with a compensating shift of the puncture windings,
\begin{equation}\label{eq:WindingTradeSuper}
w_2 \;\longrightarrow\; w_2-1\,,\qquad w_1\;\longrightarrow\; w_1+1\,,\qquad
n\;\longrightarrow\; n + \varepsilon\,,
\end{equation}
where $n\equiv n_{\psi_1}-n_{\psi_2}$. Under this deformation, the combinations
\begin{equation}
w\equiv w_1+w_2\,,
\qquad N\equiv n_{\psi_1}-n_{\psi_2}+ \varepsilon w_2\, ,
\end{equation}
are manifestly invariant, and hence so is the action \eqref{eq:InstantonActionAdSMixedSuper} (see Figure~\ref{fig:ClosedPathsTauAdSsuper}). These two quantities may therefore be taken as the independent topological data. Equivalently, one may use the freedom \eqref{eq:WindingTradeSuper} to choose a representative with $w_2=0$, in which case $w$ and $N$ coincide with $w_1$ and $n$, respectively. In terms of the invariant data, the instanton action reduces to
\begin{equation}\label{eq:InstantonActionAdSMixedSuperReduced}
S_{\rm inst}= 2\pi i \left(w\kappa - N q_e \right)\,,
\end{equation}
which has exactly the same structure found already in \eqref{eq:InstantonActionAdSMixed}, now continued via $\beta\to i\kappa$. Thus, just as in the sub-extremal regime of Section~\ref{ss:SubExtremal}, a single instanton winding $w$, together with the effective winding number $N$ around punctures, suffices to label all topologically inequivalent super-extremal sectors. The apparent need for a second winding $w_2$ is merely an artifact of describing the cycles relative to the displaced punctures and is removed by the equivalence \eqref{eq:WindingTradeSuper}.

\subsection{The extremal case}\label{ss:Extremal}

\subsubsection{Real proper time solutions}\label{sss:InterferenceSolsExtremal}
\begin{figure}[t!]
	\begin{center}
		\includegraphics[scale=1]{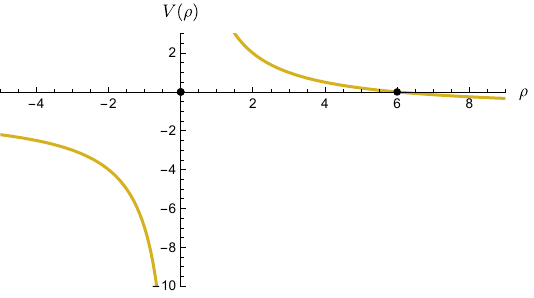}
		\caption{\small Effective potential $V(\rho)$ controlling the extremal dynamics in AdS$_2$ with real proper time. The red black denote the turning points and we set $q_e=\tilde{m}=3$, $p_t=1$.} 
		\label{fig:EffPotentialRealTauAdSExt}
	\end{center}
\end{figure}

Finally, we turn to the remaining possibility, namely the \emph{extremal} regime, characterized by saturation of the coupling relation $mR=|q_e|$. As in the previous subsections, we obtain the corresponding worldline instanton maps by directly integrating the first-order differential equation following from the Hamiltonian constraint \eqref{eq:DiffEqAdS2}, after imposing $|q_e|=mR$:
\begin{equation}\label{eq:DiffEqAdS2Extremal}
\frac{R^2\,d\rho}{\rho\,\sqrt{p_t^2\rho^2-2p_t q_e\rho}}=d\tau\, ,
\end{equation}
whose solutions are
\begin{subequations}\label{eq:RealTauInstantonSolsExtremal}
\begin{align}
p_t\,\rho(\tau) &= \frac{2q_e}{\,1-R^{-4}q_e^2\,(\tau+\tau_0)^2}\, ,
\\[4pt]
p_t\,t(\tau) &= p_t t_0 -\frac{2q_e^2\,(\tau+\tau_0)}{R^2\left[\,1-R^{-4}q_e^2\,(\tau+\tau_0)^2\right]}\, ,
\end{align}
\end{subequations}
with $t_0\in\mathbb R$ an integration constant, as may be readily verified. As before, the constant $\tau_0$ is generically complex. Notice that the extremal maps \eqref{eq:RealTauInstantonSolsExtremal} are \emph{rational} in $\tau+\tau_0$, rather than trigonometric or hyperbolic. This reflects the degenerate limit separating the sub- and super-extremal regimes, in which $\beta, \kappa \to 0$ and the analytic structure of the effective potential changes qualitatively (cf.\ Figure~\ref{fig:EffPotentialRealTauAdSExt}).

%%%%%%%%%%%%%%%%%%%%%%%%%%%%%%%%%%%%%%%%%%%%%
%%%%%%%%%%%%%%%%%%%%%%%%%%%%%%%%%%%%%%%%%%%%%

%%%%%%%%%%%%%%%%%%%%%%%%%%%%%%%%%%%%%%%%%%%%%
\begin{figure}[t!]
    \centering
\subcaptionbox{\label{sfig:TcoordinateReTauAdSExt}}{
		\includegraphics[scale=1]{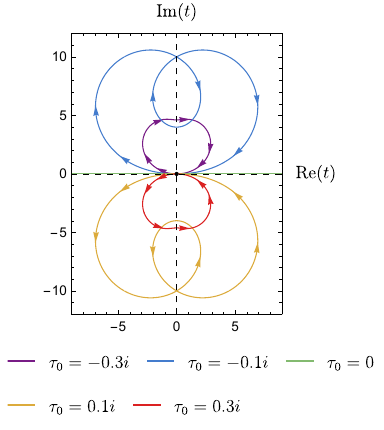}}
        \hspace{0.6cm}
        \subcaptionbox{\label{sfig:RhocoordinateReTauAdSExt}}{\includegraphics[scale=1]{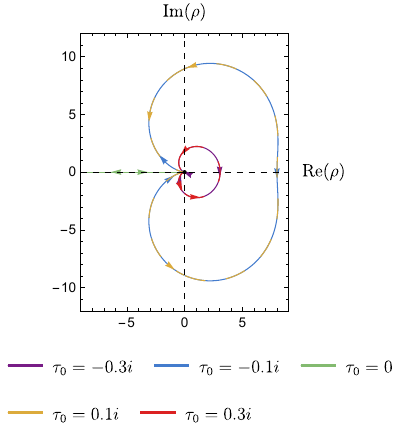}}
		\caption{\small The solutions for $\rho(\tau)$ and $t(\tau)$ in real proper time are both closed over $\tau\in(-\infty,\infty)$. The full trajectories in the $\rho$-plane are confined to the region between the boundary $\rho=0$ and the non-trivial turning point, which they never enclose; their behavior under $\mathrm{Im}(\tau_0)\to-\mathrm{Im}(\tau_0)$ is symmetric, retracing the same $\rho$-path with reversed orientation, whereas the $t$-component is reflected across the real axis. We plot various cases with $\tilde m=5$, $q_e=5$, $p_t=R=1$, $t_0=0$, and different values of $\tau_0$.}
		\label{fig:TrajectoriesRealTauAdSExt}
\end{figure}
%%%%%%%%%%%%%%%%%%%%%%%%%%%%%%%%%%%%%%%%%%%%%

\medskip

When evolving in real proper time, the solutions \eqref{eq:RealTauInstantonSolsExtremal} closely resemble those of the super-extremal case in the corresponding interference-phase (Section~\ref{sss:InterferenceSolsSuperExtremal}). In particular, as shown in Figure~\ref{fig:TrajectoriesRealTauAdSExt}, they form closed trajectories in both the complex $\rho$- and $t$-planes only over the infinite range $\tau\in \mathbb{R}$, which start and end at the point $(t,\rho)=(t_0,0)$ as $\tau\to\pm\infty$.\footnote{The motion now closes also in the complex $t$-plane---in contrast to its super-extremal analogue---because the two endpoints $t_\pm$ coalesce at $t_0$ when $|q_e|=mR$, cf.\ Figure~\ref{sfig:TcoordinateReTauAdSSuperExt}.}

\medskip

As for the dependence on the integration constant $\tau_0$, the paths are generically complex, except for the special values $\tau_0\in\mathbb R$, for which they collapse to purely real, open segments in both the space and time coordinates. For generic $\tau_0$, the motions in the complex $\rho$-plane start and end at the boundary locus $\rho=0$ and never enclose the nonzero turning point $\rho=2 \tilde{m}/p_t$. They moreover behave symmetrically under $\tau_0 \to \bar{\tau}_0$, with this involution preserving the $\rho$-trajectory while reversing its orientation. Along the time direction, by contrast, the sign of $\mathrm{Im}(t(\tau))$ is correlated with that of $\mathrm{Im}(\tau_0)$, so that the $t$-path is reflected across the real axis under the same transformation.

\subsubsection{Generalization}\label{ss:MixedSolutionsExtAdS}
%%%%%%%%%%%%%%%%%%%%%%%%%%%%%%%%%%%%%%%%%%%%%
\begin{figure}[t!]
    \centering
\subcaptionbox{\label{sfig:TcoordinateImTauAdSExt}}{
		\includegraphics[scale=1]{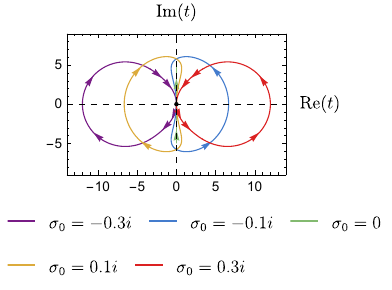}}
        \hspace{0.6cm}
        \subcaptionbox{\label{sfig:RhocoordinateImTauAdSExt}}{\includegraphics[scale=1]{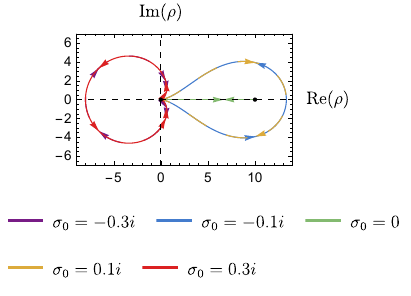}}
		\caption{\small All solutions to the imaginary proper-time EOM for $\rho(\sigma)$ and $t(\sigma)$ close only over the full range $\sigma \in (-\infty, \infty)$. They are generically complex, except for those with $\sigma_0\in \mathbb{R}$, which describe the familiar Euclidean instantons. We show various representative trajectories with $\tilde{m}=5$, $q_e=5$, $p_t =R=1$, $t_0=0$, and different values of the integration constant $\sigma_0$.} 
		\label{fig:TrajectoriesImaginaryTauAdSExt}
\end{figure}
%%%%%%%%%%%%%%%%%%%%%%%%%%%%%%%%%%%%%%%%%%%%%
We now extend the analysis beyond the interference phase by allowing $\tau$ to take generic complex values. For simplicity, we first consider the purely \emph{instanton-phase} solutions, for which the evolution occurs in imaginary proper time. Upon Wick rotating \eqref{eq:RealTauInstantonSolsExtremal} via $\tau \to -i\sigma$ with $\sigma \in \mathbb{R}$, and defining $\sigma_0=i \tau_0$ as usual, we find
\begin{subequations}\label{eq:InstantonSolsAdSExt}
\begin{equation}
p_t \rho (\sigma)= \frac{2q_e}{1+R^{-4}q_e^2 (\sigma+\sigma_0)^2}\, ,
\end{equation}
\begin{equation}
p_t t(\sigma)= p_t t_0 +\frac{2 i q_e^2(\sigma+\sigma_0)}{R^2 \left[1+R^{-4}q_e^2 (\sigma+\sigma_0)^2\right]}\, .
\end{equation}
\end{subequations}
We plot several representative trajectories for $\rho(\sigma)$ and $t(\sigma)$ in Figure~\ref{fig:TrajectoriesImaginaryTauAdSExt} for different values of the integration constant $\sigma_0$. These motions are closed and generically complex for $\sigma_0 \in \mathbb{C}$. However, when $\sigma_0 \in \mathbb{R}$, $\rho(\sigma)$ becomes purely real while $t(\sigma)-t_0$ becomes purely imaginary, corresponding to the familiar Euclidean instanton solutions \cite{Castellano:2025yur}. Notice that the periodicity found for super-extremal particles in Section~\ref{ss:MixedSolutionsSuperAdS} is pushed to infinite imaginary proper time in the extremal limit. Indeed, the corresponding period $2\pi R^2/\kappa$ diverges as $\kappa\to 0$. We also note that, in the complex $\rho$-plane, the paths are invariant up to orientation under $\sigma_0 \to \bar{\sigma}_0$, whereas the $t$-trajectories are reflected across the line $t_0+ i \mathbb{R}$. In addition, as $\sigma_0 \to \pm i\infty$, the paths shrink towards $\rho = 0$ and $t = t_0$, respectively, in the two projections. Finally, the trajectories in both the spatial and temporal projections fall into two distinct families, with transitions between them occurring whenever the configuration-space trajectory becomes singular at $\sigma = 0$. From \eqref{eq:InstantonSolsAdSExt}, this happens when
\begin{equation}
\sigma_0 = \pm \frac{i R^2}{q_e}\, .
\end{equation}

\medskip

Next, we consider the general case by analytically continuing \eqref{eq:RealTauInstantonSolsExtremal} to the entire complex $\tau$-plane. Without loss of generality, we set $t_0=\tau_0=0$, so that
\begin{equation}\label{eq:InstantonSolsExtremalNoTau0}
p_t \rho (\tau)=\frac{2q_e}{\,1-R^{-4}q_e^2\,\tau^2}\, ,\qquad  p_t t(\tau)= -\frac{2q_e^2\,\tau}{R^2\left(\,1-R^{-4}q_e^2\,\tau^2\right)}\, .
\end{equation}
As in the Minkowski, sub-extremal, and super-extremal analyses of Sections~\ref{ss:MixedSolutionsMink}, \ref{ss:MixedSolutionsSubAdS} and \ref{ss:MixedSolutionsSuperAdS}, we now provide a prescription for constructing and classifying all topologically inequivalent extremal instanton sectors, each characterized by its worldline action. To this end, we consider closed contours in the complex $\tau$-plane whose images under \eqref{eq:InstantonSolsExtremalNoTau0} describe the corresponding spacetime trajectories.

\smallskip

\begin{figure}[H]
    \centering
\subcaptionbox{\label{sfig:NoLoopTauAdSExt}}{
		\includegraphics[scale=1]{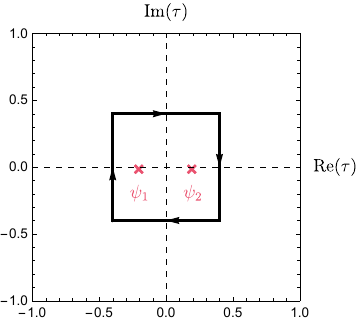}}
        \hspace{0.6cm}
        \subcaptionbox{\label{sfig:LoopTauAdSExt}}{\includegraphics[scale=1]{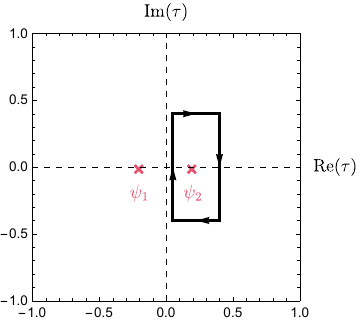}}
		\caption{\small We obtain different instanton maps by constructing closed loops in complex proper-time $\tau$. These are grouped into distinct sectors depending on the effective winding number $n=n_{\psi_1}-n_{\psi_2}\in\mathbb Z$ of the $\tau$-contour, which controls the worldline action through \eqref{eq:InstantonActionAdSMixedExt}. To make the plots we chose $\tilde m=5$, $q_e=5$, $p_t=R=1$, with $\textbf{(a)}$ $n=0$ and $\textbf{(b)}$ $n=1$.}
		\label{fig:ClosedPathsTauAdSExt}
\end{figure}

\begin{figure}[H]
    \centering
\subcaptionbox{\label{sfig:NoLoopTAdSExt}}{
		\includegraphics[scale=1]{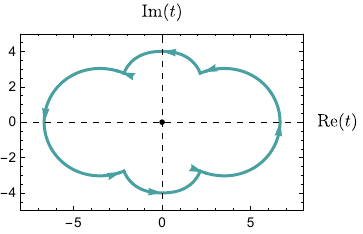}}
        \hspace{0.6cm}
        \subcaptionbox{\label{sfig:NoLoopRhoAdSExt}}{\includegraphics[scale=1]{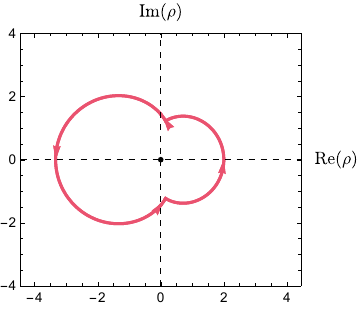}}
		\caption{\small Time $\textbf{(a)}$ and space $\textbf{(b)}$ trajectories corresponding to the
		path shown in Figure~\ref{sfig:NoLoopTauAdSExt}.}
		\label{fig:ImageNoLoopSpacetimeAdSExt}
\end{figure}

\begin{figure}[H]
    \centering
\subcaptionbox{\label{sfig:LoopTAdSExt}}{
		\includegraphics[scale=1]{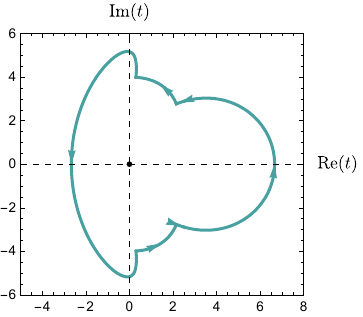}}
        \hspace{0.6cm}
        \subcaptionbox{\label{sfig:LoopRhoAdSExt}}{\includegraphics[scale=1]{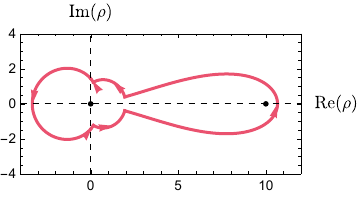}}
		\caption{\small Time $\textbf{(a)}$ and space $\textbf{(b)}$ motions associated with the
		$n=1$ path in Figure~\ref{sfig:LoopTauAdSExt}.}
		\label{fig:ImageLoopSpacetimeAdSExt}
\end{figure}

An important difference relative to the previous cases now arises: the extremal solutions exhibit no finite periodicity---the period $2\pi R^2/\kappa$ of the super-extremal regime having diverged as $\kappa\to0$. Consequently, there is no longer a periodic identification of the complex proper-time plane, and cyclic spacetime motions are obtained simply from contours that close in the $\tau$-plane itself. Therefore, the relevant contour classes are characterized by their winding numbers around two \emph{punctures} located at
\begin{equation}\label{eq:PuncturesExt}
\tau=\psi_{1} = -\frac{R^2}{q_e}\,,\qquad \tau=\psi_{2} = \frac{R^2}{q_e}\,,
\end{equation}
which, as usual, correspond to the points where the spacetime maps \eqref{eq:InstantonSolsExtremalNoTau0} become singular. As in the sub- and super-extremal cases, the two punctures carry opposite weights in the worldline action, so that a contour encircling both with equal winding has a vanishing net contribution. This defines equivalence classes of paths according to their winding data; see Figures~\ref{fig:ClosedPathsTauAdSExt}--\ref{fig:ImageLoopSpacetimeAdSExt}. One then finds that every closed contour yields a worldline action of the form
\begin{equation}\label{eq:InstantonActionAdSMixedExt}
S_{\rm inst}= \oint p_\rho\, d\rho = -2\pi i\, (n_{\psi_1}-n_{\psi_2})\,q_e\,,
\end{equation}
where $n_{\psi_i}\in\mathbb Z$ denote the winding numbers around the punctures \eqref{eq:PuncturesExt}, taken to be positive (negative) for counterclockwise (clockwise) orientation in the complex $\tau$-plane. As before, the action is independent of the chosen representative and depends only on these winding data.

\medskip

The structure of \eqref{eq:InstantonActionAdSMixedExt} makes the degenerate nature of the extremal limit manifest. Whereas the sub- and super-extremal actions \eqref{eq:InstantonActionAdSMixed} and \eqref{eq:InstantonActionAdSMixedSuperReduced} each contain a term proportional to the instanton winding number $w$, with coefficients $\beta$ and $\kappa$, respectively, this contribution vanishes at extremality, consistent with $\beta,\kappa\to0$. The absence of a finite period thus removes the instanton winding altogether, leaving the action entirely determined by the winding numbers $n_{\psi_i}$ around the two punctures \eqref{eq:PuncturesExt}. Equivalently, from the perspective of the radial momentum $p_\rho$, only the $q_e$-dependent contribution associated with the point at infinity in the complex $\rho$-plane survives (cf.\ Section~\ref{sss:CauchyAdSSubExtremal}), while the term in the action previously associated with the residue at the boundary $\rho=0$ vanishes in the extremal limit.
%%%%%%%%%%%%%%%%%%%%%%%%%%%%%%%%%%%%%%%%%%%%%
%%%%%%%%%%%%%%%%%%%%%%%%%%%%%%%%%%%%%%%%%%%%%
%%%%%%%%%%%%%%%%%%%%%%%%%%%%%%%%%%%%%%%%%%%%%
%%%%%%%%%%%%%%%%%%%%%%%%%%%%%%%%%%%%%%%%%%%%%
%%%%%%%%%%%%%%%%%%%%%%%%%%%%%%%%%%%%%%%%%%%%%
%%%%%%%%%%%%%%%%%%%%%%%%%%%%%%%%%%%%%%%%%%%%%

\section{Instanton Solutions in dS\(_2\)}\label{s:dS2}

\begin{figure}
	\centering
	%\fbox{\rule{0pt}{2in}\rule{0.8\textwidth}{0pt}}
	\includegraphics[scale=0.9]{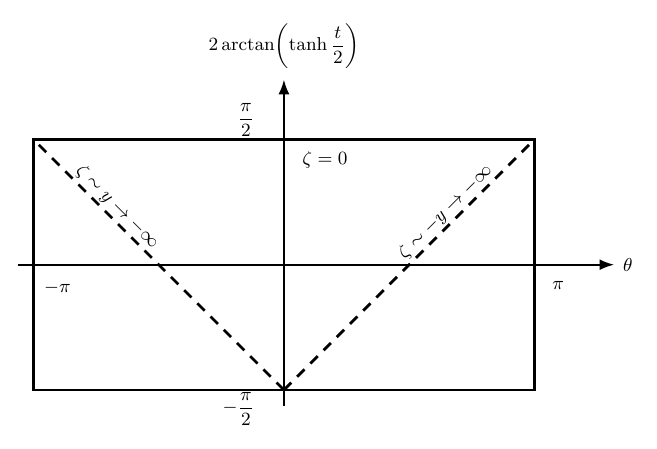}
	\caption{Penrose diagram for two-dimensional de Sitter space. The triangular region bounded by dashed lines represents the flat slicing \eqref{eq:dS2Metric}. See Section~\ref{ss:GlobaldS} for the global~analysis.}
	\label{fig:dS2Penrose}
\end{figure}

We now study the motion of a charged particle under the influence of a homogeneous background electric field \(E\) in two-dimensional de Sitter spacetime (\(\mathrm{dS}_2\)). This is the de Sitter analogue to the problem considered in Section~\ref{s:AdS2}, and much of the solution method can be readily adapted to this case if we choose the right coordinate patch to work in. Specifically, consider the causal future \(J^+(p)\) of a point \(p \in \mathcal{I}^-\) located at the past-infinity conformal boundary of \(\mathrm{dS}_2\). As can be seen from the Penrose diagram in Figure~\ref{fig:dS2Penrose}, this patch only covers half of the spacetime. However, it has the advantage that we can put the metric into a form similar to that of \(\mathrm{AdS}_2\), yielding
\begin{equation}
	\label{eq:dS2Metric}
	ds^2 = \frac{R^2}{\zeta^2}\left(-d \zeta^2 + d y^2\right),
\end{equation}
where \(R\) is the de Sitter radius, \(\zeta \in (-\infty, 0)\) is a temporal coordinate, and \(y \in (-\infty, \infty)\) denotes the spatial coordinate.\footnote{This coordinate patch is usually called the \textit{flat slicing} of de Sitter in reference to the spatial part.} Comparing to the \(\mathrm{AdS}_2\) metric,
\begin{equation}
	ds^2_{\text{AdS}_2} = \frac{R^2}{\rho^2}\left(-d t^2 + d \rho^2\right),
\end{equation}
we note that we can obtain the \(\mathrm{dS}_2\) coordinates from \(\mathrm{AdS}_2\) via the substitutions
\begin{equation}
	\label{eq:dS2AdS2Mapping}
	R_{\mathrm{AdS}_2} \longleftrightarrow - i R_{\mathrm{dS}_2}, \quad \rho \longleftrightarrow -\zeta, \quad t \longleftrightarrow y \, .
\end{equation}
These will be useful as a crosscheck, and we will make note of the analogies between the solutions on both spacetimes throughout this section. For now, we only need to worry about one more aspect in the setup: the background electric field. As in the previous cases, we wish to include a homogeneous $\U(1)$ gauge field strength in this spacetime, namely
\begin{equation}\label{eq:GaugeFieldFlatdS2}
    F = E\, \omega_{\text{dS}_2}=E \sqrt{-\det g}\, d\zeta \wedge d y= E \left(\frac{R}{\zeta}\right)^2 d\zeta \wedge d y\, .
\end{equation}
A convenient gauge choice is therefore \(A = -ER^2\zeta^{-1} \, d y\). Consequently, matching~\eqref{eq:GaugeFieldPoincareAdS} under the mapping in~\eqref{eq:dS2AdS2Mapping} also requires flipping the sign of \(E\) when comparing the gauge potentials---or, equivalently, the field strengths.

%%%%%%%%%%%%%%%%%%%%%%%%%%%%%%%%%%%%%%%%%%%%%
%%%%%%%%%%%%%%%%%%%%%%%%%%%%%%%%%%%%%%%%%%%%%

\medskip

For a particle of charge \(e\), the worldline action \eqref{eq:worldlineAction} in the \(h(\tau)=1\) gauge reduces to
\begin{equation}\label{SwldS}
 S_{\rm wl} = \int_\gamma d\tau\,\left(- \frac{m^2}{2}-\frac{R^2\dot \zeta^2}{2 \zeta^2}+\frac{R^2\dot y^2}{2 \zeta^2} -\frac{eER^2\dot y}{\zeta}  \right)\, ,
\end{equation}
from which one derives the associated EOM~\eqref{eq:EOMNambuGotoReal},
\begin{equation}
\ddot{\zeta} - \frac{\dot{\zeta}^2+\dot{y}^2}{\zeta} = - eE \dot{y}\, ,\qquad \ddot{y} - 2 \frac{\dot{\zeta} \dot{y}}{\zeta} = - eE \dot{\zeta}\, .
\end{equation}
Note that these equations have the exact same functional form as those in~\eqref{eq:EOMAdS2}, except for the replacements described in \eqref{eq:dS2AdS2Mapping} and \(E \leftrightarrow -E\). Hence, to solve them we can proceed in the same manner as in Section~\ref{sss:solutionsSubAdS}. We will use the Hamiltonian constraint, which reads
\begin{equation}
    \label{eq:HamiltonianConstraintdS2}
    g_{\mu\nu}\dot{x}^\mu \dot{x}^\nu = -m^2 \iff \frac{R^2}{\zeta^2}(-\dot{\zeta}^2+\dot{y}^2) = -m^2\,,
\end{equation}
and the fact that \(p_y\), the momentum conjugate to \(y\),
\begin{equation}\label{eq:DefPy}
    p_y = \frac{\partial L}{\partial \dot{y}} = \frac{R^2}{\zeta^2} \dot{y}-\frac{eE R^2}{\zeta}\, ,
\end{equation}
is a conserved quantity. Solving for \(\dot{y}\) and substituting the result into \eqref{eq:HamiltonianConstraintdS2} yields the following differential equation:
\begin{equation}
	\label{eq:zetaODE}
\dot{\zeta}^2 = \frac{\zeta^2}{R^4} \left[(p_y \zeta + eE R^2)^2+(m R)^2\right]\, .
\end{equation}
Keeping notation consistent with the previous section, we then define \(\tilde{m} = m R\), \(q_e = e E R^2\), and \(\Omega = \sqrt{\tilde{m}^2+q_e^2}\). In terms of these variables, the solutions to \eqref{eq:zetaODE} can be written as
\begin{equation}
    \label{eq:zetaSolutiondS2}
    p_y \zeta(\tau) = -\frac{\Omega^2}{q_e+\tilde{m} \sinh \left[R^{-2}\Omega(\tau + \tau_0)\right]}\, ,
\end{equation}
with \(\tau_0 \in \mathbb{C}\) an arbitrary integration constant. Substituting this result into \eqref{eq:DefPy}, we further obtain, for \(y_0\) some real reference constant, the solutions
\begin{equation}
    \label{eq:ySolutiondS2}
    p_y y(\tau) = p_y y_0 -\frac{\tilde{m} \Omega \cosh \left[R^{-2}\Omega(\tau + \tau_0)\right]}{q_e+\tilde{m} \sinh \left[R^{-2}\Omega(\tau + \tau_0)\right]}\, .
\end{equation}
In the process of solving \eqref{eq:zetaODE}, due to the extraction of square roots, one can pick between two different branches of solutions. These are simply related by a proper-time reversal \(\tau\to-\tau\), possibly accompanied by redefinitions of the integration constants. Since later on we consider arbitrary contours in the complex \(\tau\)-plane, the particular forms chosen in \eqref{eq:zetaSolutiondS2} and \eqref{eq:ySolutiondS2} entail no loss of generality. The explicit expressions also show that \(\zeta(\tau)\) and \(y(\tau)\) are periodic, with purely \emph{imaginary} period \(2\pi i R^2/\Omega\). This mirrors the situation in the super-extremal \(\mathrm{AdS}_2\) case (cf.\ Section~\ref{ss:Superextremal}), where the corresponding period is \(2\pi i R^2/\kappa\), and is one of several similarities between de Sitter and super-extremal Anti-de Sitter space.

\smallskip

\begin{figure}[!t]
	\begin{center}
		\includegraphics[scale=1]{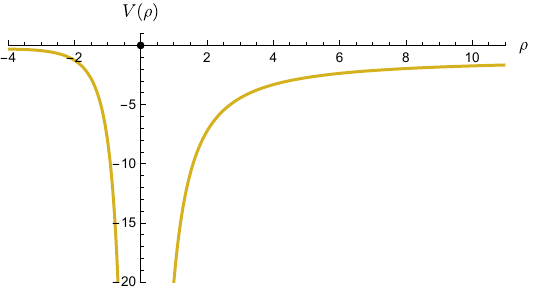}
		\caption{\small Effective potential $V(\zeta)$ governing the $\zeta$-dynamics in dS$_2$ for real proper time. The black dot marks the future-infinity boundary at $\zeta=0$. We set $q_e=3$, $\tilde{m}=2$, $p_y=1$.} 
		\label{fig:EffPotentialdS}
	\end{center}
\end{figure}

Before investigating the behavior of the trajectories more closely, we introduce two sets of distinguished points: \(\zeta_\pm\) and \(y_\pm\). These are meant to be analogous to the special points \(\rho_\pm\) and \(t_\pm\) defined in \(\mathrm{AdS}_2\). To this end, consider first the momentum conjugate to \(\zeta\),
\begin{equation}\label{eq:DefPzeta}
    p_\zeta = \frac{\partial L}{\partial \dot{\zeta}} = -\frac{R^2}{\zeta^2} \dot{\zeta}\, .
\end{equation}
Using this definition, we can rewrite the equation of motion for \(\zeta\) in terms of an effective potential \(V(\zeta)\), as follows:
\begin{equation}\label{eq:EffectivePotentialdS}
    p_\zeta^2 - \frac{1}{\zeta^2}[(p_y \zeta + q_e)^2+\tilde{m}^2] = 0 \iff p_\zeta^2 + V(\zeta) = 0\, .
\end{equation}
The turning points \(p_y \zeta_\pm = -q_e \pm i \tilde{m}\) are defined as the solutions to \(V(\zeta)=0\). We plot the shape of this potential for real proper time evolution in Figure~\ref{fig:EffPotentialdS}. Finally, the remaining points \(p_y y_\pm = p_yy_0 \pm \Omega\) are constructed as the limit of \(p_y y\) when \(\tau \to \mp \infty\).

%%%%%%%%%%%%%%%%%%%%%%%%%%%%%%%%%%%%%%%%%%%%%
%%%%%%%%%%%%%%%%%%%%%%%%%%%%%%%%%%%%%%%%%%%%%

\subsection{Real proper time evolution}\label{ss:dS2RealProperTime}

\begin{figure}[!h]
    \begin{center}
    \subcaptionbox{\label{sfig:dS2Interference-Zeta}}{
		\includegraphics[scale=1]{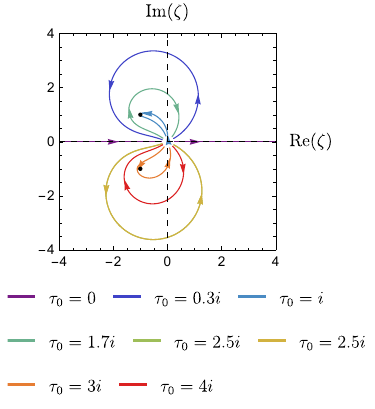}}
        \hspace{0.6cm}
        \subcaptionbox{\label{sfig:dS2Interference-Y}}{\includegraphics[scale=1]{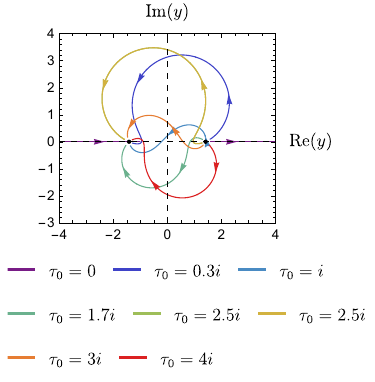}}
        \caption{\small The behavior of \eqref{eq:zetaSolutiondS2} and \eqref{eq:ySolutiondS2} for real proper time. We plot the contours and their orientation for various values of \(\tau_0\). Sending \(\tau_0 \mapsto \bar{\tau}_0\) corresponds to reflecting the solutions across the real axis for both the \(\zeta\) and \(y\) coordinates. As \(|\mathrm{Im}(\tau_0)|\) is increased, the \(\zeta\)-trajectory undergoes a periodic sequence of shrinking, growing, and reflection across the real axis, with period~\(T=2\pi R^2/\Omega\). The \(y\)-trajectory follows an analogous cycle of deformations with the same period. Here, we set \(\tilde{m} = q_e = p_y = R = 1\) and \(y_0 = 0\).} 
        \label{fig:dS2Interference-Tau0-various}
    \end{center}
\end{figure}

When we restrict \(\tau\) to the real line in \eqref{eq:zetaSolutiondS2} and \eqref{eq:ySolutiondS2}, while allowing \(\tau_0\) to remain complex, the resulting spacetime trajectories define the \textit{interference phase}. For \(\tau_0 \in \mathbb{R}\), these trajectories are open and purely real, as illustrated in Figure~\ref{fig:dS2Interference-Tau0-various}. By contrast, for \(\mathrm{Im}(\tau_0)\neq 0\), they become complex and bounded in the complexified spacetime. In particular, as \(\tau\) ranges over the real line, \(\zeta(\tau)\) traces an airfoil-like closed loop\footnote{Strictly speaking, the paths do not include the future-infinity boundary \(\zeta = 0\) (which would be necessary to close the loop) because this point is never reached in finite (proper) time.} enclosing one of the two turning points \(\zeta_\pm\). This curve approaches the point \(\zeta=0\) as \(\tau\to \pm \infty\). On the other hand, the path traced by \(y(\tau)\) is an open trajectory connecting the two limit points \(y_\pm\). For the special values
\begin{equation}
\tau_0 \in \mathbb{R} + \frac{\pi i R^2}{\Omega}\left(\mathbb{Z}+\frac{1}{2}\right)\, ,
\end{equation}
the area enclosed by the $\zeta$-trajectories shrinks to zero, and the corresponding paths degenerate into arcs connecting $\zeta=0$ to one of the turning points $\zeta_\pm$. These trajectories remain complex for $q_e\neq 0$, since
\begin{equation}
\text{Re} (\zeta_\pm) =-\frac{q_e}{p_y}\, .
\end{equation}
Here, we can again contrast the generic situation with the super-extremal Anti-de Sitter case, wherein the degenerate trajectories lie entirely along the real axis. However, when the gauge field is turned off by setting $q_e = 0$, the turning points become purely imaginary, and the corresponding $\zeta$-trajectories run along the imaginary axis. In the same limit, the interference-phase solutions for $y(\tau)$ become purely real. These features are also illustrated in Figure~\ref{fig:dS2Interference-Tau0-various}.

%%%%%%%%%%%%%%%%%%%%%%%%%%%%%%%%%%%%%%%%%%%%%
%%%%%%%%%%%%%%%%%%%%%%%%%%%%%%%%%%%%%%%%%%%%%

Finally, the worldline action \eqref{SwldS}, when evaluated on any of these trajectories, diverges, precluding their use in the saddle-point expansion of the one-loop quantum-mechanical path integral. Nevertheless, traces of these solutions reappear in the bona fide instantons obtained by complexifying \(\tau\); see Section~\ref{ss:MixedSolutionsdS2} below. Intuitively, choosing suitable contours in the complex \(\tau\)-plane can stitch together pairs of the interference phase solutions---for example, trajectories related by reflection across the real axis---thereby closing the open \(y\)-trajectories and regularizing the divergent portion of the path, which in this case lies near \(\zeta=0\). This picture is only heuristic, however: the real-proper-time solutions should more properly be regarded as ill-behaved special cases of the general framework developed in what follows.

\subsection{Generalization}\label{ss:MixedSolutionsdS2}

We now examine the spacetime trajectories obtained by allowing \(\tau\) to take complex values. As before, a particularly simple choice is evolution in imaginary proper time---the \textit{instanton phase}---implemented by the Wick rotation \(\tau \mapsto -i \sigma\), with $\sigma \in \mathbb{R}$. The corresponding solutions for \(\zeta\) and \(y\) are
\begin{subequations}\label{eq:deSitterSolsImaginaryTau}
\begin{equation}
p_y \zeta(\sigma) = \frac{\Omega^2}{-q_{\text{e}} +i \tilde{m}\sin \left[R^{-2} \Omega\left(\sigma +\sigma _0\right)\right]}\, ,
\end{equation}
\begin{equation}
p_y y (\sigma) = p_y y_0+\frac{\tilde{m} \Omega \cos\left[R^{-2} \Omega\left(\sigma +\sigma _0\right)\right]}{-q_{\text{e}}+i \tilde{m} \sin \left[R^{-2} \Omega\left(\sigma +\sigma _0\right)\right]}\, .
\end{equation}
\end{subequations}
The imaginary periodicity in \(\tau\) becomes a real periodicity in \(\sigma\), with period given by $2\pi R^2/\Omega$. Accordingly, the solutions \eqref{eq:deSitterSolsImaginaryTau} describe genuinely closed trajectories and, unlike their real-proper-time counterparts, can a priori contribute individually to the worldline path integral.

\medskip

\begin{figure}[!t]
    \begin{center}
     \subcaptionbox{\label{sfig:dS2Instanton-Zeta}}{
		\includegraphics[scale=0.915]{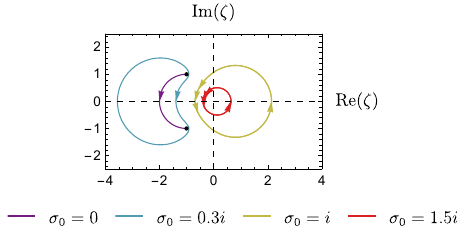}}
        \hspace{0.6cm}
        \subcaptionbox{\label{sfig:dS2Instanton-Y}}{\includegraphics[scale=0.915]{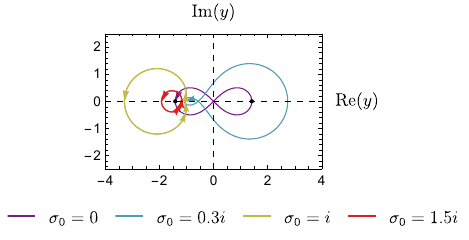}}
        \caption{\small Instanton-phase solutions for \(\zeta(\sigma)\) and \(y(\sigma)\). We plot the contours and their orientations for several values of \(\sigma_0\). The transformation \(\sigma_0 \mapsto \bar{\sigma}_0\) reverses the orientation of the $\zeta$-trajectories, while reflecting the $y$-paths across the line $y_0+i \mathbb{R}$. Notice the transition at \(|\text{Im}(\sigma_0)| = R^2\arsinh(|q_e|/\tilde{m})/\Omega\): the \(\zeta\)-contour changes from enclosing both turning points \(\zeta_\pm\) to enclosing neither, while the \(y\)-trajectory changes from a figure-eight shape to a shrinking loop around one of the limiting points $y_\pm$. Here, we set \(\tilde{m} = q_e = p_y = 1\) and \(y_0 = 0\).}
        \label{fig:dS2Instanton-Sigma0-various}
    \end{center}
\end{figure}

Turning to the dependence on $\sigma_0$, when $\sigma_0\in\mathbb{R}$, 
$\zeta(\sigma)$ traces an arc connecting the two turning points $\zeta_\pm$, 
while $y(\sigma)$ traces a figure-eight trajectory connecting $y_\pm$. When 
$\operatorname{Im}(\sigma_0)\neq 0$, the $\zeta$-arc deforms into a closed 
curve enclosing both turning points, while one of the two lobes of the 
figure-eight trajectory in the $y$-plane expands. Representative examples of these behaviors are shown in 
Figure~\ref{fig:dS2Instanton-Sigma0-various}. As 
$\lvert\operatorname{Im}(\sigma_0)\rvert$ is increased further, the 
$\zeta$-contour eventually ceases to enclose the turning points and instead 
contracts around $\zeta=0$. Meanwhile, the $y$-trajectory contracts around 
one of the limiting points $y_\pm$. More precisely,
\begin{equation}
y(\sigma)\to y_\mp
\qquad\text{as}\qquad
\operatorname{Im}(\sigma_0)\to\pm\infty\, .
\end{equation}

\smallskip

The transition between the two topologically distinct configurations of the 
$\zeta$-contour occurs when the common denominator of the solutions vanishes 
for some real value of $\sigma$, namely when \(\left\lvert\operatorname{Im}(\sigma_0)\right\rvert = \phi\), with
\begin{equation}
\phi \equiv \frac{R^2}{\Omega}
\arsinh\left(\frac{|q_e|}{\tilde m}\right)\, .
\end{equation}
The corresponding instanton phase action \(S_\text{inst}\) is moreover sensitive to which family is selected by \(\sigma_0\). Specifically, for \(\sigma \in [0, 2\pi w_\sigma R^2/\Omega]\) (i.e., the winding number is \(w_\sigma \in \mathbb{Z}\)), we find
\begin{equation}\label{eq:dSInstantonActionFamilies}
	S_\text{inst}=
	\begin{cases}
		2\pi i w_\sigma \bigl(\Omega - |q_e|\bigr)\,, & |\mathrm{Im}(\sigma_0)| \in [0, \phi) \, ,\\
		2\pi i w_\sigma \Omega\, , & |\mathrm{Im}(\sigma_0)| \in (\phi,\infty) \, .
	\end{cases}
\end{equation}
For $w_\sigma=-1$ the first case above reproduces the dS$_2$ pair-production exponent of \cite{Garriga:1994bm,Kim:2008xv}.

%%%%%%%%%%%%%%%%%%%%%%%%%%%%%%%%%%%%%%%%%%%%%
%%%%%%%%%%%%%%%%%%%%%%%%%%%%%%%%%%%%%%%%%%%%%

As was the case in Anti-de Sitter space, the above treatment can be generalized to arbitrary \(\tau\)-contours in the complex plane. We start by recalling that the solutions \eqref{eq:zetaSolutiondS2} and \eqref{eq:ySolutiondS2} are periodic under \(\tau \mapsto \tau +2\pi i R^2/\Omega\), so we may also quotient the complex $\tau$-plane by this identification and regard the resulting proper-time space as an infinite cylinder \(\mathbb{R}\times \mathbf{S}^1\). We then analogously define two families of punctures---represented by \(\phi_1\) and \(\phi_2\)---associated with the poles of the spacetime functions \(\zeta(\tau)\) and \(y(\tau)\). These are located at
\begin{equation}\label{eq:PuncturesdeSitter}
	\phi_1 = - \frac{R^2}{\Omega} \arsinh\left(\frac{q_e}{\tilde{m}}\right), \qquad \text{and} \qquad \phi_2 =  \frac{R^2}{\Omega} \arsinh\left(\frac{q_e}{\tilde{m}}\right) + \frac{\pi i R^2}{\Omega} \, .
\end{equation}
Note that we have set \(\tau_0 = 0\) because the freedom for \(\tau\) to traverse any contour in the complex plane renders \(\tau_0\) effectively redundant. Furthermore, we notice that, in contrast to the super-extremal Anti-de Sitter case, these punctures are aligned neither along the real nor along the imaginary direction (cf.\ \eqref{eq:PuncturesSuperExtremal}).

\smallskip

\begin{figure}[H]
    \centering

    \subcaptionbox{\label{sfig:dS2-Tau-1}}{
		\includegraphics[scale=1]{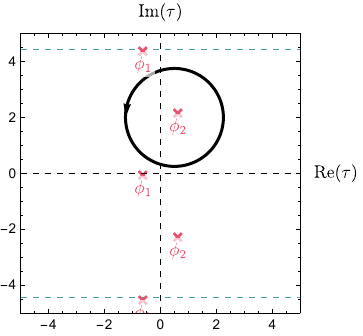}}
        \hspace{0.6cm}
        \subcaptionbox{\label{sfig:dS2-Tau-2}}{\includegraphics[scale=1]{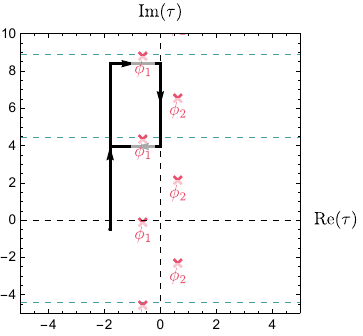}}
        \caption{\small We obtain different instanton trajectories by constructing closed loops in the complex proper-time cylinder. These are grouped into topologically distinct sectors depending on the integers $w=w_1+w_2$ and \(N=n_{\phi_1}-n_{\phi_2} + \mathrm{sgn}(q_e) w_2\) of the $\tau$-contour. To make the plots, we chose $\tilde{m}=q_e=p_y = R = 1$, $y_0=0$, as well as $\textbf{(a)}$ $w=0$, \(N=1\) and $\textbf{(b)}$ $w=1$, \(N=-1\). In the second case, the two horizontal segments exactly cancel once periodicity is enforced, resulting in a contour equivalent to two separate loops with \(w_1=2\) and \(w_2=-1\).}

    \label{fig:dS2-Tau-1-2}
\end{figure}

\begin{figure}[H]
    \centering

    \subcaptionbox{}{
		\includegraphics[scale=1]{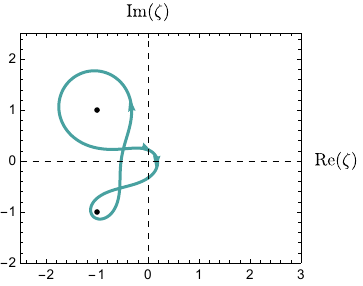}}
        \hspace{0.6cm}
        \subcaptionbox{}{\includegraphics[scale=1]{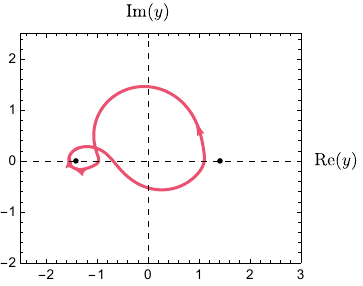}}
        \caption{\small Time $\textbf{(a)}$ and space $\textbf{(b)}$ trajectories corresponding to the path shown in Figure \ref{sfig:dS2-Tau-1}.}
    \label{fig:dS2-Trajectory-1}
\end{figure}

\begin{figure}[H]
    \centering

    \subcaptionbox{\label{sfig:LoopdS2Instanton-Zeta}}{
		\includegraphics[scale=1]{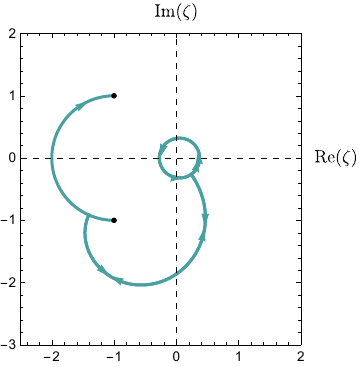}}
        \hspace{0.6cm}
        \subcaptionbox{\label{sfig:LoopdS2Instanton-Y}}{\includegraphics[scale=1]{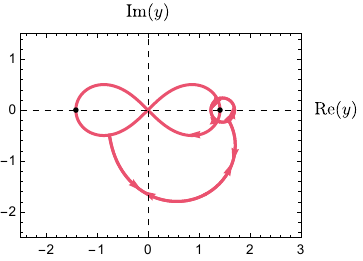}}
        \caption{\small Time $\textbf{(a)}$ and space $\textbf{(b)}$ trajectories corresponding to the path shown in Figure \ref{sfig:dS2-Tau-2}.}
    \label{fig:dS2-Trajectory-2}
\end{figure}

The action obtained from an arbitrary \(\tau\)-contour can then be decomposed in terms of four integers associated with how the path wraps around the punctures, as defined in Table~\ref{tab:definitions}. The resulting expression parallels \eqref{eq:InstantonActionAdSMixedSuper}\footnote{There is an overall sign difference between \eqref{eq:InstantonActionAdSMixedSuper} and \eqref{eq:dS2Action}, which can be traced back to the fact that the map \eqref{eq:dS2AdS2Mapping} induces the replacement \(p_\rho \dot{\rho} \mapsto - p_\zeta \dot{\zeta}\) in the Lagrangian.}:
\begin{equation}
    \label{eq:dS2Action}
    S_\text{inst} = -2\pi i \left[(w_1+w_2)\,\Omega - (n_{\phi_1}-n_{\phi_2} + \mathrm{sgn}(q_e) w_2)\, q_e\right]\, .
\end{equation}
%
%%%%%%%%%%%%%%%%%%%%%%%%%%%%%%%%%%%%%%%%
\begin{table}[H]
\centering
\caption{Definitions of the relevant winding numbers in de Sitter.}
\label{tab:definitions}
\begin{tabular}{cc}
\toprule
Quantity & Description \\
\midrule
\(w_1\) & Winding along the imaginary direction away from \(\phi_1\) and \(\phi_2\)\\
\(w_2\) & Winding along the imaginary direction in-between \(\phi_1\) and \(\phi_2\)\\
\(n_{\phi_1}\) & Counterclockwise winding number around \(\phi_1\)\\
\(n_{\phi_2}\) & Counterclockwise winding number around \(\phi_2\) \\
\bottomrule
\end{tabular}
\end{table}
%%%%%%%%%%%%%%%%%%%%%%%%%%%%%%%%%%%%%%%%
Several examples of complex proper-time paths as well as the corresponding spacetime trajectories are shown in Figures~\ref{fig:dS2-Tau-1-2},~\ref{fig:dS2-Trajectory-1},~and~\ref{fig:dS2-Trajectory-2}. From the arrangement of the punctures on the cylinder, we can also see that the four integers in \eqref{eq:dS2Action} are not fully independent: deforming a contour that winds in-between the two punctures (i.e., with \(w_2 \neq 0\)) to a contour that winds away from them (\(w_1 \neq 0\)) requires sweeping across \(\phi_1\) or \(\phi_2\), hence picking up a unit of winding around it, depending on which direction the deformation goes. This is the same type of deformation described around \eqref{eq:WindingTradeSuper} for the super-extremal regime of \(\mathrm{AdS}_2\). Thus, the action reduces to the analogous expression in terms of two \emph{topological} invariants
\begin{equation}
    S_\text{inst} = -2\pi i \left(w \Omega - N q_e\right),
\end{equation}
where \(w=w_1+w_2\) and \(N=n_{\phi_1}-n_{\phi_2}+\mathrm{sgn}(q_e) w_2\). This reduction can also be understood through the derivation of \eqref{eq:dS2Action} from Cauchy's residue theorem, as done in the next section.

%%%%%%%%%%%%%%%%%%%%%%%%%%%%%%%%%%%%%%%%%%%%%%%%%%%%%%%%%%%%%%%%%%%%%

%%%%%%%%%%%%%%%%%%%%%%%%%%%%%%%%%%%%%%%%%%%%%%%%%%%%%%%%%%%%%%%%%%%%%
%%%%%%%%%%%%%%%%%%%%%%%%%%%%%%%%%%%%%%%%%%%%%%%%%%%%%%%%%%%%%%%%%%%%%
%%%%%%%%%%%%%%%%%%%%%%%%%%%%%%%%%%%%%%%%%%%%%%%%%%%%%%%%%%%%%%%%%%%%%
%%%%%%%%%%%%%%%%%%%%%%%%%%%%%%%%%%%%%%%%%%%%%%%%%%%%%%%%%%%%%%%%%%%%%
\subsection{Instanton action from Cauchy's residue theorem}\label{ss:CauchydS}

As we have emphasized so far, the worldline action is controlled by the topological properties of the contours in \(\tau\). Nonetheless, it is insightful to analyze how the same result emerges purely from the spacetime trajectories themselves, since these give rise to the usual instanton picture. To evaluate the desired integral
\begin{equation}
S_{\rm inst} = \oint p_\mu \dot x^\mu\, d\tau
= \oint p_\zeta\, d\zeta + \oint p_y\, dy
= \oint p_\zeta\, d\zeta\,,
\label{eq:Sinst-ds-contour}
\end{equation}
we will primarily make use of the fact that
\begin{equation}\label{eq:PzetaVSV(-zeta)}
    p_\zeta = \sqrt{-V(\zeta)} = \frac{p_y}{\zeta}\sqrt{(\zeta-\zeta_+)(\zeta-\zeta_-)}\, ,
\end{equation}
with an appropriate choice of sign and placement of branch cuts. This expression yields the following expansions around zero and infinity:
\begin{subequations}
\begin{equation}
    \label{eq:dS2V0}
 \sqrt{-V(\zeta)} = \frac{\Omega}{\zeta} + \mathcal{O}(1)\qquad \text{as}\quad \zeta\to 0\, ,
\end{equation}
\begin{equation}
    \label{eq:dS2Vinf}
\sqrt{-V(\zeta)} = p_y + \frac{q_e}{\zeta} + \mathcal{O}\left(\frac{1}{\zeta^2}\right)\qquad \text{as}\quad \zeta\to \infty\, .
\end{equation}
\end{subequations}
Depending on the specific \(\zeta\)-trajectory followed by the particle, we will need to attach the appropriate signs to the expansions. We first consider \(\tau\)-contours that describe the instanton phase solutions. These are simply vertical lines starting at some initial value \(\tau_i\) and wrapping along the periodic direction: \(\tau \in [\tau_i, \tau_i + 2\pi i w R^2/\Omega]\). There are two distinguished cases, \nameref{par:dS2Res1}~and~\nameref{par:dS2Res2}, corresponding to the windings \(w_2\) and \(w_1\), respectively.

\paragraph{(1)}\label{par:dS2Res1} If \(|\mathrm{Re}(\tau_i)| \in [0, \phi)\), then the \(\zeta\)-component of the instanton solutions encloses the two turning points \(\zeta_\pm\), meaning that we can compute the action by adding the residues both at infinity and zero. For concreteness, we take \(\mathrm{Re}(\tau_i)>0\), which makes \(\zeta\) run counterclockwise. Then, from the residue theorem we find
\begin{equation}
    \oint p_\zeta\, d \zeta = - 2\pi i \big(\text{Res}_{\zeta=0}\left( p_\zeta\right) + \text{Res}_{\zeta=\infty}\left( p_\zeta\right) \big)\, .
\end{equation}
Regardless of the values of \(p_y\) and \(q_e\), the expansion \eqref{eq:dS2V0} has the appropriate signs, giving
\begin{equation}
    \text{Res}_{\zeta=0}\left( p_\zeta\right) = \Omega \, .
\end{equation}
The overall sign of \eqref{eq:dS2Vinf}, however, depends on the sign of \(q_e\) and whether we are letting \(\zeta \to \infty\) by making \(\tau\) approach \(\phi_1\) or \(\phi_2\). If \(q_e>0\), then \(\phi_2\) lies to the right of \(\phi_1\) and we can approach \(\phi_2\) while preserving the counterclockwise orientation, with no additional sign. If instead \(q_e<0\), then we need to approach \(\phi_1\) so as to keep the orientation counterclockwise, and we pick up an overall minus sign in~\eqref{eq:dS2Vinf}. In summary, as \(\zeta\to\infty\) we find
\begin{equation}
    p_\zeta = \mathrm{sgn}(q_e)\left(p_y+\frac{q_e}{\zeta} + \mathcal{O}(\zeta^{-2})\right)\, ,
\end{equation}
which gives the residue \(-|q_e|\) at \(\zeta=\infty\). If we had chosen \(\mathrm{Re}(\tau_i)<0\), the orientation of the \(\zeta\)-trajectory would reverse, but the calculation would have been analogous. Therefore, winding around \(w\in \mathbb{Z}\) times, the instanton action is
\begin{equation}
    S_{\rm inst} = -2\pi i w \left(\Omega-|q_e| \right)\, .
\end{equation}

\paragraph{(2)}\label{par:dS2Res2} If \(|\mathrm{Re}(\tau_i)| \in (\phi,\infty) \), the instanton solutions do not enclose any turning points, but rather shrink towards zero, such that we can compute the action from the residue at $\zeta=0$. The computation proceeds in exactly the same manner, and we find
\begin{equation}
    S_{\rm inst} = -2\pi i w \Omega \, .
\end{equation}

Finally, we compute the action associated with enclosed punctures. The behavior of \(\zeta\) is the same regardless of whether \(\tau\) encloses \(\phi_1\)~or~\(\phi_2\): one loop around each of the turning points \(\zeta_\pm\) and a third one around \(\zeta=0\) that is traversed in the opposite orientation. One can combine the first two into a single loop which reduces to the situation of case \nameref{par:dS2Res1} above. The action is then computed in the same manner and the \(-2\pi i \Omega\) terms cancel across the two remaining loops, due to their different orientations. The overall sign depends only on which puncture is enclosed, giving rise to the claimed expression
\begin{equation}
    S_{\rm inst} = 2\pi i n q_e \, ,
\end{equation}
where \(n = n_{\phi_1}-n_{\phi_2}\). This is the last piece needed to verify \eqref{eq:dS2Action}, concluding our spacetime analysis and confirming agreement with the results obtained by evaluating the contour integrals directly in the proper-time plane. We note that the need for careful case splitting and accounting for various signs in the spacetime picture further reinforces the perspective that the proper-time plane encodes most of the information about the complex instantons and the value of their action.

\pagebreak

\section{Global (Anti-)de Sitter Analysis}\label{s:Global(A)dS}

The analysis of Sections~\ref{s:AdS2} and \ref{s:dS2} was carried out in Poincar\'e (and flat-slicing) coordinates, which cover only a portion of the full (Anti-)de Sitter spacetime. This restriction, however, is not innocuous once the trajectories are complexified: as we detail below, the map to global coordinates is multi-valued, so a worldline that closes in a single patch need not remain closed on the full spacetime---or, more precisely, on its universal cover. Moreover, the punctures that organize the local classification are tied to the trajectories reaching the edges of the Poincar\'e (or flat-slicing) patch, so it is natural to ask how much of this structure survives once some of these boundaries are no longer present. We therefore repeat the instanton analysis directly in global coordinates, treating AdS$_2$ and dS$_2$ in turn, and determine which of the complex saddles found earlier descend to genuine instantons on the universal cover. As we shall see, the outcome is a classification that parallels---but refines---the one obtained locally, with the differences between the two spacetimes reflecting their distinct global structures.

\subsection{AdS$_2$}\label{ss:GlobalAdS}

\begin{figure}[t!]
    \centering
		\includegraphics[scale=0.45]{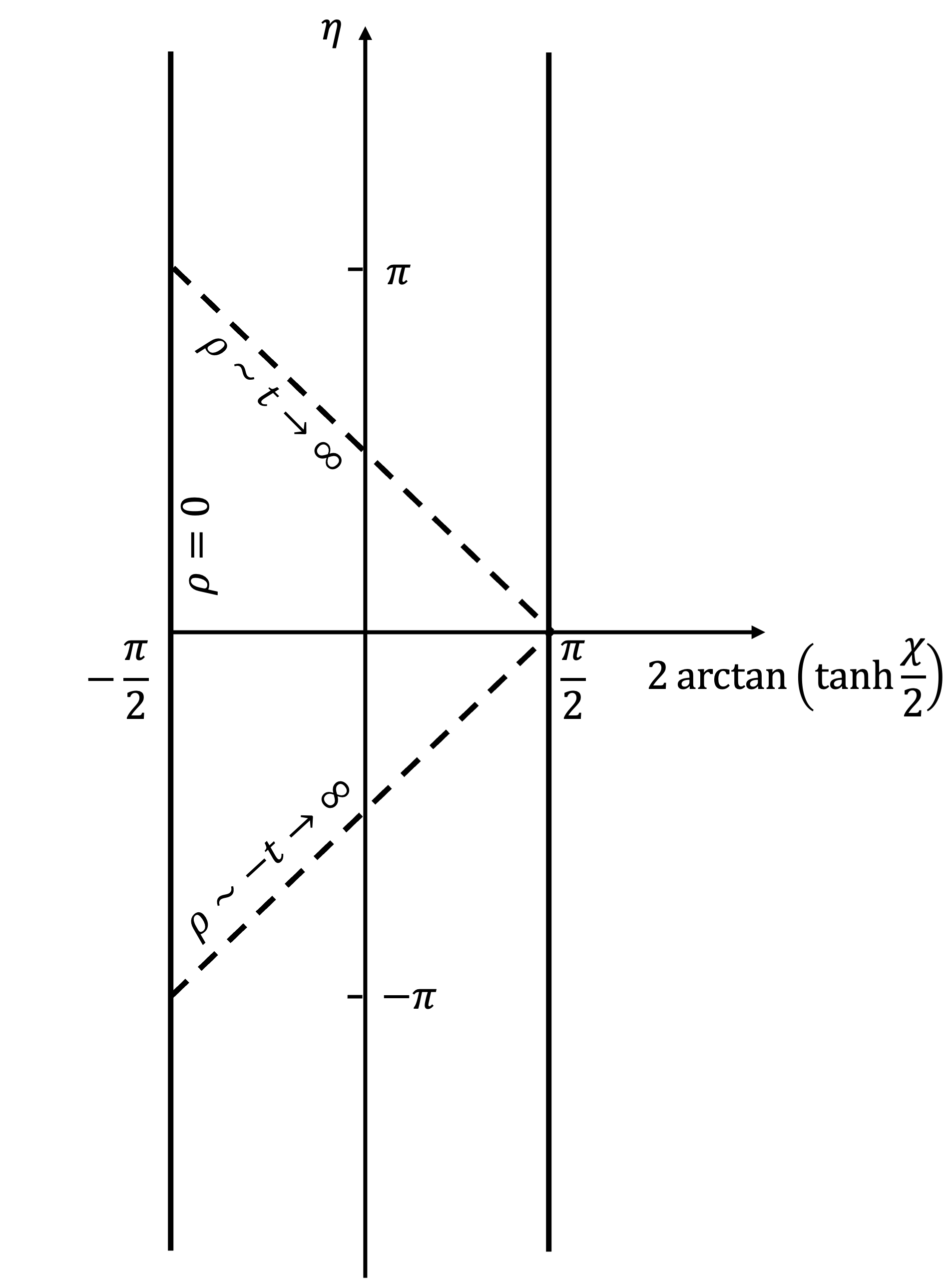}

		\caption{\small A single Poincar\'e patch of AdS$_2$, corresponding to the triangular region bounded by the dashed null lines (the Poincar\'e horizons). Note that it covers the range $\eta \in (-\pi, \pi)$. In the hyperboloid \eqref{eq:HyperboloidEmbeddingAdS2} one further has $\eta \sim \eta + 2\pi$, whereas on the universal cover considered here $\eta \in (-\infty, \infty)$.} 
		\label{Fig:patch}
\end{figure}

In Section~\ref{s:AdS2}, we derived all worldline instanton solutions in 2D Anti-de Sitter space, which we parametrize using Poincar\'e coordinates, in terms of which the line element and gauge field read as in \eqref{eq:BackgroundAdS2}. For real $(t,\rho)$, these coordinates cover only a portion of the entire manifold as $t$ varies continuously from $-\infty$ to $+\infty$ and $\rho$ from $0$ to $+\infty$. To describe the full spacetime, it is convenient to view AdS$_2$ as the one-sheeted
hyperboloid
\begin{equation}\label{eq:HyperboloidEmbeddingAdS2}
-(X^0)^2 - (X^1)^2 + (X^2)^2 = -R^2\, ,
\end{equation}
embedded in $\mathbb{R}^{2,1}$ with metric $\eta_{AB}=\mathrm{diag}(-1,-1,1)$. The Poincar\'e patch (cf.\ Figure~\ref{Fig:patch}) is recovered by the parametrization
\begin{equation}\label{eq:PoincareMap}
X^0 = \frac{R}{\rho}\,t\, ,\qquad
X^1+X^2 = \rho-\frac{t^2}{\rho}\, ,\qquad
X^1-X^2 = \frac{R^2}{\rho}\, ,
\end{equation}
which pulls the ambient metric back to \eqref{eq:PoincareMetricAdS} and, with $t\in\mathbb{R}$, $\rho>0$, covers only half of the hyperboloid constructed via~\eqref{eq:HyperboloidEmbeddingAdS2}. 

\noindent Throughout this work, we consider the \emph{universal cover} of global AdS$_2$. This is obtained via
\begin{equation}\label{eq:GlobalMap}
X^0 = R\cosh\chi\,\sin\eta\, ,\qquad
X^1 = R\cosh\chi\,\cos\eta\, ,\qquad
X^2 = R\sinh\chi\, ,
\end{equation}
under which the metric and gauge field read
\begin{equation}\label{eq:globalmetricAdS2}
ds^2=R^2\left(-\cosh^2\chi\,d\eta^2+d\chi^2\right)\, ,\qquad
A=-ER^2\sinh\chi\,d\eta\, ,
\end{equation}
with $\chi\in(-\infty,\infty)$. Notice that, since $(X^0,X^1)$ depend on $\eta$ only through $(\sin\eta,\cos\eta)$, the embedding naturally identifies $\eta\sim\eta+2\pi$. Nevertheless, on the universal cover this identification is removed and $\eta\in(-\infty,\infty)$. Composing \eqref{eq:PoincareMap} and \eqref{eq:GlobalMap}, the two charts are related directly by
\begin{equation}\label{eq:PoincareToGlobal}
\tan\eta=\frac{2Rt}{R^2+\rho^2-t^2}\, ,\qquad
\sinh\chi=\frac{\rho^2-t^2-R^2}{2R\rho}\, ,
\end{equation}
whose inversion for $(\eta,\chi)$ hence involves the functions $\arctan$ and $\arsinh$.

%%%%%%%%%%%%%%%%%%%%%%%%%%%%%%%%%%%%%%%%%%%%%
%%%%%%%%%%%%%%%%%%%%%%%%%%%%%%%%%%%%%%%%%%%%%

\smallskip

For real $t$ and $\rho$, reaching the Poincar\'e horizon requires a lightcone coordinate $x^\pm=t\pm \rho$ to run off to infinity---$x^+\to +\infty$ at fixed $x^-$, or vice versa---so that the patch boundary cannot be crossed at finite $(t,\rho)$. The situation becomes more subtle once the coordinates are complexified. In that case, analytic continuation through complex values of $(t,\rho)$ can connect different patches without the lightcone coordinates having to diverge along the complexified trajectory. More importantly, the inverse map \eqref{eq:PoincareToGlobal} from $(t,\rho)$ to $(\eta,\chi)$ is multi-valued.\footnote{Concretely, the branch points of the map \eqref{eq:PoincareToGlobal} lie where $\sinh\chi=\pm i$, which in lightcone coordinates factorizes as $(x^+\mp iR)(x^-\pm iR)=0$, i.e., the four null loci $x^+=\pm iR$ or $x^-=\pm iR$. These sit at finite but complex $(t,\rho)$, so a complexified trajectory can wind around them---picking up the monodromy \eqref{etaminuspi} and \eqref{chitaunpm}---without any lightcone coordinate diverging. This is the mechanism underlying the failure of closure discussed below; see e.g., Section~\ref{sec:global-sub-extremal} for details.} Consequently, even if $(t,\rho)$ return to their original values after traversing a closed contour, the analytically continued global coordinates $(\eta,\chi)$ need not lie on the same branch, and hence need not correspond to the same point on the universal covering space. Thus, a trajectory that appears closed in complex Poincar\'e coordinates may fail to be closed when lifted to the complex global chart. The complex instantons found in Section~\ref{s:AdS2} must therefore be reexamined from the present global perspective. As we will see, the latter refines the precise structure of the instanton solutions found therein.

\medskip

Hence, we first set up the worldline problem directly in the global chart. In terms of the global coordinates \eqref{eq:globalmetricAdS2}, the worldline action \eqref{eq:worldlineAction} is given by
\begin{equation}\label{SwlAdSglobal}
 S_{\rm wl} =  \int_\gamma d\tau\, \left(-\frac{m^2 }{2}-\frac{R^2\cosh^2\chi\,\dot\eta^2}{2}+\frac{R^2\dot\chi^2}{2} -q_e  \sinh\chi\,\dot\eta\right)\, ,
\end{equation}
where $q_e= eE R^2$, and whose EOM read 
\begin{equation}\label{eq:EOMGlobalAdS}
\begin{aligned}
    &\Ddot{\eta} \cosh^2 \chi + 2 \dot{\eta} \dot{\chi} \cosh{\chi} \sinh \chi+ \frac{q_e}{R^2} \dot{\chi} \cosh \chi=0\, ,\\
    &\Ddot{\chi} + \dot{\eta}^2 \sinh \chi \cosh \chi + \frac{q_e}{R^2} \dot{\eta} \cosh \chi=0\, .
\end{aligned}
\end{equation} 
These must be supplemented, as usual, with the Hamiltonian constraint
\begin{equation}
\label{H0AdSglobal}
-R^2\cosh^2\chi\,\dot\eta^2+R^2\dot\chi^2+m^2=0\ .
\end{equation}
Similarly to the Poincar\'e case, the momentum conjugate to the time coordinate,
\begin{equation}
\label{peta}
    p_\eta=-R^2\cosh^2\chi\,\dot\eta-q_e\sinh\chi\, ,
\end{equation}
is conserved. Thus, the Hamiltonian condition \eqref{H0AdSglobal} can be recast as a first-order differential equation for $\chi(\tau)$,
\begin{equation}
    p_\chi=R^2\dot\chi=\sqrt{-V(\chi)}\, ,
    \qquad
    V(\chi)=\tilde{m}^2-
    \left(p_\eta\,\operatorname{sech}\chi+q_e\tanh\chi\right)^2\, .
    \label{eq:global-super-potential}
\end{equation}

In the following, we will study the sub-extremal, super-extremal, and extremal regimes separately, mimicking the analysis of Section~\ref{s:AdS2}.

\subsubsection{Sub-extremal particles}
\label{sec:global-sub-extremal}

\begin{figure}[t!]
\begin{center}
\includegraphics[scale=1]{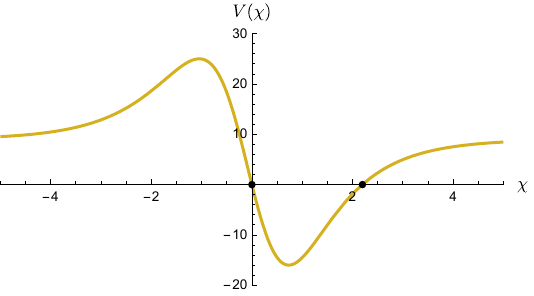}
\caption{\small Effective potential $V(\chi)$ governing the sub-extremal motion in global AdS$_2$ coordinates. The dots at $\chi=\log\left[(p_\eta\pm\Delta)/(\tilde m-q_e)\right]$ denote the turning points. We set $q_e=4$ and $\tilde m=p_\eta=5$.}
\label{fig:EffPotentialGlobalAdSSubExtremal}
\end{center}
\end{figure}

For the sub-extremal case, $\tilde m>|q_e|$, the effective potential (\ref{eq:global-super-potential}) is shown in Figure~\ref{fig:EffPotentialGlobalAdSSubExtremal}, taking $\tilde m=p_\eta=5$ and $q_e=4$ as an illustrative example. For complex $\chi$, there are infinitely many roots of $V(\chi)$, located at
\begin{equation}\label{eq:GlobalTurningPoints}
\chi^{(\pm)}_+
\in 2\pi i\mathbb{Z}+\log \left(\frac{p_\eta\pm\Delta}{\tilde m-q_e}\right)\, ,
\qquad
\chi^{(\pm)}_-
\in \pi i\left(2\mathbb{Z}+1 \right)+\log \left(\frac{p_\eta\pm\Delta}{\tilde m+q_e}\right)\, ,
\end{equation}
where $\Delta=\sqrt{p_\eta^2-\beta^2}$ with $\beta=\sqrt{\tilde m^2-q_e^2}$.

%%%%%%%%%%%%%%%%%%%%%%%%%%%%%%%%%%%%%%%%%%%%%
%%%%%%%%%%%%%%%%%%%%%%%%%%%%%%%%%%%%%%%%%%%%%

\smallskip

By integrating (\ref{eq:global-super-potential}), we obtain the solution
\begin{equation}\label{eq:SinhChiSolAdS}
\sinh\chi(\tau)
=\frac{p_\eta q_e-\tilde m\Delta\sin\left[\frac{\beta(\tau+\tau_0)}{R^2}\right]}{\beta^2}\, ,
\end{equation}
with $\tau_0\in \mathbb C$ the integration constant. Substituting (\ref{eq:SinhChiSolAdS}) into (\ref{peta}) then yields
\begin{equation}
\label{etatauglobalsub}
\tan{\eta(\tau)}
=\frac{\tilde m\beta\cos\left[\frac{\beta(\tau+\tau_0)}{R^2}\right]}
{\tilde m p_\eta\sin\left[\frac{\beta(\tau+\tau_0)}{R^2}\right]-q_e\Delta}\, .
\end{equation}
Alternatively, the same solutions can be obtained directly from \eqref{eq:RealTauInstantonSolsSubExtremal} under the coordinate transformation (\ref{eq:PoincareToGlobal}), which gives
\begin{subequations}\label{eq:InstantonSolsSubExtremalGlobalNoT0}
\begin{equation}\label{eq:TanEtaSolSubExtremalGlobalNoT0}
\tan{\eta(\tau)}
=\frac{\tilde m\cos \left(\frac{\beta\tau}{R^2}\right)}
{\tilde m\cosh\lambda\sin\left(\frac{\beta\tau}{R^2}\right)-q_e\sinh\lambda}\, ,
\end{equation}
\begin{equation}\label{eq:SinChiSolSubExtremalGlobalNoT0}
\sinh{\chi(\tau)}
=\frac{q_e\cosh\lambda-\tilde m\sinh\lambda\sin\left(\frac{\beta\tau}{R^2}\right)}{\beta}\, ,
\end{equation}
\end{subequations}
where we have defined\footnote{Note that $\lambda$ is real for positive $p_t$, whereas for negative $p_t$ (anti-particles) it acquires an imaginary part~$i\pi$.}
\begin{equation}
\lambda=\log \left(\frac{Rp_t}{\beta}\right)\, .
\end{equation}
These agree with (\ref{eq:SinhChiSolAdS}) and (\ref{etatauglobalsub}) upon shifting $\tau\to\tau+\tau_0$ and identifying
\begin{equation}
p_\eta=\beta\cosh\lambda\, , \qquad \Delta=\beta\sinh\lambda\ .
\end{equation}
From this it follows that $p_\eta^2\geq\beta^2$,\footnote{This condition sets a lower bound on the particle's energy in global AdS$_2$. It can equivalently be understood as a consequence of the Hamiltonian constraint \eqref{H0AdSglobal} together with the underlying $\SL(2,\mathbb{R})$ symmetry of the problem \cite{Castellano:2025yur,Castellano:2025rvn}.} so that the arguments of the logarithms in (\ref{eq:GlobalTurningPoints}) are purely real.

\medskip

We now study in detail the behavior of these solutions in global AdS. To do so, it is convenient to first express them in the form
\begin{equation}
\label{etachi}
\begin{split}
\eta(\tau)&=\frac{1}{2i}\big[\log(D+iN)-\log(D-iN)\big]\, ,\\
\chi(\tau)&=\log\left(z+\sqrt{z^2+1}\right)\, ,
\end{split}
\end{equation}
where
\begin{equation}
\label{ND}
N=\tilde m\cos\left[\frac{\beta(\tau+\tau_0)}{R^2}\right]\, ,\qquad
D=\tilde m\cosh\lambda\sin\left[\frac{\beta(\tau+\tau_0)}{R^2}\right]-q_e\sinh\lambda\, ,
\end{equation}
and
\begin{equation}
z(\tau)=\sinh\chi=\frac{q_e\cosh\lambda-\tilde m\sinh\lambda\sin\left[\frac{\beta(\tau+\tau_0)}{R^2}\right]}{\beta}\, .
\end{equation}
These quantities satisfy the useful identity
\begin{equation}\label{eq:NDidAdS2}
N^2+D^2=\beta^2\left(z^2+1\right)\, .
\end{equation}
It follows immediately that the branch points of $\eta(\tau)$, which occur when either logarithm in (\ref{etachi}) becomes singular, namely at $D=\pm iN$, coincide with the branch points of $\chi(\tau)$, located at $z=\pm i$.\footnote{The branch points are equivalently characterized by $\cosh^2 \chi =0$, or by $\sec^2 \eta =0$. That these coincide follows trivially from the embedding \eqref{eq:GlobalMap}, whence
\begin{equation}
\cosh^2 \chi = \frac{(X^0)^2+(X^1)^2}{R^2}\, ,\qquad  \sec^2 \eta = \frac{(X^0)^2+(X^1)^2}{(X^1)^2}\, ,\notag
\end{equation}
so that $\cosh^2 \chi=0 \iff \sec^2 \eta=0 \iff (X^0)^2=-(X^1)^2$, i.e., $(X^2)^2=-R^2$ on the hyperboloid \eqref{eq:HyperboloidEmbeddingAdS2}.}

%%%%%%%%%%%%%%%%%%%%%%%%%%%%%%%%%%%%%%%%%%%%%
\begin{figure}[t!]
    \centering
\subcaptionbox{\label{etataufig}}{
		\includegraphics[scale=0.925]{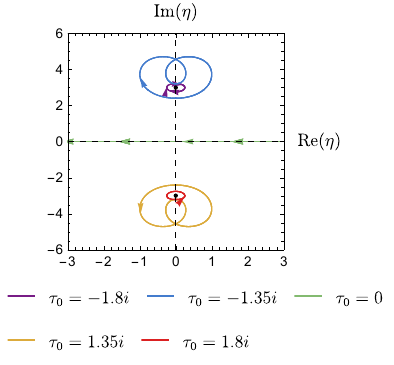}}
        \hspace{0.6cm}
        \subcaptionbox{\label{chitaufig}}{\includegraphics[scale=0.925]{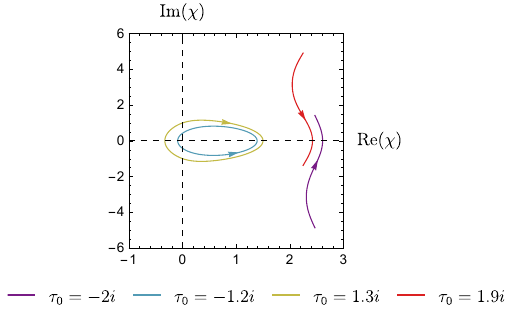}}
		\caption{\small Solutions to the real proper-time EOM for $\eta(\tau)$ and $\chi (\tau)$. In the universal covering space, they are generically not closed. Their behavior with respect to $\text{Im}(\tau_0)$ is symmetric, such that changing the sign preserves the trajectory in $\chi$-plane but reverses its orientation, whereas for the $\eta$-component the paths get reflected across the real axis. We plot various cases with $\tilde{m}=3$, $q_e=2$, $R=1$, $\lambda=0.1$, and different values of $\tau_0$.} 
		\label{fig:TrajectoriesRealTauGlobalAdS}
\end{figure}
%%%%%%%%%%%%%%%%%%%%%%%%%%%%%%%%%%%%%%%%%%%%%

\medskip

For the \emph{interference phase}, we show the solutions (\ref{etachi}) in Figure~\ref{fig:TrajectoriesRealTauGlobalAdS} for different values of the integration constant $\tau_0$. The trajectories are generically complex, except when $\tau_0\in\mathbb{R}$ \cite{Castellano:2025rvn}. Since the functions $N$, $D$, and $z$ appearing in (\ref{etachi}) are periodic in $\tau$ with period $2\pi R^2/\beta$, one might naively expect the corresponding trajectories in the complex $\eta$- and $\chi$-planes to exhibit the same periodicity. However, this is generically not the case because of the non-trivial monodromy exhibited by the solutions (\ref{etachi}), as we analyze below.

\smallskip

In the complex $\eta$-plane, the trajectories are not closed for sufficiently small $|\text{Im} (\tau_0)|$; the $\text{Im} (\tau_0)=0$ trajectory shown in Figure~\ref{fig:TrajectoriesRealTauGlobalAdS} provides a simple example. As $|\text{Im} (\tau_0)|$ increases past a critical value determined by the branch points discussed below, the trajectories become closed. Interestingly, they then develop a single self-intersection and, as $\tau_0\to\pm i\infty$, eventually localize around
\begin{equation}
\eta_\pm=\mp\arctan\left(\frac{i}{\cosh\lambda}\right)\, ,
\end{equation}
approaching circular paths whose radii shrink to zero.

\smallskip

In the complex $\chi$-plane, the trajectories behave symmetrically under $\text{Im} (\tau_0)\to-\text{Im} (\tau_0)$. Their closure properties are, however, opposite to those in the $\eta$-plane: the trajectories are closed only for sufficiently small $|\text{Im} (\tau_0)|$. As $|\text{Im} (\tau_0)|$ increases past a critical value determined by the branch points discussed below, they become open, as illustrated, for example, by the $\tau_0=-2i$ and $\tau_0=1.9i$ trajectories in Figure~\ref{fig:TrajectoriesRealTauGlobalAdS}.

\medskip

\begin{figure}[t!]
    \centering
\subcaptionbox{\label{etasigmafig}}{
		\includegraphics[scale=1]{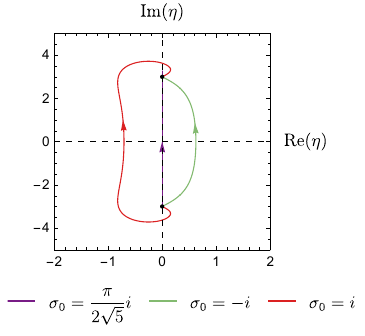}}
        \hspace{0.6cm}
        \subcaptionbox{\label{chisigmafig}}{\includegraphics[scale=1]{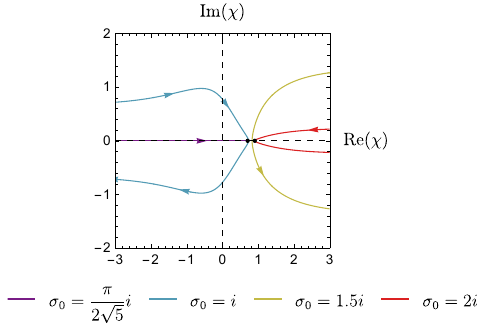}}
		\caption{\small The solutions for $\chi(\sigma)$ and $\eta(\sigma)$ in imaginary proper-time $\sigma \in (-\infty, \infty)$ are open. We plotted various representative cases with $\tilde{m}=3$, $q_e=2$, $R=1$, $\lambda=0.1$, and different values of $\sigma_0$.} 
		\label{fig:TrajectoriesImaginaryTauGlobalAdS}
\end{figure}

For the \emph{instanton phase}, the proper-time evolution proceeds along the imaginary direction. The corresponding spacetime trajectories can therefore be obtained from those in (\ref{etachi}) by performing the Wick rotation $\tau\to-i\sigma$, with $\sigma \in \mathbb{R}$. In Figure~\ref{fig:TrajectoriesImaginaryTauGlobalAdS}, we display several representative solutions for $\chi(\sigma)$ and $\eta(\sigma)$ at different values of the integration constant $\sigma_0=i\tau_0$. In both complex planes, the trajectories are open and extend over the full range $\sigma\in(-\infty,\infty)$. They are generically complex; however, for the special values
\begin{equation}
\sigma_0 \in \mathbb{R}+\frac{i\pi R^2}{\beta}\left(\mathbb{Z}+\frac{1}{2}\right)\, ,
\end{equation}
the trajectories lie entirely along the real axis in the $\chi$-plane and along the imaginary axis in the $\eta$-plane. In the $\chi$-plane, these special trajectories begin at the conformal boundary, $|\chi|\to\infty$, reach the nearer of the two real turning points among the infinite set \eqref{eq:GlobalTurningPoints}, and then return to the same component of the boundary.

%%%%%%%%%%%%%%%%%%%%%%%%%%%%%%%%%%%%%%%%%%%%%
%%%%%%%%%%%%%%%%%%%%%%%%%%%%%%%%%%%%%%%%%%%%%

\medskip

For general contours in the complex $\tau$-plane, we need to analyze the monodromy of the solution (\ref{etachi}) carefully. Without loss of generality, let us take $\tau_0=0$. As mentioned before, $\eta(\tau)$ and $\chi(\tau)$ have the same branch points, although their monodromies around these points are different. To characterize their locations, it is convenient to define
\begin{equation}
\epsilon\equiv\operatorname{sgn}\left(\text{Re} (\lambda)\right)\, ,\qquad
\theta\equiv\arctan\left(\frac{q_e}{\beta}\right)\, ,\qquad
L\equiv\arcosh\left|\coth \lambda\right|\, .
\end{equation}
The branch points associated with $D=\pm iN$ are located at
\begin{equation}
\label{taukpm}
\tau_k^{(\pm)}
=\frac{R^2}{\beta}
\left[
k\pi+(-1)^k\epsilon\theta\pm iL
\right]\, ,
\qquad k\in\mathbb{Z}\, .
\end{equation}
Note that these form two horizontal arrays in the $\tau$-plane, related by complex conjugation. At these points, $z(\tau)$ takes the values
\begin{equation}
\label{ypmi}
z\left(\tau_k^{(\pm)}\right)=\mp i\epsilon(-1)^k\, ,
\end{equation}
confirming explicitly that they coincide with the branch points $z=\pm i$ of $\chi(\tau)$.

\smallskip

Let us now examine the corresponding monodromies. Around $\tau_k^{(+)}$, the function $D-iN$ has a simple zero, whereas $D+iN$ remains nonzero. Therefore, upon encircling $\tau_k^{(+)}$ once counterclockwise,
\begin{equation}
\log(D-iN)\to\log(D-iN)+2\pi i\, ,
\end{equation}
and hence
\begin{equation}
\eta\to\eta-\pi\, .
\end{equation}
Similarly, around $\tau_k^{(-)}$, it is $D+iN$ that has a simple zero, giving
\begin{equation}
\label{etapluspi}
\eta\to\eta+\pi\, .
\end{equation}
In other words, the two families of branch points induce the monodromies
\begin{equation}
\label{etaminuspi}
\tau_k^{(\pm)}:\qquad
\eta\to\eta\mp\pi\, .
\end{equation}

There is, in addition, a non-trivial monodromy associated with the periodic direction of the complex $\tau$-plane. Consider a positively oriented contour extending over one period, $\tau\to\tau+2\pi R^2/\beta$, and lying between the upper branch points $\tau_k^{(+)}$ and the lower branch points $\tau_k^{(-)}$. Along this contour, $D+iN$ winds once clockwise around the origin, whereas $D-iN$ winds once counterclockwise. Consequently,
\begin{equation}
\log(D+iN)\to\log(D+iN)-2\pi i\, ,
\qquad
\log(D-iN)\to\log(D-iN)+2\pi i\, ,
\end{equation}
which gives the additional monodromy
\begin{equation}
\label{etaperiodmonodromy}
 \eta\to\eta-2\pi\qquad \text{as}\quad \tau\to\tau+\frac{2\pi R^2}{\beta}\, .
\end{equation}
This is consistent with the non-closure observed for the corresponding interference-phase trajectories; see, e.g., Figure~\ref{fig:TrajectoriesRealTauGlobalAdS}.

\smallskip

The monodromy of the spatial coordinate $\chi$ is instead associated with the square root in (\ref{etachi}). Encircling either branch point once changes the sign of $\sqrt{z^2+1}$, and one finds
\begin{equation}
\label{chitaunpm}
\tau_k^{(\pm)}:\qquad
\chi\to-\chi\mp i\pi\epsilon(-1)^k\, .
\end{equation}
Note that whether the second term is $+\pi i$ or $-\pi i$ depends on whether $z=i$ or $z=-i$ at the corresponding branch point, respectively, cf.\ (\ref{ypmi}). Consequently, after encircling the same branch point twice, $\chi$ returns to its original value, whereas the global time undergoes the shift $\eta\to\eta\mp2\pi$. Although $\tan\eta$ and $\sinh\chi$ are seen to be unchanged, the point is not the same on the universal covering space of AdS$_2$.\footnote{Interestingly, had we considered the periodic hyperboloid defined via the surface constraint \eqref{eq:HyperboloidEmbeddingAdS2}---rather than its universal cover---even encircling a single puncture would yield a closed trajectory, since the combined monodromy \eqref{etaminuspi}--\eqref{chitaunpm} leaves the embedding point \eqref{eq:GlobalMap} invariant.}

\medskip

To compute the worldline action associated with a given contour $\gamma$ in the complex $\tau$-plane, we use
\begin{equation}\label{eq:ActionSumMomentaGlobal}
S_{\rm wl}
=\int_\gamma d\tau\left(p_\chi\,\dot{\chi}+p_\eta\,\dot{\eta}\right)\, .
\end{equation}
Since $p_\eta$ is conserved, this can be written as
\begin{equation}
S_{\rm wl}
=p_\eta\Delta_\gamma\eta+\int_\gamma d\tau\,p_\chi\dot{\chi}\, ,
\end{equation}
where $\Delta_\gamma\eta$ denotes the total monodromy of $\eta$ along $\gamma$. For a genuine instanton in global AdS$_2$, $\eta$ must return to its original value on the universal covering space, so that $\Delta_\gamma\eta=0$. In that case,
\begin{equation}
S_{\rm inst}
=\oint d\tau\,p_\chi\dot{\chi}\, .
\end{equation}
Nevertheless, it is useful to retain both terms in (\ref{eq:ActionSumMomentaGlobal}) while analyzing a general contour, since their contributions around the branch points simplify considerably, as we will see below.

\smallskip

Using $p_\chi=R^2\dot{\chi}$ and $z=\sinh\chi$, the on-shell Lagrangian can be written as
\begin{equation}
\label{Linst}
L_{\rm inst}(\tau)
\equiv p_\chi\dot{\chi}+p_\eta\dot{\eta}
=R^2\frac{\dot{z}^{\,2}}{1+z^2}+\frac{p_\eta}{1+\tan^2 \eta} \frac{d}{d\tau} \tan \eta\, .
\end{equation}
Notice that, even though $\eta(\tau)$ and $\chi(\tau)$ themselves are multi-valued, $L_{\rm inst}(\tau)$ is a single-valued meromorphic function of $\tau$. It is moreover periodic under
\begin{equation}
\tau\to\tau+\frac{2\pi R^2}{\beta}\, ,
\end{equation}
and has simple poles at the branch points $\tau_k^{(\pm)}$ introduced in (\ref{taukpm}). Furthermore, since
\begin{equation}
\tau_{k+2}^{(\pm)}
=\tau_k^{(\pm)}+\frac{2\pi R^2}{\beta}\, ,
\end{equation}
it is natural to regard the complex $\tau$-plane modulo this periodicity as a cylinder, with four branch points $\tau_k^{(\pm)}$, $k=0,1$, in each fundamental domain.

\smallskip

Let $w\in\mathbb{Z}$ denote the winding number of $\gamma$ around the periodic direction of the $\tau$-cylinder. To fix this definition unambiguously, we choose as the reference periodic cycle any contour running once in the direction of increasing real $\tau$ and lying between the upper branch points $\tau_{0,1}^{(+)}$ and the lower branch points $\tau_{0,1}^{(-)}$.\footnote{This is in contrast to the choice adopted in Sections~\ref{ss:Superextremal} and \ref{s:dS2}, where the reference $w$-cycle is fixed so that all punctures lie to one side of it. Switching between the two choices does not affect the form of the instanton action \eqref{eq:GlobalActionWindings}; it is reflected only in the monodromy induced on $\eta$ and $\chi$, and hence in their closure conditions.} We further denote by $n_k^{(\pm)}\in\mathbb{Z}$ the counterclockwise winding numbers of $\gamma$ around the individual branch points $\tau_k^{(\pm)}$, with $k=0,1$. With respect to this choice of basis, any closed contour on the punctured $\tau$-cylinder can be uniquely characterized by the integers $w$ and $n_k^{(\pm)}$. Notice that a periodic contour lying above or below the branch points need not have $n_k^{(\pm)}=0$ with respect to this choice of basis: deforming it to the reference periodic cycle generally generates additional windings around the branch points.

\begin{figure}[H]
    \centering

		\includegraphics[scale=1]{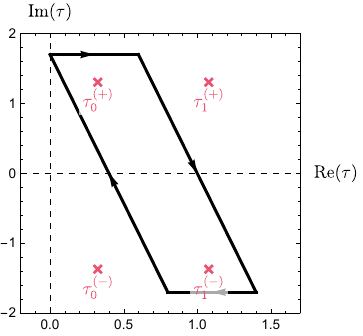}
        \hspace{0.4cm}
      
		\caption{\small We obtain different candidate instanton maps by constructing closed contours in the complex proper-time cylinder. These are grouped into distinct sectors depending on various integers, namely $w$ and $n_k^{(\pm)}$, with $k=0,1$. The latter control the value of the worldline action via eq. \eqref{eq:GlobalActionWindings}. To make the plots, we set $\tilde{m}=3$, $q_e=2$, $R=1$, $\lambda=0.1$, and $w=n_0^{(-)}=n_1^{(+)}=0, n_0^{(+)}=n_1^{(-)}=1$.} 
		\label{fig:ClosedPathsTauGlobalAdS}
\end{figure}

\begin{figure}[H]
    \centering
\subcaptionbox{\label{sfig:NoLoopEtaGlobalAdS}}{
		\includegraphics[scale=1]{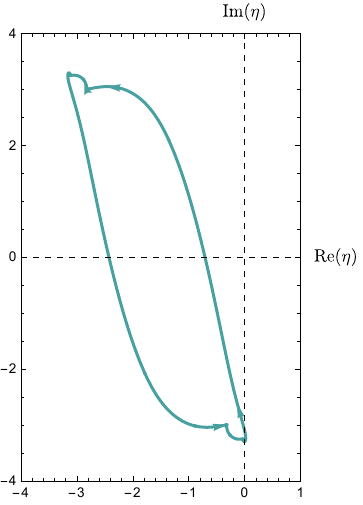}}
        \hspace{0.6cm}
        \subcaptionbox{\label{sfig:NoLoopChiGlobalAdS}}{\includegraphics[scale=1]{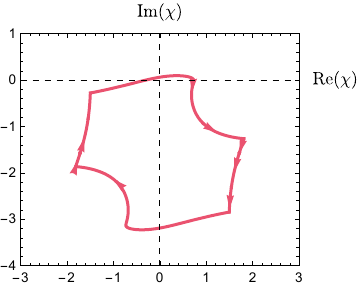}}
		\caption{\small Time $\textbf{(a)}$ and space $\textbf{(b)}$ trajectories corresponding to the path shown in Figure \ref{fig:ClosedPathsTauGlobalAdS}.} 
		\label{fig:ImageNoLoopSpacetimeGlobalAdS}
\end{figure}

Crucially, however, not every such closed contour corresponds to a closed path in global AdS$_2$. From the monodromy of $\eta$ derived above, we find
\begin{equation}
\Delta_\gamma\eta
=-2\pi w
-\pi\sum_{k=0}^{1}n_k^{(+)}
+\pi\sum_{k=0}^{1}n_k^{(-)}\, .
\end{equation}
Hence, closure in the global time direction requires
\begin{equation}\label{eq:GlobalEtaClosure}
2w
+\sum_{k=0}^{1}
\left(n_k^{(+)}-n_k^{(-)}\right)
=0\, .
\end{equation}
This must be supplemented by the requirement of trivial monodromy in $\chi$. A single puncture winding acts as a reflection $\chi \to -\chi \mp i\pi\epsilon(-1)^k$ (cf.\ \eqref{chitaunpm}) rather than an additive shift, so its net effect on $\chi$ is $\Delta \chi= \mp i \pi \epsilon (-1)^k-2\chi$, which cannot vanish for a regular contour. Closure in $\chi$ therefore requires an even total number of such windings, together with the vanishing of the accumulated $\pm i\pi$ shifts. Notice that this condition does depend on the order in which the punctures are wound.\footnote{For two punctures the pair of windings compose to a translation by the difference of their shifts, so closure is symmetric under exchange. Order dependence first appears at four punctures, where the net displacement is the alternating sum $\sum_{j}(-1)^{j}c_{j}$ of the shifts $c_{j}=\mp i\pi\,\epsilon(-1)^{k_{j}}$. Thus, e.g., for shifts $\{+i\pi,+i\pi,-i\pi,-i\pi\}$ the order $(+,+,-,-)$ closes while $(+,-,+,-)$ gives $-4\pi i\neq0$.} Consequently, only contours satisfying both this and the $\eta$-closure condition \eqref{eq:GlobalEtaClosure} correspond to genuine instantons on the universal cover of AdS$_2$. One example of such contours is shown in Figure \ref{fig:ClosedPathsTauGlobalAdS}, with the corresponding instanton solution plotted in Figure \ref{fig:ImageNoLoopSpacetimeGlobalAdS}.

%%%%%%%%%%%%%%%%%%%%%%%%%%%%%%%%%%%%%%%%%%%%%
%%%%%%%%%%%%%%%%%%%%%%%%%%%%%%%%%%%%%%%%%%%%%

Now, a direct evaluation along the reference periodic cycle gives
\begin{equation}
\int_0^{2\pi R^2/\beta}d\tau\,L_{\rm inst}
=-2\pi\beta\, .
\end{equation}
Moreover, using (\ref{ypmi}) together with the local behavior of the solutions near the branch points, one finds
\begin{equation}
\underset{\tau=\tau_k^{(\pm)}}{\operatorname{Res}}L_{\rm inst}(\tau)
=-\frac{q_e}{2}\epsilon(-1)^k\, ,
\qquad k=0,1\, .
\end{equation}
The worldline action associated with a general contour $\gamma$ can therefore be expressed entirely in terms of these winding numbers:
\begin{equation}\label{eq:GlobalActionWindings}
S_{\rm wl}
=-2\pi\beta\,w
-i\pi q_e\epsilon
\sum_{k=0}^{1}(-1)^k
\left(n_k^{(+)}+n_k^{(-)}\right)\, .
\end{equation}
Thus, the value of the action depends only on the homotopy class of the contour on the punctured $\tau$-cylinder, rather than on its detailed shape.

%%%%%%%%%%%%%%%%%%%%%%%%%%%%%%%%%%%%%%%%%%%%%

%%%%%%%%%%%%%%%%%%%%%%%%%%%%%%%%%%%%%%%%%%%%%

\subsubsection{Super-extremal particles}
\label{sec:global-super-extremal}

\begin{figure}[t!]
\begin{center}
\includegraphics[scale=1]{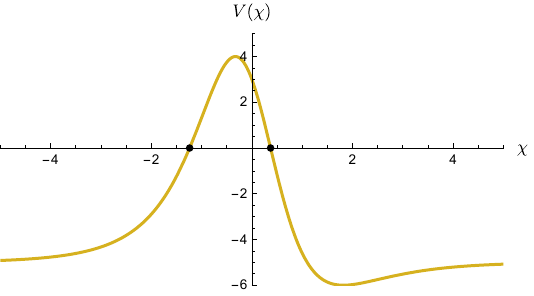}
\caption{\small Effective potential $V(\chi)$ governing the super-extremal motion in global AdS$_2$ coordinates. We set $q_e=3$, $\tilde m=2$, and $p_\eta=1$.}
\label{fig:EffPotentialGlobalAdSSuperExtremal}
\end{center}
\end{figure}

We now turn to the super-extremal case, $\tilde m<|q_e|$. Without loss of generality we take $q_e>0$. The effective potential defined in (\ref{eq:global-super-potential}) is shown in Figure~\ref{fig:EffPotentialGlobalAdSSuperExtremal}, taking $q_e=3$, $\tilde m=2$, and $p_\eta=1$ as an illustrative example. The infinitely many turning points in the complex $\chi$-plane take the same form as in (\ref{eq:GlobalTurningPoints}),
\begin{equation}
\chi^{(\pm)}_+
\in 2\pi i\mathbb{Z}+\log\left(\frac{p_\eta\pm\Delta}{\tilde m-q_e}\right)\, ,
\qquad
\chi^{(\pm)}_-
\in \pi i\left(2\mathbb{Z}+1\right)+\log\left(\frac{p_\eta\pm\Delta}{\tilde m+q_e}\right)\, ,
\end{equation}
where we introduced the quantities 
\begin{equation}
\kappa=\sqrt{q_e^2-\tilde m^2}\, ,
\qquad
\Delta=\sqrt{p_\eta^2+\kappa^2}\, ,
\end{equation}
which are the analytic continuation of those arising in the sub-extremal case; cf.\ \eqref{eq:beta-to-kappa}. 

%%%%%%%%%%%%%%%%%%%%%%%%%%%%%%%%%%%%%%%%%%%%%
%%%%%%%%%%%%%%%%%%%%%%%%%%%%%%%%%%%%%%%%%%%%%

\smallskip

The first-order equation obtained from the Hamiltonian constraint \eqref{H0AdSglobal} can be integrated explicitly. The solution for $\chi$ takes the form
\begin{equation}\label{eq:SinhChiSolAdSSuper}
\sinh\chi(\tau)
=\frac{-p_\eta q_e+\tilde m\Delta
\cosh\left[\frac{\kappa(\tau+\tau_0)}{R^2}\right]}{\kappa^2}\, ,
\end{equation}
while substituting this expression into (\ref{peta}) gives
\begin{equation}\label{eq:EtaSolAdSSuper}
\tan\eta(\tau)
=\frac{\tilde m\kappa
\sinh\left[\frac{\kappa(\tau+\tau_0)}{R^2}\right]}
{\tilde m p_\eta
\cosh\left[\frac{\kappa(\tau+\tau_0)}{R^2}\right]
-q_e\Delta}\, .
\end{equation}
To make contact with the parametrization used in the sub-extremal case, we introduce\footnote{Notice that, unlike in the sub-extremal case, here no lower bound on $|p_\eta|$ arises, since from \eqref{eq:Peta&DeltaWRTLambda} one automatically deduces that $p_\eta^2\geq -\kappa^2$. This also follows directly from the Hamiltonian constraint \cite{Castellano:2025yur, Castellano:2025rvn}.}
\begin{equation}\label{eq:Peta&DeltaWRTLambda}
p_\eta=\kappa\sinh\lambda\, ,
\qquad
\Delta=\kappa\cosh\lambda\, ,
\end{equation}
in terms of which the solutions can be written as
\begin{equation}\label{etachiSuper}
\begin{split}
\eta(\tau)
&=\frac{1}{2i}\big[\log(D+iN)-\log(D-iN)\big]\, ,\\
\chi(\tau)
&=\log\left(z+\sqrt{z^2+1}\right)\, ,
\end{split}
\end{equation}
where
\begin{equation}\label{eq:NDSuper}
N=\tilde m\sinh\left[\frac{\kappa(\tau+\tau_0)}{R^2}\right]\, ,
\qquad
D=\tilde m\sinh\lambda\cosh\left[\frac{\kappa(\tau+\tau_0)}{R^2}\right]-q_e\cosh\lambda\, ,
\end{equation}
and
\begin{equation}
z=\sinh\chi
=\frac{\tilde m\cosh\lambda\cosh\left[\frac{\kappa(\tau+\tau_0)}{R^2}\right]-q_e\sinh\lambda}{\kappa}\, .
\end{equation}
These quantities again satisfy
\begin{equation}
N^2+D^2=\kappa^2\left(z^2+1\right)\, ,
\end{equation}
so that the branch points of $\eta(\tau)$ and $\chi(\tau)$ occur at the same values of complex $\tau$.

%%%%%%%%%%%%%%%%%%%%%%%%%%%%%%%%%%%%%%%%%%%%%
\begin{figure}[t!]
\centering
\subcaptionbox{\label{etatausupfig}}{
\includegraphics[scale=1]{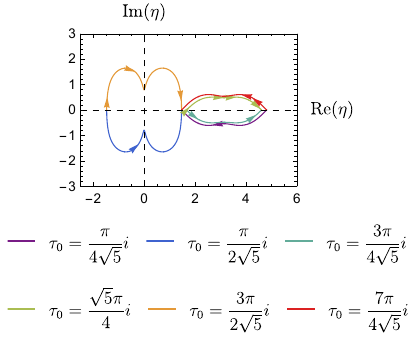}}
\hspace{0.6cm}
\subcaptionbox{\label{chitausupfig}}{
\includegraphics[scale=1]{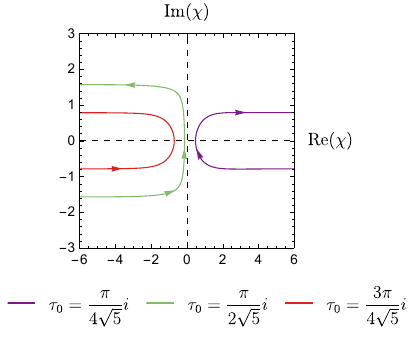}}
\caption{\small Solutions to the real proper-time EOM for $\eta(\tau)$ and $\chi(\tau)$. We show representative trajectories with $\tilde m=2$, $q_e=3$, $R=1$, $\lambda=0.1$, and different values of $\tau_0$.}
\label{fig:TrajectoriesRealTauGlobalAdSsup}
\end{figure}
%%%%%%%%%%%%%%%%%%%%%%%%%%%%%%%%%%%%%%%%%%%%%

\medskip

For the \emph{interference phase}, $\tau$ evolves along the real direction, and the corresponding trajectories are open. In particular, the hyperbolic dependence in (\ref{eq:SinhChiSolAdSSuper}) implies that there is no periodicity along the real $\tau$-direction. We show the solutions (\ref{etachiSuper}) in Figure~\ref{fig:TrajectoriesRealTauGlobalAdSsup} for different values of $\text{Im} (\tau_0)$. As $\tau$ evolves from $-\infty$ to $+\infty$, $\eta(\tau)$ approaches two distinct finite limiting values, while $\chi(\tau)$ starts at the conformal boundary, moves into the bulk, and eventually returns to the same boundary. The trajectories are generically complex, except for the special values
\begin{equation}
\tau_0\in\mathbb{R}+\frac{i\pi R^2}{\kappa}\mathbb{Z}\, ,
\end{equation}
for which both $\eta(\tau)$ and $\chi(\tau)$ are real.

\smallskip

\begin{figure}[t!]
\centering
\subcaptionbox{\label{etasigmasupfig}}{
\includegraphics[scale=1]{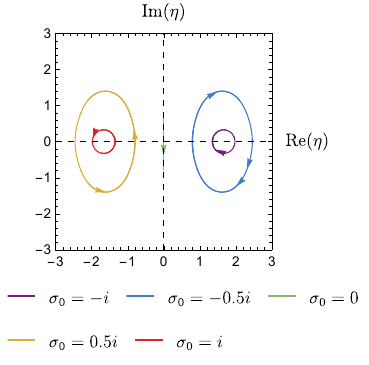}}
\hspace{0.6cm}
\subcaptionbox{\label{chisigmasupfig}}{
\includegraphics[scale=1]{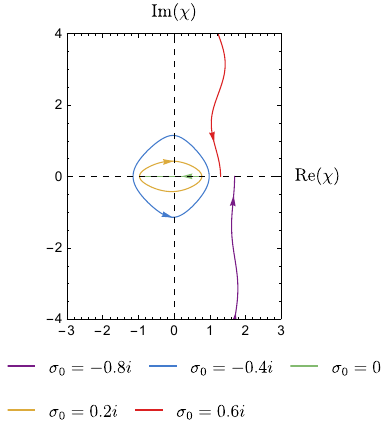}}
\caption{\small Solutions to the imaginary proper-time EOM for $\eta(\sigma)$ and $\chi(\sigma)$. We show representative trajectories with $\tilde m=2$, $q_e=3$, $R=1$, $\lambda=0.1$, and different values of $\sigma_0$.}
\label{fig:TrajectoriesImagTauGlobalAdSsup}
\end{figure}

The situation is reversed in the \emph{instanton phase}. Performing the Wick rotation $\tau=-i\sigma$ and writing $\tau_0=-i\sigma_0$, we plot the solutions $\eta(\sigma)$ and $\chi(\sigma)$ in Figure~\ref{fig:TrajectoriesImagTauGlobalAdSsup} for several values of $\sigma_0$. While the trajectories in the complex $\eta$-plane are always closed, with period $2\pi R^2/\kappa$ in the imaginary proper time $\sigma$, those in the complex $\chi$-plane are closed only for sufficiently small $|\text{Im} (\sigma_0)|$. This difference will be explained by the monodromy analysis below.

%%%%%%%%%%%%%%%%%%%%%%%%%%%%%%%%%%%%%%%%%%%%%
%%%%%%%%%%%%%%%%%%%%%%%%%%%%%%%%%%%%%%%%%%%%%

In particular, for real $\sigma_0$, $\chi(\sigma)$ is real, while $\eta(\sigma)$ is purely imaginary, up to the appropriate choice of branches. Upon Wick rotating the global time coordinate $\eta$, these trajectories therefore reduce to the familiar real Euclidean instantons. Once $|\text{Im} (\sigma_0)|$ exceeds a critical value determined by the branch points discussed below, the trajectory in the complex $\chi$-plane becomes open. As $|\text{Im} (\sigma_0)|$ increases further, the closed trajectory in the complex $\eta$-plane progressively shrinks and eventually localizes around a finite point on the real axis.

\medskip

We now analyze the monodromy of the solutions (\ref{etachiSuper}) for general contours in the complex $\tau$-plane. Without loss of generality, we set $\tau_0=0$. To characterize the branch points, we define
\begin{equation}
\theta
\equiv\arcosh \left(\frac{q_e}{\tilde m}\right)
=\log\left(\frac{q_e+\kappa}{\tilde m}\right)\, ,
\qquad
L\equiv\arccos(\tanh\lambda)\, ,
\end{equation}
with $0<L<\pi$. Introducing $r=\pm1$, the branch points satisfying $D=\pm iN$ are located at
\begin{equation}\label{eq:BranchPointsSuperPlus}
\tau_{k,r}^{(\pm)}
=\frac{R^2}{\kappa}
\left[
r\theta+i(2\pi k\pm L)
\right]\, ,
\qquad k\in\mathbb{Z}\, .
\end{equation}
The branch points therefore form two vertical arrays in the complex $\tau$-plane: the left array consists of $\tau_{k,-}^{(\pm)}$, while the right array consists of $\tau_{k,+}^{(\pm)}$. The monodromies of $\eta$ are exactly analogous to those in the sub-extremal case. Under a single counterclockwise winding around a branch point,
\begin{equation}
\tau_{k,r}^{(\pm)}:
\qquad
\eta\to\eta\mp\pi\, .
\end{equation}
For $\chi$, one instead finds
\begin{equation}\label{eq:ChiMonodromySuper}
\tau_{k,r}^{(\pm)}:
\qquad
\chi\to-\chi\pm r\pi i\, .
\end{equation}
Thus, a single winding around a branch point takes $\chi$ to the other sheet, whereas two windings restore $\chi$ to its original value.

The on-shell Lagrangian remains given by (\ref{Linst}) and is a single-valued meromorphic function of $\tau$. A further qualitative difference from the sub-extremal case is that the periodicity now lies along the imaginary direction,
\begin{equation}
\tau\sim\tau+\frac{2\pi iR^2}{\kappa}\, .
\end{equation}
It is therefore natural to regard the complex $\tau$-plane modulo this identification as a cylinder, with the imaginary direction being periodic. Each fundamental domain contains four branch points, corresponding to $r=\pm1$ and the two signs in (\ref{eq:BranchPointsSuperPlus}).

%%%%%%%%%%%%%%%%%%%%%%%%%%%%%%%%%%%%%%%%%%%%%
\begin{figure}[H]
    \centering

		\includegraphics[scale=1]{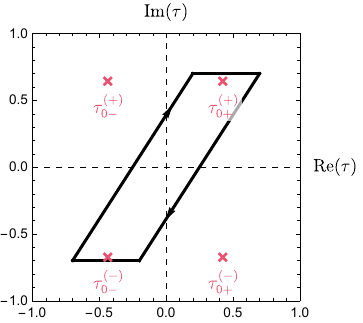}
        \hspace{0.4cm}
      
		\caption{\small We obtain different candidate instanton maps by constructing closed contours in the complex proper-time cylinder. These are grouped into distinct sectors depending on various integers, namely $w$ and $n_{r}^{(\pm)}$, with $r=\pm$. The latter control the value of the worldline action via eq. \eqref{eq:GlobalActionWindings}. To make these plots, we chose $\tilde{m}=2$, $q_e=3$, $R=1$, $\lambda=0.1$, and $w=n_{+}^{(-)}=n_{-}^{(+)}=0, n_+^{(+)}=n_-^{(-)}=1$.} 
		\label{fig:ClosedPathsTauGlobalAdSsup}
\end{figure}
%%%%%%%%%%%%%%%%%%%%%%%%%%%%%%%%%%%%%%%%%%%%%

\begin{figure}[H]
    \centering
\subcaptionbox{\label{sfig:NoLoopEtaGlobalAdSsup}}{
		\includegraphics[scale=1]{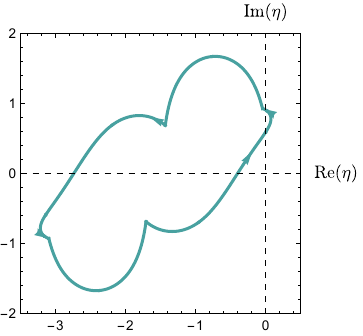}}
        \hspace{0.6cm}
        \subcaptionbox{\label{sfig:NoLoopChiGlobalAdSsup}}{\includegraphics[scale=1]{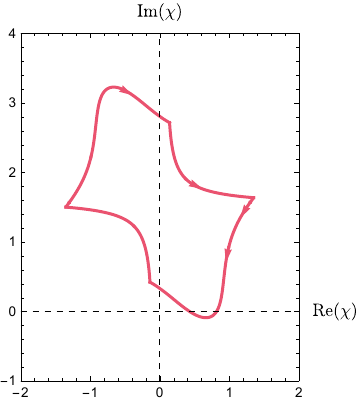}}
		\caption{\small Time $\textbf{(a)}$ and space $\textbf{(b)}$ trajectories corresponding to the path shown in Figure \ref{fig:ClosedPathsTauGlobalAdSsup}.} 
		\label{fig:ImageNoLoopSpacetimeGlobalAdSsup}
\end{figure}

Choosing as the canonical periodic cycle a contour running between the two columns of branch points, a direct evaluation over one period in the positive imaginary direction gives
\begin{equation}
\int_0^{2\pi iR^2/\kappa}
d\tau\,L_{\rm inst}
=-2\pi i(q_e-\kappa)\, .
\end{equation}
The residues of the on-shell Lagrangian at the branch points are
\begin{equation}
 \text{Res}_{\tau=\tau_{0,r}^{(\pm)}}
\left(L_{\rm inst}(\tau)\right)
=\frac{rq_e}{2}\, .
\end{equation}
Let $w\in\mathbb{Z}$ denote the winding number around the periodic direction of the cylinder, with positive $w$ corresponding to increasing $\operatorname{Im}(\tau)$, and let $n_r^{(\pm)}$ denote the counterclockwise winding numbers around the four branch points in a fundamental domain. The action associated with a general contour is then
\begin{equation}\label{eq:GlobalActionWindingsSuper}
S_{\rm wl}
=-2\pi i(q_e-\kappa)w
+i\pi q_e
\sum_{r=\pm1}
r\left(n_r^{(+)}+n_r^{(-)}\right)\, .
\end{equation}
The periodic cycle itself induces no monodromy in $\eta$. Therefore, closure in the global time direction requires the total monodromy generated by the branch-point windings to vanish, namely
\begin{equation}\label{eq:GlobalEtaClosureSuper}
\sum_{r=\pm1}
\left(n_r^{(+)}-n_r^{(-)}\right)
=0\, .
\end{equation}
As in the sub-extremal case, this condition must be supplemented by trivial monodromy in $\chi$. Since the monodromies (\ref{eq:ChiMonodromySuper}) act nontrivially on $\chi$, the corresponding closure condition depends not only on the winding numbers but also on the order in which the branch points are encircled. One example of trajectories satisfying these conditions and the corresponding solutions are shown in Figure \ref{fig:ClosedPathsTauGlobalAdSsup} and \ref{fig:ImageNoLoopSpacetimeGlobalAdSsup}.

%%%%%%%%%%%%%%%%%%%%%%%%%%%%%%%%%%%%%%%%%%%%%
%%%%%%%%%%%%%%%%%%%%%%%%%%%%%%%%%%%%%%%%%%%%%

The standard instanton-phase contour $\tau=-i\sigma$, with $\sigma$ increasing over one period, corresponds to $w=-1$ and vanishing branch-point windings. Its action is therefore
\begin{equation}\label{eq:SuperExtremalGlobalInstantonAction}
S_{\rm inst}
=2\pi i(q_e-\kappa)
=2\pi i\left(q_e-\sqrt{q_e^2-\tilde m^2}\right)\, .
\end{equation}
Thus, unlike in the sub-extremal regime, a closed instanton already arises from evolution purely along the imaginary proper-time direction.

\subsubsection{Extremal particles}
\label{sec:global-extremal}

We finally turn to the extremal case, $\tilde m=q_e$. This regime is qualitatively distinct from the non-extremal cases discussed above, since the characteristic scale $q_e^2-\tilde m^2$ vanishes at extremality. In particular, the imaginary proper-time period $2\pi R^2/\kappa$ of the super-extremal solutions \eqref{eq:SinhChiSolAdSSuper} and \eqref{eq:EtaSolAdSSuper} diverges as the extremal limit is approached.

\begin{figure}[t!]
\begin{center}
\includegraphics[scale=1]{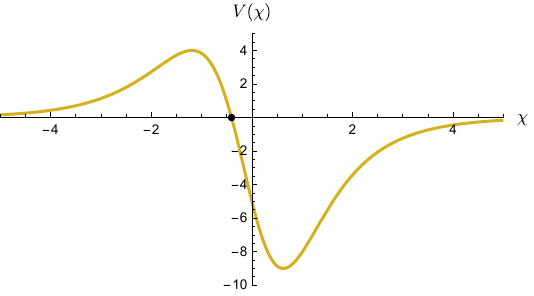}
\caption{\small Effective potential $V(\chi)$ governing the extremal particlemotion in global AdS$_2$ coordinates. We set $\tilde m=q_e=2$ and $p_\eta=1$.}
\label{fig:EffPotentialGlobalAdSExtremal}
\end{center}
\end{figure}

The effective potential (\ref{eq:global-super-potential}) is shown in Figure~\ref{fig:EffPotentialGlobalAdSExtremal}. For $p_\eta\neq0$, there is a single finite real turning point, while the second real turning point present in the non-extremal cases is pushed to infinity in the extremal limit. In the complex $\chi$-plane, however, there remain infinitely many turning points,
\begin{equation}
\chi_*^{(1)}
\in2\pi i\mathbb{Z}+\log\left(\frac{q_e}{p_\eta}\right)\, ,
\qquad
\chi_*^{(2)}
\in(2\mathbb{Z}+1)\pi i+\log\left(\frac{p_\eta}{q_e}\right)\, ,
\end{equation}
where we have taken $q_e>0$ as above.

%%%%%%%%%%%%%%%%%%%%%%%%%%%%%%%%%%%%%%%%%%%%%
%%%%%%%%%%%%%%%%%%%%%%%%%%%%%%%%%%%%%%%%%%%%%

Integrating the differential equation (\ref{eq:global-super-potential}) directly at extremality, $\tilde m=q_e$, we obtain
\begin{equation}\label{eq:SinhChiSolAdSExtremal}
z=\sinh\chi(\tau)
=\frac{p_\eta q_e}{2}\left(\frac{\tau+\tau_0}{R^2}\right)^2
+\frac{q_e^2-p_\eta^2}{2p_\eta q_e}\, ,
\end{equation}
where $\tau_0$ is an integration constant. Substituting this expression into (\ref{peta}) then gives
\begin{equation}\label{eq:EtaSolAdSExtremal}
\tan\eta(\tau)
=\frac{2q_e^2p_\eta(\tau+\tau_0)R^2}
{q_e^2p_\eta^2(\tau+\tau_0)^2-(p_\eta^2+q_e^2)R^4}\, .
\end{equation}
These expressions can also be retrieved by taking the $\kappa\to0$ limit of the super-extremal solutions, provided the integration constants are scaled appropriately.

The branch structure of the extremal solutions can again be made explicit by writing
\begin{equation}\label{eq:etachiExtremal}
\begin{split}
\eta(\tau)
&=\frac{1}{2i}\big[\log(D+iN)-\log(D-iN)\big]\, ,\\
\chi(\tau)
&=\log\left(z+\sqrt{z^2+1}\right)\, ,
\end{split}
\end{equation}
where
\begin{equation}\label{eq:NDExtremal}
N=2q_e^2p_\eta\frac{\tau+\tau_0}{R^2}\, ,
\qquad
D=q_e^2p_\eta^2\left(\frac{\tau+\tau_0}{R^2}\right)^2-p_\eta^2-q_e^2\, .
\end{equation}
These quantities satisfy the identity
\begin{equation}
N^2+D^2
=4q_e^2p_\eta^2\left(z^2+1\right)\, ,
\end{equation}
so that the branch points of $\eta(\tau)$, characterized by $D=\pm iN$, coincide with those of $\chi(\tau)$, located at $z=\pm i$. Thus, as in the non-extremal cases, the two global coordinates develop branch points at the same values of the complex proper time $\tau$.

\smallskip
%%%%%%%%%%%%%%%%%%%%%%%%%%%%%%%%%%%%%%%%%%%%%
\begin{figure}[t!]
    \centering
\subcaptionbox{\label{etatauextfig}}{
		\includegraphics[scale=1]{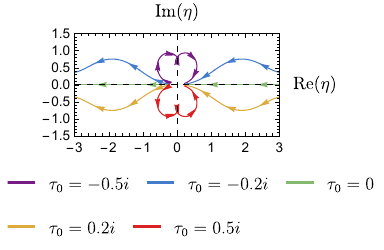}}
        \hspace{0.6cm}
        \subcaptionbox{\label{chitauextfig}}{\includegraphics[scale=1]{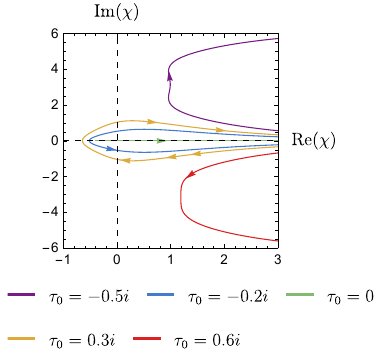}}
		\caption{\small Solutions to the real proper-time EOM for $\eta(\tau)$ and $\chi (\tau)$. In the universal covering space, they are generically not closed. Their behavior with respect to $\text{Im}(\tau_0)$ is symmetric, such that changing the sign preserves the trajectory in $\chi$-plane but reverses its orientation, whereas for the $\eta$-component the paths get reflected across the real axis. We plot various cases with $\tilde{m}=q_e=2$, $R=1$, $p_\eta=3$, and different values of $\tau_0$.} 
		\label{fig:TrajectoriesRealTauGlobalAdSext}
\end{figure}
%%%%%%%%%%%%%%%%%%%%%%%%%%%%%%%%%%%%%%%%%%%%%

For the \emph{interference phase}, the proper time $\tau$ evolves along the real direction. When $\tau_0\in\mathbb{R}$, both $\chi(\tau)$ and $\eta(\tau)$ can be chosen to be real, and the resulting trajectories are open and non-periodic. For generic $\tau_0\in\mathbb{C}$, the trajectories instead become complex. Unlike in the sub- and super-extremal cases, however, no finite proper-time periodicity emerges: the quadratic dependence on $\tau+\tau_0$ in (\ref{eq:SinhChiSolAdSExtremal}), together with the corresponding rational dependence in (\ref{eq:EtaSolAdSExtremal}), precludes any nonzero finite shift of $\tau$ under which the solutions are periodic. Several representative trajectories are shown in Figure~\ref{fig:TrajectoriesRealTauGlobalAdSext}.

\begin{figure}[t!]
    \centering
\subcaptionbox{\label{etasigmaextfig}}{
		\includegraphics[scale=1]{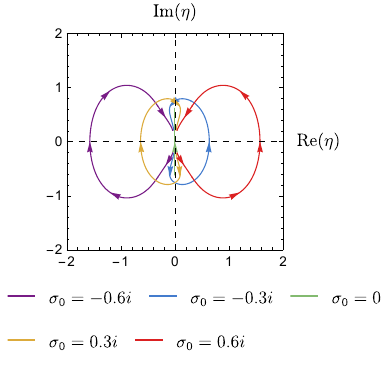}}
        \hspace{0.6cm}
        \subcaptionbox{\label{chisigmaextfig}}{\includegraphics[scale=1]{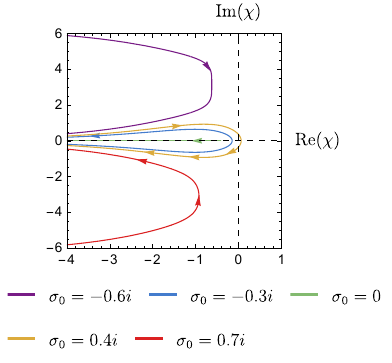}}
		\caption{\small The solutions for $\chi(\sigma)$ and $\eta(\sigma)$ in imaginary proper-time $\sigma \in (-\infty, \infty)$ are open. We plotted various representative cases with $\tilde{m}=q_e=2$, $R=1$, $p_\eta=3$, and different values of $\sigma_0$.} 
		\label{fig:TrajectoriesImaginaryTauGlobalAdSext}
\end{figure}

The \emph{instanton-phase} solutions are obtained instead by setting $\tau=-i\sigma$ and $\tau_0=-i\sigma_0$. In Figure~\ref{fig:TrajectoriesImaginaryTauGlobalAdSext}, we plot these solutions for several values of $\sigma_0$. For real $\sigma_0$, $\chi(\sigma)$ can be chosen to be real, while $\eta(\sigma)$ is purely imaginary on the corresponding branches. More precisely, as $\sigma$ varies from $-\infty$ to $+\infty$, $\eta(\sigma)$ starts at the origin and oscillates along the imaginary axis, with turning points
\begin{equation}
\eta_\pm
=\pm i\,\artanh\left(\frac{q_e}{\sqrt{p_\eta^2+q_e^2}}\right)\, ,
\end{equation}
before returning to $\eta=0$. Meanwhile, $\chi(\sigma)$ starts from the conformal boundary at $\chi\to-\infty$, reaches the finite turning point $\chi_*^{(1)}$, and then returns to the same boundary.

For small nonzero $|\operatorname{Im}(\sigma_0)|$, both projections become complex and deform so as to enclose their respective $\sigma_0=0$ representatives. Once $|\operatorname{Im}(\sigma_0)|$ exceeds a critical value determined by the branch points discussed below, their qualitative behavior changes. The trajectory in the complex $\eta$-plane becomes a closed loop lying on one side of the imaginary axis, whereas $\chi(\sigma)$ undergoes a change of branch as $\sigma$ evolves from $-\infty$ to $+\infty$. This transition will be understood more precisely from the monodromy analysis below.

%%%%%%%%%%%%%%%%%%%%%%%%%%%%%%%%%%%%%%%%%%%%%
%%%%%%%%%%%%%%%%%%%%%%%%%%%%%%%%%%%%%%%%%%%%%

For general contours in the complex $\tau$-plane, setting $\tau_0=0$ and introducing $r=\pm1$, the two roots of $D=\pm iN$ are
\begin{equation}\label{eq:BranchPointsExtremalPlus}
\tau_r^{(\pm )}
=R^2\left(\frac{r}{q_e}\pm \frac{i}{p_\eta}\right)\, .
\end{equation}
At these points,
\begin{equation}
z\left(\tau_r^{(\pm )}\right)=\pm ri\, .
\end{equation}
Their monodromies therefore take exactly the same local form as in the super-extremal case:
\begin{equation}
\label{eq:ChiMonodromyExtremal}
\tau_r^{(\pm)}:
\qquad
\eta\to\eta\mp\pi\,, \quad \chi\to-\chi\pm ri\pi\, .
\end{equation}

Unlike in the non-extremal cases, however, there is no finite periodic direction in the complex $\tau$-plane. The on-shell Lagrangian
\begin{equation}
L_{\rm inst}(\tau)
=R^2\frac{\dot z^{\,2}}{1+z^2}+p_\eta\dot\eta\, ,
\end{equation}
is a meromorphic function with only the four finite branch-point poles (\ref{eq:BranchPointsExtremalPlus}). Their residues are
\begin{equation}
 \text{Res}_{\tau=\tau_r^{(\pm)}}
\left(L_{\rm inst}(\tau)\right)
=\frac{rq_e}{2}\, .
\end{equation}
If $n_r^{(\pm)}$ denotes the counterclockwise winding number of a contour around each branch point, its action is therefore
\begin{equation}\label{eq:GlobalActionWindingsExtremal}
S_{\rm wl}
=i\pi q_e
\sum_{r=\pm1}
r\left(n_r^{(+)}+n_r^{(-)}\right)\, .
\end{equation}
Closure in the global time direction requires
\begin{equation}\label{eq:GlobalEtaClosureExtremal}
\sum_{r=\pm1}
\left(n_r^{(+)}-n_r^{(-)}\right)
=0\, ,
\end{equation}
while closure in $\chi$ must again be imposed using the ordered product of the reflections (\ref{eq:ChiMonodromyExtremal}). Only contours satisfying both conditions describe genuine closed trajectories on the universal covering space of AdS$_2$.

It is instructive to compare this result with the super-extremal regime. As $\tilde m\to q_e$ from below,
\begin{equation}
\kappa=\sqrt{q_e^2-\tilde m^2}\to0\, ,
\qquad
\frac{2\pi R^2}{\kappa}\to\infty\, .
\end{equation}
Thus, the canonical periodic instanton of the super-extremal theory is pushed to infinite proper time at extremality. Nevertheless, its action has the finite limiting value
\begin{equation}\label{eq:ExtremalInstantonActionLimit}
\lim_{\tilde m\to q_e^-}
S_{\rm inst}
=2\pi iq_e\, .
\end{equation}
The extremal configuration should therefore be understood with some care: at exactly $\tilde m=q_e$, the finite-period instanton degenerates into a trajectory extending to infinity in the complex proper-time plane, even though its action remains finite when approached from the super-extremal side.

\subsection{dS$_2$}\label{ss:GlobaldS}

In Section~\ref{s:dS2}, we considered the worldline instanton problem associated with charged massive particles evolving in 2D de Sitter spacetime threaded by a homogeneous electric field. Moreover, to simplify the analysis---and to compare the results with the AdS$_2$ solutions derived in Section~\ref{s:AdS2}---we employed the so-called flat-slicing coordinates, in terms of which the line element and gauge field read as in \eqref{eq:dS2Metric} and \eqref{eq:GaugeFieldFlatdS2}, respectively. However, as already stated there, these coordinates only cover a portion of the global dS$_2$ spacetime (cf.\ Figure~\ref{fig:dS2Penrose}). Consequently, our aim in what follows will be to reevaluate this problem from the perspective of the full de Sitter spacetime.

\smallskip

\begin{figure}[H]
\begin{center}
\includegraphics[scale=0.7]{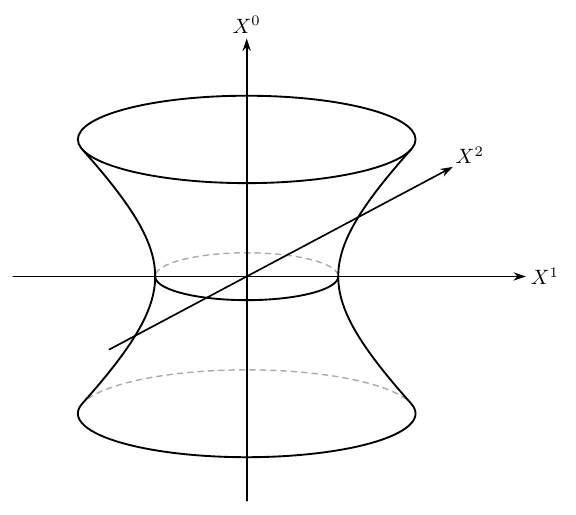}
\caption{\small Depiction of dS$_2$ as a hyperboloid embedded in $\mathbb{R}^{1,2}$ via \eqref{eq:HyperboloidEmbeddingdS2}.}
\label{fig:HyperboloiddS2}
\end{center}
\end{figure}

Accordingly, let us define global two-dimensional de Sitter as the one-sheeted hyperboloid embedded in 3D Minkowski space (cf.\ Figure \ref{fig:HyperboloiddS2}) via
\begin{equation}\label{eq:HyperboloidEmbeddingdS2}
-(X^0)^2 + (X^1)^2 + (X^2)^2 = R^2\, ,
\end{equation}
with ambient metric $\eta_{AB}=\mathrm{diag}(-1,1,1)$. The above surface constraint can be easily solved via the parametrization
\begin{equation}\label{eq:deSitterHyperboloid}
    X^0 = R \sinh t \, , \quad X^1 = R \cosh t\, \cos \theta \, , \quad X^2 = R \cosh t\, \sin \theta \, ,
\end{equation}
where the coordinates \(t\in\mathbb{R}\) and \(\theta \in [0, 2\pi)\) are known as \textit{global coordinates}, since they cover the entire manifold. The resulting line element is obtained by pulling back $\eta_{A B}$ via \eqref{eq:deSitterHyperboloid}:
\begin{equation}\label{eq:dSMetricGlobal}
    ds^2 = R^2 \left(-d{t}^2 + \cosh^2t\, d{\theta}^2\right)\,.
\end{equation}
Furthermore, since $(X^1,X^2)$ depend on $\theta$ only through $(\sin\theta,\cos\theta)$, the embedding naturally identifies $\theta\sim\theta+2\pi$, which is a genuine symmetry of de Sitter space. Unlike the timelike compact direction of global AdS$_2$ (Section~\ref{ss:GlobalAdS}), which we unwrapped by passing to the universal cover, this spatial identification is physical and we retain it throughout.

\smallskip

%%%%%%%%%%%%%%%%%%%%%%%%%%%%%%%%%%%%%%%%%%%%%
%%%%%%%%%%%%%%%%%%%%%%%%%%%%%%%%%%%%%%%%%%%%%

On the other hand, the flat-slicing patch can be retrieved from \eqref{eq:HyperboloidEmbeddingdS2} by defining instead
\begin{equation}\label{eq:flatslicing}
    X^0 = \frac{\zeta^2-R^2}{2\zeta} - \frac{y^2}{2\zeta}\, , \quad X^1 = -\frac{\zeta^2+R^2}{2\zeta} +\frac{y^2}{2\zeta}\, , \quad X^2 = -\frac{y R}{\zeta}\, ,
\end{equation}
with \(\zeta \in (-\infty, 0)\) and \(y \in (-\infty, \infty)\). The corresponding pull-back metric is therefore \eqref{eq:dS2Metric}, which we repeat here for the reader's convenience:
\begin{equation}
    \label{eq:dS2ConformallyFlatMetric}
    ds^2 = \frac{R^2}{\zeta^2} \left(-d{\zeta}^2+d{y}^2 \right)\, .
\end{equation}
Notice that the flat slicing covers exactly half of the full spacetime---since from \eqref{eq:flatslicing} we have that $X^0+X^1=-R^2/\zeta\geq 0$---and geometrically it corresponds to the causal future of a given point at past infinity.

\medskip

As in the AdS$_2$ case, this restriction becomes significant once the coordinates are complexified. For real $(\zeta,y)$ the boundary of the flat-slicing patch---at future infinity $\zeta\to0^-$ and past infinity $\zeta\to-\infty$---cannot be reached at finite values of the coordinates. Upon complexification, however, the map relating the flat slicing to the global chart $(t,\theta)$ becomes multi-valued, so that a trajectory which closes in the complexified flat slicing need not close once lifted to global de Sitter. We must therefore reexamine the dS$_2$ instantons of Section~\ref{s:dS2} from the present global viewpoint. Because the compact direction is spatial and physically identified, closure in global dS$_2$ requires only that $(t,\theta)$ return to their original values modulo $\theta\sim\theta+2\pi$---a milder condition than in AdS$_2$, where closure had to hold on the universal cover. As we shall see, this is what permits certain complex trajectories to descend to genuine instantons in de Sitter even when their AdS$_2$ analogues fail to close.

\smallskip

Let us finally point out that, in addition to the spatial identification $\theta\sim\theta+2\pi$, the global parametrization \eqref{eq:deSitterHyperboloid} is periodic under $t\to t+2\pi i$. This imaginary-time periodicity reflects the fact that the Euclidean section of global dS$_2$---obtained via the Wick rotation $t\to iT$---is the round two-sphere of radius $R$, on which $T$ is an angular coordinate of period $2\pi$; any other possibility would introduce a conical defect at the poles. The associated period $2\pi R$ in Euclidean proper time is the inverse Gibbons--Hawking temperature $T_{\rm dS}=1/(2\pi R)$ of the de Sitter static patch \cite{Gibbons:1977mu}, and it will play a role in the monodromy analysis below.

\subsubsection{Solutions to the worldline EOM}
\label{sss:SolsGlobaldS2}

\begin{figure}[t!]
	\begin{center}
		\includegraphics[scale=1]{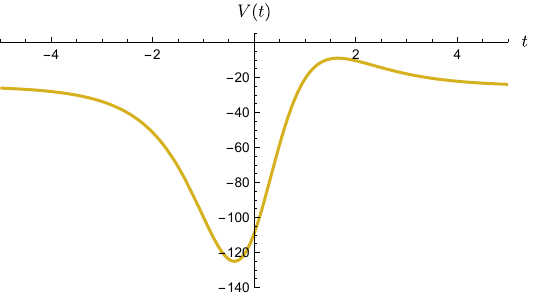}
		\caption{\small Effective potential $V(t)$ controlling the temporal dynamics in global dS$_2$ coordinates. Notice that, unlike in AdS$_2$, there are no real turning points. We set $q_e=4$, $m R=3$, and $p_\theta=10$.} 
		\label{fig:EffPotentialGlobaldS}
	\end{center}
\end{figure}

We start by first setting up the problem directly in the global chart. In terms of the global coordinates \eqref{eq:dSMetricGlobal}, the worldline action \eqref{eq:worldlineAction} is given by
\begin{equation}\label{SwldSglobal}
 S_{\rm wl} =  \int_\gamma d\tau\, \left(-\frac{m^2 }{2}-\frac{R^2\dot t^2}{2}+\frac{R^2\cosh^2t\,\dot\theta^2}{2} +q_e  \sinh t\,\dot\theta\right)\, ,
\end{equation}
where we have included some homogeneous $\U(1)$ gauge field of the form
\begin{equation}
    F = E\, \omega_{\text{dS}_2}= E R^2 \cosh t\, dt \wedge d \theta\, ,
\end{equation}
in a gauge with $A=E R^2\sinh t\, d\theta$, and we also defined $q_e= eE R^2$. The corresponding EOM now become
\begin{equation}\label{eq:EOMGlobaldS}
\begin{aligned}
    &\Ddot{\theta} \cosh^2 t + 2 \dot{\theta} \dot{t} \cosh{t} \sinh t+ \frac{q_e}{R^2} \dot{t} \cosh t=0\, ,\\
    &\Ddot{t} + \dot{\theta}^2 \sinh t \cosh t + \frac{q_e}{R^2} \dot{\theta} \cosh t=0\, .
\end{aligned}
\end{equation} 
As always, these must be supplemented with the Hamiltonian constraint
\begin{equation}\label{eq:HamiltonianConstGlobaldS}
  R^4 \dot{t}^2- 
\left( p_\theta \sech t - q_e \tanh{t} \right)^2-m^2 R^2 = 0 \iff p_t^2+V(t)=0\, ,
\end{equation}
with $p_\theta =R^2 \dot{\theta}\cosh^2 t+q_e \sinh t$, and $p_t=-R^2 \dot{t}$. In the second expression above, we introduced the effective potential
\begin{equation}\label{eq:EffPotentialGlobaldS}
 V(t)=- 
\left( p_\theta \sech t - q_e \tanh{t} \right)^2-\tilde{m}^2\, ,
\end{equation}
which governs the dynamics of the global time coordinate $t$; see Figure~\ref{fig:EffPotentialGlobaldS}. Before proceeding, let us note the direct relation between the analogous AdS$_2$ problem studied in Section~\ref{ss:GlobalAdS} and the global de Sitter analysis considered herein. Indeed, under the map
\begin{equation}\label{eq:MapGlobalAdStodS}
R_{\rm{AdS}_2}\leftrightarrow-iR_{\rm{dS}_2}\,,
\qquad
\chi\leftrightarrow t\,,
\qquad
\eta\leftrightarrow\theta\,,
\end{equation}
which also implies $q_e \to -q_e$, the relevant equations of motion and their solutions can be obtained from their Anti-de Sitter counterparts, as one may easily verify. In the following, we derive the general solutions to the complexified equations of motion \eqref{eq:EOMGlobaldS} by direct integration of the Hamiltonian constraint rather than solving the second-order equations. We then examine their main properties, leaving a close analysis of their worldline action for Section~\ref{sss:ClassificationGlobaldS2}.

%%%%%%%%%%%%%%%%%%%%%%%%%%%%%%%%%%%%%%%%%%%%%
%%%%%%%%%%%%%%%%%%%%%%%%%%%%%%%%%%%%%%%%%%%%%

\medskip

To obtain the instanton profiles, we proceed by integrating the first-order Hamiltonian constraint \eqref{eq:HamiltonianConstGlobaldS}, which may be rewritten as
\begin{equation}\label{eq:DiffEqGlobaldS}
R^4\dot{z}^2 =
\left(p_\theta - q_e z\right)^2+\tilde{m}^2\left(1+z^2\right)\, ,
\end{equation}
where $z=\sinh t$. Equivalently,
\begin{equation}\label{eq:globaldSturning-points}
R^4\dot z^{2}
=\Omega^2(z-z_+)(z-z_-)\, ,
\qquad
z_\pm=\frac{p_\theta q_e\pm i\tilde{m}\Delta}{\Omega^2}\, .
\end{equation}
Here, we introduced the dimensionless quantities
\begin{equation}\label{eq:OmegaDeltaGlobaldS}
\Omega=\sqrt{q_e^2+\tilde{m}^2}\, ,
\qquad
\Delta=\sqrt{p_\theta^2+\Omega^2}\, .
\end{equation}
The roots $z_\pm$ are the turning points of the global de Sitter motion. In contrast to the Anti-de Sitter case, the $z_\pm$ are now
complex conjugates for any real $p_\theta$, since the discriminant of \eqref{eq:DiffEqGlobaldS} is $-4\tilde{m}^2\Delta^2<0$. Consequently, no analogue of the sub-extremal/super-extremal distinction of the AdS$_2$ problem arises here: the qualitative structure of the solutions is insensitive to the ratio $R|eE|/m$, exactly as in the flat-slicing analysis of Section~\ref{s:dS2}.

\smallskip

Integrating \eqref{eq:DiffEqGlobaldS} yields
\begin{equation}\label{eq:globalSoldS}
\sinh t(\tau)=z(\tau)=\frac{p_\theta q_e-i\tilde{m}\Delta\cosh\left[ \frac{\Omega}{R^2}(\tau+\tau_0)\right]}{\Omega^2}\, ,
\end{equation}
with $\tau_0\in\mathbb{C}$ arbitrary, while the conserved momentum $p_\theta$ fixes the angular profile,
\begin{equation}\label{eq:GlobalTheta}
\tan\theta(\tau)
=-\frac{\tilde{m}\Omega\sinh\left[ \frac{\Omega}{R^2}(\tau+\tau_0)\right]}
{iq_e\Delta+p_\theta\tilde{m}\cosh\left[ \frac{\Omega}{R^2}(\tau+\tau_0)\right]}\, .
\end{equation}
These are the global-coordinate analogues of the flat-slicing solutions
\eqref{eq:zetaSolutiondS2}--\eqref{eq:ySolutiondS2} and, via \eqref{eq:MapGlobalAdStodS}, of the global AdS$_2$ solutions \eqref{eq:SinhChiSolAdS} and \eqref{etatauglobalsub}. As there, it is convenient to write
\begin{equation}\label{thetaTdS}
\theta(\tau)=\frac{1}{2i}\big[\log(D+iN)-\log(D-iN)\big]\, ,
\qquad
t(\tau)=\log\!\left(z+\sqrt{z^2+1}\right)\, ,
\end{equation}
with
\begin{equation}\label{eq:NDdS}
D=-\frac{iq_e\Delta+p_\theta\tilde m\cosh\left[ \frac{\Omega}{R^2}(\tau+\tau_0)\right]}{\Omega}\, ,
\qquad
N=\tilde m\sinh\left[ \frac{\Omega}{R^2}(\tau+\tau_0)\right]\, .
\end{equation}
They obey $N^2+D^2=-\Omega^2(z^2+1)$---the de Sitter counterpart of the AdS$_2$ identity \eqref{eq:NDidAdS2}---such that the branch points of $\theta(\tau)$ at $D=\pm iN$ coincide with those of $t(\tau)$ at $z=\pm i$.

\medskip

\begin{figure}[t!]
    \centering
    \subcaptionbox{\label{sfig:TcoordinateReTauGlobaldS}}{
        \includegraphics[scale=1]{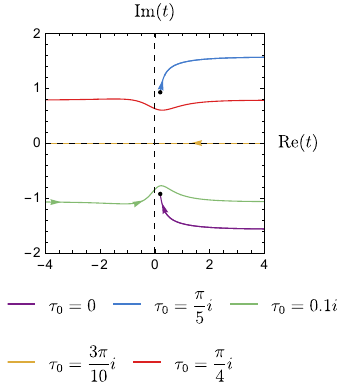}}
    \hspace{0.6cm}
    \subcaptionbox{\label{sfig:ThetacoordinateReTauGlobaldS}}{
        \includegraphics[scale=1.1]{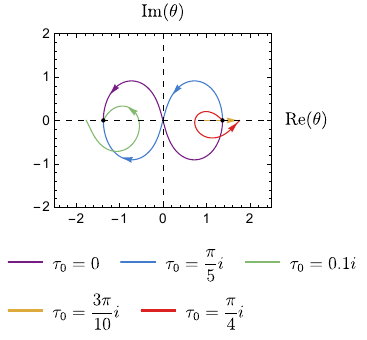}}
    \caption{Solutions to the real proper-time dS$_2$ EOM for $t(\tau)$ and $\theta(\tau) \in (-\pi, \pi]$. We show various representative trajectories with $\tilde m=4$, $q_e=3$, $p_\theta=R=1$, and different values of $\tau_0$.}
    \label{fig:globaldSRealTauProfiles}
\end{figure}

Having obtained the general profiles \eqref{eq:globalSoldS}, \eqref{eq:GlobalTheta}, and \eqref{thetaTdS}, we now characterize their qualitative behaviour. In the \emph{interference phase}, where $\tau$ evolves along the real axis, Figure~\ref{fig:globaldSRealTauProfiles} shows the $t$- and $\theta$-projections for various representative values of the integration constant $\tau_0$. The trajectories are always open, extending over the full range $\tau \in (-\infty, \infty)$, and are generically complex; the only exception occurs on the lattice
\begin{equation}
    \tau_0\in \mathbb{R} +\frac{i\pi R^2}{\Omega}\left( \mathbb{Z}+\frac12\right)\, ,
\end{equation}
where both $t(\tau)$ and $\theta(\tau)$ are real. These are the familiar charged geodesics of dS$_2$. 
 
%%%%%%%%%%%%%%%%%%%%%%%%%%%%%%%%%%%%%%%%%%%%%
%%%%%%%%%%%%%%%%%%%%%%%%%%%%%%%%%%%%%%%%%%%%%

\medskip
 
\begin{figure}[t!]
    \centering
    \subcaptionbox{\label{sfig:TcoordinateImTauGlobaldS}}{
        \includegraphics[scale=1]{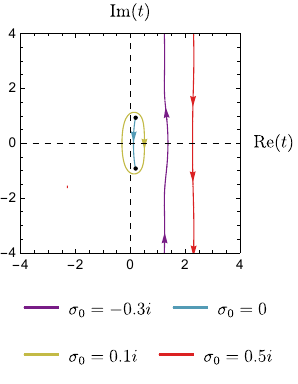}}
    \hspace{0.6cm}
    \subcaptionbox{\label{sfig:ThetacoordinateImTauGlobaldS}}{
        \includegraphics[scale=1.2]{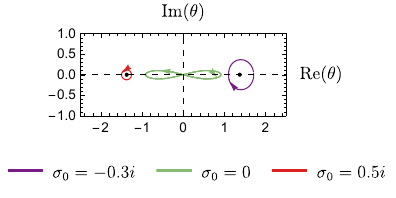}}
    \caption{Solutions to the imaginary proper-time dS$_2$ EOM for $t(\sigma)$ and $\theta(\sigma) \in (-\pi, \pi]$. We show various representative cases with $\tilde m=4$, $q_e=3$, $p_\theta=R=1$, and different $\sigma_0$.}
    \label{fig:globaldSImaginaryTauProfiles}
\end{figure}

On the other hand, the \emph{instanton phase} solutions are obtained by Wick-rotating the proper time variable via $(\tau+\tau_0)=-i(\sigma+\sigma_0)$, with $\sigma\in \mathbb{R}$. We plot the resulting spacetime trajectories in Figure~\ref{fig:globaldSImaginaryTauProfiles}, for various representative values of the complex constant $\sigma_0$. We note that, despite the periodicity 
\begin{equation}
    \sigma\sim\sigma+\frac{2\pi R^2}{\Omega}\, ,
\end{equation}
of the $\sinh t$ and $\tan \theta$ solutions (cf.\ \eqref{eq:globalSoldS} and \eqref{eq:GlobalTheta}), the analogous profiles for $t(\sigma)$ and $\theta(\sigma)$ are not necessarily periodic. In fact, whilst the spatial projections are always closed---regardless of the value of $\sigma_0$---those associated with the global time coordinate are only so for sufficiently small $|\text{Im} (\sigma_0)|$. Once $|\text{Im} (\sigma_0)|$ exceeds a certain critical value determined by the branch points discussed below, the trajectory in the complex $t$-plane becomes open. Still, since the monodromy exhibited by the latter is $t \sim t+2\pi i$ and this coincides precisely with the Euclidean time coordinate identification, we conclude that also the $t$-projection yields a periodic motion. As $|\text{Im} (\sigma_0)|$ is further increased, the path traced in the complex $\theta$-plane
progressively shrinks and eventually localizes around the special points $\theta_\pm = \arctan (\pm \Omega/p_\theta)$.
 
%%%%%%%%%%%%%%%%%%%%%%%%%%%%%%%%%%%%%%%%%%%%%
%%%%%%%%%%%%%%%%%%%%%%%%%%%%%%%%%%%%%%%%%%%%%

\subsubsection{Classification of instanton sectors and worldline action}\label{sss:ClassificationGlobaldS2}

We now classify the topologically inequivalent instanton sectors and compute their worldline actions, following the same strategy as in Section~\ref{ss:GlobalAdS}. For general contours in the complex $\tau$-plane, we must in addition keep track of the monodromy exhibited by the solution \eqref{thetaTdS}. Without loss of generality we henceforth set $\tau_0=0$.

\smallskip

Since $N$, $D$ and $z$ depend on $\tau$ only through $\cosh\Theta(\tau)$ and $\sinh\Theta(\tau)$, with $\Theta=\Omega\tau/R^2$, the quantities appearing inside the logarithms in \eqref{thetaTdS} are periodic under
\begin{equation}\label{eq:dSperiod}
\tau\to\tau+\frac{2\pi i R^2}{\Omega}\, .
\end{equation}
In contrast to the sub-extremal AdS$_2$ case---and mirroring both the super-extremal AdS$_2$ and the flat-slicing dS$_2$ analyses---it is now the \emph{imaginary} $\tau$-direction that becomes periodic. We may therefore regard the complex $\tau$-plane, modulo \eqref{eq:dSperiod}, as a cylinder whose compact direction is purely imaginary.

The branch points of \eqref{thetaTdS}---where both $D=\pm iN$ and $z=\pm i$---are the solutions of $\cosh\Theta(\tau)=(p_\theta q_e\mp i\Omega^2)/(i\tilde m\Delta)$, which are located at
\begin{equation}\label{eq:dSGlobalPunctures}
\tau_{r, k}^{(\pm)}=\frac{R^2}{\Omega}
\left[
r\Upsilon+i(2\pi k-r \epsilon \frac{\pi}{2}\pm L)
\right]\, ,
\qquad
k\in \mathbb{Z}\, ,
\end{equation}
where $r=\pm 1$, and we defined the quantities
\begin{equation}
\epsilon\equiv\operatorname{sgn}\left(p_\theta q_e\right)\, ,\qquad
\Upsilon\equiv\arsinh\left(\frac{|q_e|}{\tilde{m}}\right)\, ,\qquad
L\equiv \arcsin\left(\frac{|p_\theta|}{\Delta}\right)-\frac{\pi}{2}\, .
\end{equation}
Note that these form two infinite arrays in the complex $\tau$-plane, which are no longer symmetric with respect to the imaginary $\tau$-axis. Encircling them induces the monodromies
\begin{equation}\label{eq:dSmonodromies}
\tau_{r,k}^{(\pm)}:\qquad
\theta\to\theta\mp \text{sgn} (p_\theta)\pi\, ,
\qquad
t\to-t\pm r \epsilon i\pi\, ,
\end{equation}
the first following from the logarithms in \eqref{thetaTdS} and the second from the square root, exactly as for $\eta$ and $\chi$ in AdS$_2$. Crucially, and in contrast to the timelike $\eta$-circle of global AdS$_2$, the angle $\theta$ is here a genuine, physically identified coordinate, so that closure requires only returning to the same $\theta$ \emph{modulo} $2\pi$. We also note that the combined monodromy resulting from winding around any single puncture is such that the embedding coordinates defined by \eqref{eq:deSitterHyperboloid} are left invariant, even though both $t$ and $\theta$ are not. Indeed, under \eqref{eq:dSmonodromies} we have
\begin{equation}\label{eq:dSmonodromiesCoshTCos/SinTheta}
\sin \theta\to-\sin \theta\, ,
\qquad \cos \theta\to-\cos \theta\, ,\qquad
\cosh t\to-\cosh t\, ,\qquad \sinh t\to\sinh t\, .
\end{equation}
which preserves the combinations $X^0=R \sinh t$, $X^1=R \cosh t \cos\theta$, and $X^2=R\cosh t \sin \theta$.

\smallskip

Let $w\in\mathbb{Z}$ denote the winding number of $\gamma$ around the compact direction of the cylinder, defined so that all punctures lie on one side of it, and $n_r^{(\pm)}\in\mathbb{Z}$ its winding numbers around the branch points $\tau_{r, 0}^{(\pm)}$, with $r=\pm 1$. The on-shell Lagrangian $L_{\rm inst}=p_t\dot t+p_\theta\dot\theta$ is a single-valued meromorphic function of $\tau$, periodic under \eqref{eq:dSperiod}, with simple poles at $\tau_{r, 0}^{(\pm)}$. A direct evaluation over one period gives
\begin{equation}\label{eq:dSPeriodIntegral}
\int_{\tau}^{\tau+2\pi iR^2/\Omega}d\tau\,L_{\rm inst}=-2\pi i\,\Omega\, ,
\end{equation}
while the residues at the branch points read
\begin{equation}\label{eq:dSResidues}
\mathrm{Res}_{\tau=\tau_{r, 0}^{(\pm)}}\left(L_{\rm inst}(\tau)\right)=-\frac{|q_e|}{2}r\, .
\end{equation}
The on-shell action of a general contour therefore depends only on its homotopy class,
\begin{equation}\label{eq:dSActionWindings}
S_{\rm inst}=-2\pi i\,\Omega\,w
-i\pi |q_e|\sum_{r=\pm 1}r \left(n_r^{(+)}+n_r^{(-)}\right)\, ,
\end{equation}
in complete parallel with the AdS$_2$ result \eqref{eq:GlobalActionWindings}, now continued via $\beta\to i\Omega$. Furthermore, since every closed contour in the punctured cylinder is such that the resulting global trajectory is closed in the hyperboloid, every possible winding arrangement provides, a priori, a genuine instanton saddle for the worldline theory.

\section{Conclusions and Outlook}\label{s:Conclusions}

In this work, we have carried out a systematic study of the complex worldline instantons governing the non-perturbative corrections to the Wilsonian effective action due to massive charged fields in two-dimensional maximally symmetric spacetimes threaded by a homogeneous electric field. Working within the first-quantized worldline formalism, we found that the semiclassical saddles relevant to the one-loop effective action generically live in a complexified configuration space, and that a complete characterization thereof requires promoting the proper-time parameter $\tau$ itself to a complex variable. This viewpoint naturally organizes the solutions into two qualitatively distinct classes \cite{Dumlu:2011cc}: an \emph{instanton phase}, evolving in imaginary proper time, whose action typically governs the exponential suppression of non-perturbative effects such as Schwinger pair production, and an \emph{interference phase}, evolving in real proper time, which encodes the relative phases responsible for quantum interference between saddles. Furthermore, a recurrent theme has been that neither phase is complete on its own: it is only upon analytically continuing $\tau$ over the full complex plane that the two are unified into a single family of trajectories with a common topological classification.

\smallskip

The central technical result is a prescription for constructing and classifying all topologically inequivalent instanton sectors. The periodicity of the solutions makes the complexified proper time a coordinate on a cylinder, and the closed contours on it fall into homotopy classes labeled by a finite set of integers: winding numbers around the compact direction and around a collection of punctures where the spacetime maps become singular. The worldline action depends only on this topological data, and neither on the detailed shape of the contour nor on the continuous integration constants---often related to the Noether charges associated with the problem at hand. We further recast the on-shell action as a sum over residues of the (non-conserved) conjugate momentum, thereby making manifest which analytic data control each contribution: in Anti-de Sitter space, for instance, the pole at the conformal boundary $\rho=0$ yields the term proportional to the instanton winding, whereas the point at infinity in the radial plane supplies the electric-charge contribution. This residue reformulation provides an independent derivation of the topological result and clarifies the physical origin of ~term.

\medskip

A qualitatively new feature emerges in curved spacetimes. In Anti-de Sitter space, the behavior of the solutions is controlled by the dimensionless ratio $R|eE|/m$, which distinguishes the sub-extremal, extremal, and super-extremal regimes \cite{Castellano:2025yur}. We found that these are smoothly connected by the analytic continuation $\beta = \sqrt{\tilde{m}^2-q_e^2}\to i\kappa=i \sqrt{q_e^2-\tilde{m}^2}$ (with $\beta,\kappa\to0$ at extremality), under which the roles of real and imaginary proper time are effectively exchanged and the trigonometric maps of one regime become the hyperbolic maps of the other. A notable consequence is that the sub-extremal regime is the only one to admit instantons whose action has a non-zero real part---consistent with the purely real non-perturbative contribution to the effective action found by the exact computation of \cite{Castellano:2026ojb}. Physically, recalling that each saddle is weighted by $e^{iS_{\rm inst}}$, the imaginary part of the action governs the exponential suppression while any real part contributes an interference phase; the sub-extremal regime is thus the only one whose instantons carry a non-trivial phase. The extremal case sits at the degenerate boundary between the two: the finite period diverges, the instanton winding drops out of the action altogether, and only the charge-dependent residue survives. By contrast, in Minkowski and de Sitter space the qualitative structure of the instantons is insensitive to this or other similar ratios. Indeed, we found that the de Sitter analysis closely mirrors the super-extremal Anti-de Sitter one, a similarity that can be traced to the analytic map relating the two backgrounds; see also \cite{Kim:2008xv}. Finally, we highlighted a subtlety specific to the curved-space problem: a trajectory that closes in Poincar\'e (or flat-slicing) coordinates need not close once lifted to the \emph{universal covering} of the global spacetime. The reason is that the map from Poincar\'e to global coordinates is multi-valued: encircling a single branch point acts by a non-trivial monodromy---for instance $\chi \to i \pi-\chi, \eta \to \eta+\pi$ in AdS$_2$---which leaves the embedding coordinates (and hence $\tan \eta$ and $\sinh \chi$) invariant. As a result, the trajectory appears closed in the periodically identified global chart, yet it might shift the global time $\eta$ and/or the spatial coordinate $\chi$ by a nonzero amount on the covering space. A careful treatment of this monodromy is therefore required to decide which complexified trajectories descend to genuine closed instantons on the universal covering space.

\medskip

Several future directions naturally suggest themselves. First, while we have classified the complex saddles and computed their worldline actions, we have not addressed the \emph{selection problem}: given a physical observable, which of these saddles actually contribute, and with what weights? Answering this requires embedding our classification within the framework of Picard--Lefschetz theory \cite{Witten:2010zr,Cherman:2014ofa,Tanizaki:2014xba,Behtash:2015zha,Behtash:2015loa} and the associated analysis of Stokes phenomena, which would determine the relevant integration cycle and the interference pattern between saddles. We regard this as the most pressing open question, and we intend to return to it, together with the computation of the one-loop fluctuation determinants \cite{Dunne:2005sx,Dunne:2006st}, in future work \cite{PicardLefschetz}.

\smallskip

Second, our analysis has been restricted to two-dimensional backgrounds with a homogeneous field. It would therefore be interesting to extend it to higher-dimensional spaces---for instance to $\mathrm{AdS}_2\times \mathbf{S}^2$ and $\mathrm{dS}_2\times \mathbf{S}^2$ geometries \cite{Castellano:2025ljk,Castellano:2025yur, Castellano:2026ojb, Castellano:2026pwj,Chen:2023swn}, which arise as near-horizon regions of charged black holes and in cosmological contexts, respectively---and less symmetric settings---including non-homogeneous $\U (1)$ gauge backgrounds \cite{Dunne:2005sx,Dunne:2006ur,Dumlu:2011cc,Ilderton:2014mla,Ilderton:2015lsa,Schneider:2018huk} or space- and time-dependent gravitational fields \cite{Parker:2025jef,Hawking:1975vcx,Semren:2025dix}---where the interplay between the electric field and the spacetime curvature may enrich the structure of complex saddles. In particular, it would be important to determine how much of the topological picture uncovered here---i.e., the organization of saddles into homotopy classes on a punctured proper-time cylinder, with actions fixed by winding data---survives in more complicated gauge and gravitational backgrounds, where the additional ingredients may deform or obstruct it. Relatedly, it would be valuable to elucidate the true physical meaning of the punctures \cite{Letter}, whose location and weights govern so much of the classification yet whose interpretation in the original spacetime picture remains to be fully clarified.

\smallskip

Third, the topological organization described above is strongly reminiscent of the structure that appears in the resurgent analysis (see e.g., \cite{Dunne:2015eaa,Aniceto:2018bis} for reviews) of quantum-mechanical and quantum-field-theoretic path integrals. It would be worthwhile to make this analogy precise, and to ask whether the tower of multi-instanton sectors classified here assembles into a resurgent trans-series for the effective action. Crucially, the availability of exact pair-production results in these backgrounds---from the original Schwinger computation in flat space \cite{Schwinger:1951nm} to the closed-form rates obtained in dS$_2$ and AdS$_2$ \cite{Comtet:1984mm,Garriga:1993fh,Garriga:1994bm,Villalba:1995za,Pioline:2005pf,Anninos:2020hfj,Grewal:2021bsu}---provides a concrete benchmark against which any conjectured trans-series for the effective action could be tested. Relatedly, the fact that much of the physical information---the value of the action, the number and weights of the contributing saddles---is encoded in the analytic data of the proper-time plane, rather than in the detailed spacetime trajectories, suggests that the proper-time representation may be the more economical starting point for such an analysis.

\medskip

We hope to report on some of these questions in the near future.

\section*{Acknowledgements}
	
We acknowledge valuable conversations with Lars Aalsma, Christian Aoufia, Jos\'e Calder\'on-Infante, Gonzalo Casas, \'Alvaro Herra\'ez, Kurt Hinterbichler, Guanda Lin, Puxin Lin, Emil Martinec, Georges Obied, Savdeep Sethi, Gary Shiu, Houri-Christina Tarazi, Cumrun Vafa, and Claire Zukowski. A.C. is also grateful to Dieter L\"ust, Carmine Montella, and especially Matteo Zatti for very useful discussions and  collaboration on related topics. J.C. would like to thank Westlake University and B(eijing)IMSA for hospitality during the completion of this work. A.C., E.K., and J.L.S. thank UC Santa Cruz for hospitality during the final stages of this project. The work of A.C. was supported by a Kadanoff and Associate KICP fellowships, by NSF grants PHY-2014195 and PHY-2412985, by a grant from
the Simons Foundation (602883,CV), and by the DellaPietra Foundation. A.C. is also grateful to Teresa Lobo for her continuous encouragement and support. We used Claude and ChatGPT to numerically verify some of our computations and to help edit several sentences in the text. All analytical work in the paper was carried out originally and independently by the authors, who take full responsibility for its contents.

\appendix

\section{Complex Shifts in 2D (Anti-)de Sitter as $\SL(2, \mathbb{C})$ Transformations}\label{ap:SL2C&ComplexShifts}

The aim of this appendix is to relate the invariance of the action for both AdS$_2$ and dS$_2$ instanton solutions (cf.\ Sections~\ref{s:AdS2} and \ref{s:dS2}) under generic constant shifts of the complex proper-time parameter $\tau$ to the conformal isometries of the underlying spacetime. Indeed, the isometry group of 2D (Anti-)de Sitter is given by $\SO(2,1)\cong \SL(2, \mathbb{R})$. However, upon complexifying the configuration space, the relevant transformations are naturally realized in $\SL(2, \mathbb{C})$, as we now show. We treat each case in turn, focusing on the sub-extremal regime for AdS. 

\paragraph{Anti-de Sitter.} We start by writing the embedding map \eqref{eq:RhoSolutionRealTau} and \eqref{eq:TSolutionRealTau} as follows
\begin{equation}\label{eq:Defupm}
 u_+=i \beta\, \frac{\tilde{m} z+\mu}{\tilde{m} z-\mu}\, ,\qquad u_-=i \beta\, \frac{\tilde{m} z-\bar{\mu}}{\tilde{m} z+\bar{\mu}}\, ,
\end{equation}
where $\beta = \sqrt{\tilde{m}^2-q_e^2}$ and we introduced shifted lightcone coordinates $x^\pm(\tau)=t_0+ u_\pm(\tau)/p_t$. Above, we also defined the complex quantities
\begin{equation}
 z=e^{i R^{-2} \beta\,\left(\tau+\tau_0\right)}\, ,\qquad \mu= \beta+i q_e\, .
\end{equation}
This implies that the classical motion along each lightcone coordinate is given by a fractional-linear (i.e., Möbius) transformation of $z(\tau)$, namely $u_\pm = \mathcal{N}_\pm \cdot z$ with (cf.\ \eqref{eq:SL2Maps})
\begin{equation}
  \mathcal{N}_+=\left(\begin{array}{cc}
        i \tilde{m}\beta\ & i \mu \beta \\
        \tilde{m}\ &  -\mu
    \end{array}\right)\, ,\qquad \mathcal{N}_-=\left(\begin{array}{cc}
        i \tilde{m}\beta\ & -i \bar{\mu}\beta \\
        \tilde{m}\ &  \bar{\mu}
    \end{array}\right)\, .
\end{equation}
Next, we note that the shift $\tau_0\to \tau_0 - i \alpha$, with $\alpha \in \mathbb{R}$, acts as $z \to \lambda z$, where $\lambda=\exp{\left(R^{-2} \alpha \beta\right)}$, namely as the dilation matrix in $\SL(2, \mathbb{C})$
\begin{equation}\label{eq:DilationMatrix}
  D_\lambda=\left(\begin{array}{cc}
        \sqrt{\lambda} & 0 \\
        0 &  \frac{1}{\sqrt{\lambda}}
    \end{array}\right)\, .
\end{equation}
Hence, under such rescalings we have that
\begin{equation}
  u_\pm \to u_\pm'=\mathcal{N}_\pm \cdot (\lambda z) \implies u_\pm'=\mathcal{A}_\pm \cdot u_\pm\, ,
\end{equation}
where
\begin{equation}\label{eq:ConformalTransfAdS}
  \mathcal{A}_\pm=\mathcal{N}_\pm \cdot D_\lambda \cdot\mathcal{N}_\pm^{-1}= \frac{1}{2 \sqrt{\lambda}}\left(\begin{array}{cc}
        \lambda+1\ & i \beta (\lambda-1) \\
        -i \frac{\lambda-1}{\beta}\ &  \lambda+1
    \end{array}\right)\, .
\end{equation}
In particular, we notice that, since $\mathcal{A}_+=\mathcal{A}_- \in \SL(2, \mathbb{C})$, the corresponding transformations become a \emph{symmetry} of the classical worldline theory \eqref{SwlAdS} when we allow for a complexified configuration space, i.e., for $x^\pm \in \mathbb{C}$. This consequently explains the observation made in Section~\ref{sss:solutionsSubAdS} regarding the invariance of the instanton action under such constant shifts in $\tau$.

\smallskip

Furthermore, from this perspective, one can also easily understand why the instanton solutions tend to localize around $\rho=0$ and $t= t_\pm = t_0\pm i \beta/p_t$ whenever we take $\tau_0 \to \mp i \infty$. Indeed, the latter can be readily recognized to be the fixed points of $\mathcal{A}_\pm$.\footnote{\label{fnote:AdSfp}One quick way to see this is by noting that $z \to \lambda z$ fixes $z= 0, \infty$, thereby yielding $u_\pm (z \to 0)=-i \beta$ and $u_\pm (z \to \infty)=i \beta$ as the fixed points under complex shifts of the form $\tau_0\to \tau_0 - i \alpha$.}

\paragraph{De Sitter.} The analysis in this case proceeds along identical lines. Using the lightcone coordinates $x^\pm=y\pm\zeta$, the metric takes the conformally flat form
\begin{equation}
	ds^2=\frac{4R^2}{(x^+-x^-)^2}\,dx^+dx^-\, ,
\end{equation}
related to its AdS counterpart by \eqref{eq:dS2AdS2Mapping}, under which $x^\pm_{\rm AdS}\mapsto x^\mp_{\rm dS}$ and $R^2_{\rm{AdS}_2}\to -R^2_{\rm{dS}_2}$. Consequently, we can write the spacetime embedding maps \eqref{eq:zetaSolutiondS2} and \eqref{eq:ySolutiondS2} in terms of $x^\pm(\tau)=y_0+v_\pm(\tau)/p_y$ as follows
\begin{equation}\label{eq:Defvpm}
 v_+=-\Omega\,\frac{\tilde{m}\,w+\nu_-}{\tilde{m}\,w-\nu_-}\, ,\qquad
 v_-=-\Omega\,\frac{\tilde{m}\,w-\nu_+}{\tilde{m}\,w+\nu_+}\, ,
\end{equation}
where $\Omega=\sqrt{\tilde{m}^2+q_e^2}$ and we defined
\begin{equation}
 w=e^{ R^{-2}\Omega\left(\tau+\tau_0\right)}\, ,\qquad
 \nu_\pm=\Omega\pm q_e\, .
\end{equation}
Comparing with \eqref{eq:Defupm}, the imaginary exponent in $z$ has been replaced by the real one in $w$, and the complex number $\mu=\beta+iq_e$ by the real pair $\nu_\pm$. As before, each lightcone coordinate is a fractional-linear transformation of $w(\tau)$, namely $v_\pm=\mathcal{M}_\pm \cdot w$ with
\begin{equation}
  \mathcal{M}_+=\left(\begin{array}{cc}
        -\Omega\,\tilde{m}\ & -\Omega\,\nu_- \\
        \tilde{m}\ &  -\nu_-
    \end{array}\right)\, ,\qquad \mathcal{M}_-=\left(\begin{array}{cc}
        -\Omega\,\tilde{m}\ & \Omega\,\nu_+ \\
        \tilde{m}\ &  \nu_+
    \end{array}\right)\, .
\end{equation}
The shift $\tau_0\to \tau_0 + \alpha$, with $\alpha \in \mathbb{R}$, acts as
$w \to \lambda w$, where $\lambda=\exp{\left(R^{-2}\alpha\,\Omega\right)}$, namely as the dilation matrix \eqref{eq:DilationMatrix}. The contrast with AdS is instructive: because the proper-time dependence enters through a real exponential rather than a phase, the symmetry-generating shift is \emph{real} in de Sitter but \emph{imaginary} in Anti-de Sitter, one more instance of the real$\leftrightarrow$imaginary exchange relating de Sitter to the super-extremal regime. Conjugating the dilation then yields
\begin{equation}
  v_\pm \to v_\pm'=\mathcal{M}_\pm \cdot (\lambda w) \implies v_\pm'=\mathcal{A}_\pm \cdot v_\pm\, ,
\end{equation}
where
\begin{equation}\label{eq:ConformalTransfdS}
  \mathcal{A}_\pm=\mathcal{M}_\pm \cdot D_\lambda \cdot\mathcal{M}_\pm^{-1}=
  \frac{1}{2 \sqrt{\lambda}}\left(\begin{array}{cc}
        \lambda+1\ & -\Omega\,(\lambda-1) \\
        -\frac{\lambda-1}{\Omega}\ &  \lambda+1
    \end{array}\right)\, .
\end{equation}
This is precisely the analytic continuation of the AdS result \eqref{eq:ConformalTransfAdS} under $\beta\to i\Omega$, induced by the map \eqref{eq:dS2AdS2Mapping}. Once again $\mathcal{A}_+=\mathcal{A}_-\in \SL(2,\mathbb{C})$, so that the shift acts as a single fractional-linear transformation on both null coordinates and is therefore a symmetry of the worldline action \eqref{SwldS}, when extended to the complexified configuration space $x^\pm\in\mathbb{C}$. This accounts for the independence of the de Sitter action on the constant $\tau_0$ (see Section~\ref{ss:MixedSolutionsdS2} for details).\footnote{This invariance holds only within a given family of solutions. Although $\mathcal{A}_\pm$ is a conformal isometry for all $\lambda$, the action \eqref{SwldS} is a contour integral fixed by the winding of the path with the punctures \eqref{eq:PuncturesdeSitter}, located at $w=\nu_-/\tilde{m}$ and $w=-\nu_+/\tilde{m}$, where $v_+$ and $v_-$ respectively develop poles. The dilation $w\to\lambda w$ preserves the action only if it does not drag the contour across such a pole. This fails at the boundary $|\mathrm{Re}(\tau_0)|=\phi$, where the winding data jumps and one recovers the transition between the two branches of \eqref{eq:dSInstantonActionFamilies}.}

\smallskip

Finally, the same fixed-point analysis explains the localization of the de Sitter instantons: the dilation fixes $w=0,\infty$, and both null coordinates take the common values\footnote{Both maps \eqref{eq:Defvpm} share the fixed points $\{+\Omega,-\Omega\}$, which is what guarantees $\mathcal{A}_+=\mathcal{A}_-$; the corresponding AdS statement is footnote~\ref{fnote:AdSfp}.}
\begin{equation}
  v_\pm(w\to 0)=+\Omega\, ,\qquad v_\pm(w\to\infty)=-\Omega\, .
\end{equation}
Since $v_\pm=\pm\Omega$ forces $\zeta=0$ together with $p_y(y-y_0)=\pm\Omega$, the trajectories collapse onto the future-infinity boundary $\zeta=0$ and localize at $y=y_\pm$, with $p_y y_\pm=p_y y_0\pm\Omega$, exactly the limiting points introduced around~\eqref{eq:EffectivePotentialdS}. These are reached as $\tau_0\to\mp\infty$ along the real axis.

\bibliography{ref.bib}
\bibliographystyle{JHEP}

\end{document}